\documentclass[12pt]{article}
\usepackage{amsmath, amsfonts,amsbsy,amssymb}
\usepackage{graphicx,psfrag,epsf}
\usepackage{natbib}
\usepackage{booktabs} 
\usepackage{bbm}
\usepackage[colorlinks=true,linkcolor=blue,citecolor=blue,urlcolor=blue]{hyperref}
\usepackage{algorithmicx}
\usepackage{algpseudocode}
\newtheorem{theorem}{Theorem}[section]

\newtheorem{lemma}{Lemma}[section]

\newcommand{\prob}{\mathbb{P}}

\usepackage{multibib}
\newcites{app}{References}% use \citeapp{} and \citepapp[][]{}
\newcommand{\blind}{0} 

\begin{document}

\def\spacingset#1{\renewcommand{\baselinestretch}%
{#1}\small\normalsize} \spacingset{1}

%%%%%%%%%%%%%%%%%%%%%%%%%%%%%%%%%%%%%%%%%%%%%%%%%%%%%%%%%%%%%%%%%%%%%%%%%%%%%%

\if0\blind
{
  \title{\bf Parameter Regimes in Partial Functional Panel Regression}

  \author{
  	Dominik Liebl\thanks{Institute for Financial Economics and Statistics, University of Bonn}, and
    Fabian Walders\thanks{BGSE \& Institute for Financial Economics and Statistics, University of Bonn}
  }
  \date{}
  \maketitle
} \fi

\if1\blind
{
  \vspace*{3cm}
  
  \begin{center}
    {\LARGE\bf Parameter Regimes in Partial Functional\\[1.5ex]Panel Regression}
\end{center}
  \medskip
} \fi

\bigskip
\begin{abstract}
A new partial functional linear regression model for panel data with time varying parameters is introduced. The parameter vector of the multivariate model component is allowed to be completely time varying while the function-valued parameter of the functional model component is assumed to change over $K$ unknown parameter regimes. 
%%%%%%%%%%%%
Consistency is derived for the suggested estimators and for the classification procedure used to detect the $K$ unknown parameter regimes. Additionally, the convergence rates of the estimators are derived under a double asymptotic differentiating between asymptotic scenarios depending on the relative order of the panel dimensions $n$ and $T$.
%%%%%%%%%%%%
The statistical model is motivated by a real data application considering the so-called ``idiosyncratic volatility puzzle'' using high frequency data from the S\&P500.
\end{abstract}

\noindent%
\vfill

\newpage

\spacingset{1}

%%%%%%%%%%%%%%%%%%%%%%%%%%%%%%%%%%%%%%%%%%%%
\section{Introduction}\label{sec:intro}
%%%%%%%%%%%%%%%%%%%%%%%%%%%%%%%%%%%%%%%%%%%%
The availability of mixed, i.e., functional and multivariate data types and the need to analyze such data types appropriately, has trigged the development of new statistical models and procedures. In this work we consider the so-called partial functional linear model for scalar responses, which combines the functional linear regression model \citep[see, e.g.,][]{HH07} with the multivariate regression model. This model was first proposed by \cite{ZXS07} and \cite{STL08}---two mixed effects modeling approaches. The first theoretical work is by \cite{S09}, who uses a functional-principal-components-based estimation procedure and derives convergence rates for the case of independent cross sectional data. Recently, the partial functional linear regression model was extended in several directions. \cite{SL12} consider the case of prediction, \cite{LDS14} and \cite{TC14} focus on quantile regression, \cite{KXYZ16} consider the case of a high-dimensional multivariate model component, \cite{PZT16} allow for varying coefficients in the multivariate model component, and \cite{WFC16} and \cite{DLXZ17} are concerned with a functional single-index model component.

Motivated by our real data application, we contribute a new partial functional linear panel regression model with time-varying parameters allowing for $K<\infty$ latent parameter regimes, which can be estimated from the data. In the theoretical part of this work we show consistency of our estimators and of our unsupervised classification procedure identifying the $K$ parameter regimes. In addition, we derive convergence rates of the regression slope estimators under a double asymptotic, for which we differentiate among different asymptotic scenarios depending on the relative order of the panel dimensions $n$ and $T$. The consideration of time-varying parameters is quite novel in the literature on functional data analysis. To the best of our knowledge, the only other work concerned with this issue is \cite{HR12}, who focus on testing the hypothesis of a time constant parameter function in the case of a classical fully-functional regression model.

Closely related to the partial functional linear model is the so-called Semi-Functional Partial Linear (SFPL) model proposed by \cite{AV06}, where the functional component consists of a nonparametric functional regression model instead of a functional linear regression model. The SFPL model is further investigated by \cite{AV08}, \cite{L11}, \cite{ZC12}, and \cite{AV13}, among others. Readers with a general interest in functional data analysis are referred to the textbooks of \cite{RS05}, \cite{FV06}, \cite{HK12} and \cite{HR15_book}.

The usefulness of our model and the applicability of our estimation procedure is demonstrated by means of a simulation study and a real data application. For the latter we consider the so-called ``idiosyncratic volatility puzzle'', an empirical phenomenon occurring in stock markets. This puzzle was first described in \cite{AHXZ06} and concerns the empirical observation that the idiosyncratic, i.e., non-systematic volatility component of stocks is typically negatively correlated with the stock returns. This observation is puzzling, since asset pricing theory predicts either no correlation, if investors hold well-diversified portfolios, or a positive correlation, if investors hold underdiversified portfolios. We use our novel model to assess time instabilities in this empirical phenomenon using high frequency stock-level data from the S\&P 500. We calculate a functional measure for the idiosyncratic volatility component, and allow its impact to vary over a latent set of time regimes. Our model allows us to consider the idiosyncratic volatility puzzle at a much less aggregated time scale than considered so far in the literature. This leads to new insights into the temporal heterogeneity in the pricing of idiosyncratic volatility in equity markets.

The remainder of this work is structured as follows. In Sections \ref{sec:MOD} and \ref{sec:EST} we introduce the model and present the estimation procedure. Section \ref{sec:ASY} contains our main assumptions and asymptotic results. Section \ref{sec:THR} discusses the practical choice of the tuning parameters involved. The finite sample performance of the estimators is explored in Section \ref{sec:SIM}. Section \ref{sec:APL} contains our real data application and Section \ref{sec:CON} a short conclusion. All proofs can be found in the online supplement supporting this article.

%%%%%%%%%%%%%%%%%%%%%%%%%%%%%%%%%%
\section{Model}\label{sec:MOD}
%%%%%%%%%%%%%%%%%%%%%%%%%%%%%%%%%%
We introduce a partial linear regression model for panel data, which allows us to model the time-varying effect of a square integrable random function $X_{it}\in L^2([0,1])$ on a scalar response $y_{it}\in \mathbb{R}$ in the presence of a random, finite dimensional explanatory variable $z_{it}\in \mathbb{R}^P$. Indexing the cross section units $i=1,\dots,n$ and time points $t=1,\dots,T$, our statistical model reads as
%%%%%%%%%%%%%%%%%
\begin{align} 
y_{it} =\rho_{t} + \int_{0}^{1} \alpha_t(s)X_{it}(s) \mathrm{d}s  + \beta_t^\top z_{it} + \epsilon_{it},\label{Mod} 
\end{align}
%%%%%%%%%%%%%%%%%
where $\rho_{t}$ is a time fixed effect, $\alpha_t\in L^2([0,1])$ is a time-varying deterministic functional parameter, $\beta_t\in\mathbb{R}^P$ is a time-varying deterministic parameter vector, and $\epsilon_{it}$ is a scalar error term with zero mean and finite but potentially time heteroscedastic variances (see also our assumptions in Section \ref{sec:ASY}).

The unknown function-valued parameters $\alpha_t$, $1\leq t \leq T$, are assumed to differ only across unknown time regimes $G_k \subset \{1,\dots,T\}$. That is, each regime $G_k$ is associated with a regime specific parameter function $A_k\in L^2([0,1])$, such that
%%%%%%%%%%%%%%%%
\begin{align} 
\alpha_t(s)\equiv A_k(s) \quad \text{if} \quad t \in G_k. \label{Ak}
\end{align}
%%%%%%%%%%%%%%%%
The regimes $G_1,\dots,G_K$ form a partition of the set of periods $\{1,\dots,T\}$ and do not have to consist of subsequent periods $t$. The number of regimes $K$ is fixed and does not depend on the number of points in time $T$. For our theoretical analysis in Section \ref{sec:ASY}, we also allow the joint and the marginal distributions of $X_{it}$, $z_{it}$ and $\epsilon_{it}$ to vary over the different regimes $G_k$.

Model \eqref{Mod} is motivated from our real data application, where $G_k$ is a collection of time points $t$, which belong to the $k$-th volatility pricing regime. The $k$-th pricing regime is characterized by the function-valued slope parameter $A_k$ describing the effect of the functional idiosyncratic volatility curve $X_{it}$ on the scalar stock price return $y_{it}$. We allow for autocorrelations between volatility curves $X_{is}$ and $X_{it}$, $s\neq t$, from the same stock $i$, but assume independence between different stocks $i$ and $j$, $i\neq j$. The latter independence assumption is justified as we consider the idiosyncratic, i.e., non-systematic volatility curves after controlling for the systematic market components of stock $i$ (see Section 7 for details).

Model \eqref{Mod} nests several different specifications. It might be the case that $K=1$ and hence $G_1=\{1,\dots,T\}$. In this situation the effect of the random function on the response is time invariant. The classical functional or the classical multivariate linear regression model are obtained if $\beta_t=0$ or $\alpha_t=0$ for all $t=1,\dots,T$.

%%%%%%%%%%%%%%%%%%%%%%%%%%%%%%%%%%%%%%%%%
\section{Estimation}\label{sec:EST}
%%%%%%%%%%%%%%%%%%%%%%%%%%%%%%%%%%%%%%%%%
Our objective is to estimate the model parameters $A_k$, $\beta_t$, and the regimes $G_1,\dots,G_K$ from a sample $\{(y_{it},X_{it},z_{it}): 1\leq i \leq n, \, 1\leq t \leq T \}$. For this purpose, we suggest a three-step estimation procedure. The first step is a pre-estimation step where Model \eqref{Mod} is fitted to the data separately for each $t=1,\dots,T$. This pre-estimation step reveals information about the regime memberships, which is used in the second step, where we apply our unsupervised classification procedure in order to estimate the regimes $G_1,\dots,G_K$. The third step is the final estimation step, in which we improve the estimation of the functional parameter $A_k$ by employing information about the regime membership gathered in step two. The general procedure is inspired by the work of \cite{VL17}, but differs from it as we consider a functional data context which demands for a different estimation procedure. In the following we explain the three estimation steps in more detail:

%%%%%%%%%%%%%%%%%%%%%%%%%%%%%%%%%%%%%
\smallskip\noindent\textbf{Step 1.} 
%%%%%%%%%%%%%%%%%%%%%%%%%%%%%%%%%%%%%
In this step, we pre-estimate the parameters $\alpha_t$ and compute the final estimates of $\beta_t$ separately for each $t=1,\dots,T$. Estimation starts from removing the fixed effect $\rho_t$ using a classical within-transformation. For this we denote the centered variables as $y_{it}^c=y_{it}-\bar y_t$, $X_{it}^c=X_{it}-\bar X_{t}$, $z_{it}^c=z_{it}-\bar z_{t}$, and $\epsilon_{it}^c=\epsilon_{it}-\bar \epsilon_{t}$, where $\bar y_t=n^{-1}\sum_{i=1}^n y_{it}$, $\bar X_{t}=n^{-1}\sum_{i=1}^n X_{it}$, $\bar z_t=n^{-1}\sum_{i=1}^n z_{it}$, and $\bar \epsilon_t=n^{-1}\sum_{i=1}^n \epsilon_{it}$. Then, the within-transformed version of Model \eqref{Mod} is
\begin{align*} 
y^c_{it}=\int_{0}^{1} \alpha_t(u)X^c_{it}(u)\mathrm{d}u  + \beta_t^\top z^c_{it} + \epsilon^c_{it}.
\end{align*}
By adapting the methodology in \cite{HH07}, we estimate the slope parameter $\alpha_t$ using $t$-wise truncated series expansions of $\alpha_t$ and $X_{it}^c$, i.e., 
%%%%%%%%%%
\begin{align*}
\alpha_t(s)&\approx\sum_{j=1}^{m_t}a_{j,t}\hat{\phi}_{j,t}(s) \quad \text{ where } \quad  a_{j,t}:=\langle \alpha_t, \hat{\phi}_{j,t} \rangle, \ 1 \leq j \leq m_t \\
\text{and} \qquad X^c_{it}(s)&=\sum_{j=1}^{n}\langle X_{it}^c, \hat \phi_{j,t} \rangle\hat \phi_{j,t}(s) \approx\sum_{j=1}^{m_t}\langle X_{it}^c, \hat \phi_{j,t} \rangle\hat \phi_{j,t}(s),
\end{align*} 
%%%%%%%%%%% 
%%%%%%%%%%%
%%%%%%%%%%%
which can be used to approximate the functional $\int_{0}^{1} \alpha_t(u)X^c_{it}(u)\mathrm{d}u$ in the regression equation by $\sum_{j=1}^{m_t} \langle X_{it}^c, \hat \phi_{j,t} \rangle a_{j,t}$. Here, $\langle \cdot, \cdot \rangle$ is the inner product in $L^2([0,1])$ and $\hat \phi_{j,t}$ denotes the eigenfunction corresponding to the $j$-th largest eigenvalue $\hat{\lambda}_{j,t}$ of the empirical covariance operator $\hat{\Gamma}_t$ of $\{X_{it}: \ 1 \leq i \leq n\}$ defined as
\begin{align*}
(\hat \Gamma_tx)(u)&:=\int_0^1\hat{K}_{X,t}(u,v)x(v)\mathrm{d}v\quad\text{for any}\quad x\in L^2([0,1])\\ 
\text{and}\quad\hat{K}_{X,t}(u,v)&:=\frac{1}{n}\sum_{i=1}^n X_{it}^c(u)X_{it}^c(v).
\end{align*}
The empirical eigenfunctions $\hat{\phi}_{j,t}$ and eigenvalues $\hat{\lambda}_{j,t}$ are defined as the solutions of the eigenequations $\int_0^1 \hat{K}_{X,t}(u,v)\hat{\phi}_{j,t}(v)\mathrm{d}v=\hat{\lambda}_{j,t}\hat{\phi}_{j,t}(u)$, where $\langle\hat{\phi}_{j,t},\hat{\phi}_{\ell ,t}\rangle=1$ for all $j=\ell$ and $\langle\hat{\phi}_{j,t},\hat{\phi}_{\ell ,t}\rangle=0$ if $j\neq \ell$, with $j,\ell\in\{1,\dots,n\}$.

Using the above approximations for $\alpha_t(s)$ and $X^c_{it}(s)$ and the orthonormality of the eigenfunctions leads to the following approximate multivariate linear regression model:
\begin{align*} 
y^c_{it}\approx \sum_{j=1}^{m_t}a_{j,t} \langle X_{it}^c, \hat \phi_{j,t} \rangle  + \beta_t^\top z^c_{it} + \epsilon^c_{it},
\end{align*}
where the approximation is due to the truncation bias. Here, $\langle X_{it}^c, \hat \phi_{j,t} \rangle$ and $z_{it}^c$ denote the regressors and $a_{j,t}$ and $\beta_t$ denote the slope coefficients which can be estimated using ordinary least squares estimators $\hat a_{j,t}$ and $\hat\beta_t$. The estimators $\hat a_{j,t}$ lead to the estimator of $\alpha_t(s)$ by 
\begin{align*}
\hat \alpha_t(s)=\sum_{j=1}^{m_t} \hat a_{j,t} \hat \phi_{j,t}(s).
\end{align*}
The closed form solutions of $\hat a_{j,t}$ and $\hat\beta_t$ can be found at the beginning of Appendix \ref{sec:APA} of the online supplement supporting this article.

For our theoretical analysis, we let $m_t=m_{t,nT}\to\infty$ as $n,T\to\infty$. In practice, the cut-off parameter $m_t$ can be chosen, for instance, by Cross Validation (CV) or by a suitable information criterion as introduced in Section \ref{sec:THR}.

Besides obtaining the final estimators $\hat\beta_t$ for $\beta_t$, this first estimation step is intended to facilitate the classification procedure in Step 2. With such classification we aim to distinguish systematically large from systematically small differences between estimated functions $\hat \alpha_t$ and $\hat \alpha_s$ across different time points $t\neq s$. For this purpose one could compare the magnitude of the differences between the functions $\hat \alpha_t$ and $\hat \alpha_s$ to an appropriate threshold. However, the estimators $\hat\alpha_t,\hat\alpha_s$ are not well suited for deriving a practically useful threshold parameter. We, therefore, suggest the following transformed estimators, for which it is straightforward to derive a practically useful threshold parameter using distributional arguments (see Section \ref{sec:THR}):
%%%%%%%%%%%%%%%
\begin{align}
\hat \alpha_t^{(\Delta)}:=\sum_{j=1}^{\underline{m}} \frac{\hat \lambda_{j,t}^{1/2}}{\hat \sigma_{\epsilon,t}} \hat a_{j,t} \hat\phi_{j,t}, \label{dalph}
\end{align}
%%%%%%%%%%%%%%%
where $\hat \sigma_{\epsilon,t}^2:=n^{-1}\sum_{i=1}^n \big(y^c_{it}-\langle \hat \alpha_t,X_{it}^c \rangle+\hat \beta_t^\top z_{it}^c \big)^2$ and $\underline{m}:=\min_{1\leq t\leq T}m_t$. The scaled estimator $\hat \alpha_t^{(\Delta)}$ allows for a simple derivation of threshold parameter (see Section \ref{sec:THR}). Alternatively, the recent inferential results in \cite{GP17} and \cite{CR18} may be used to derive a threshold parameter based on the unscaled estimator $\hat\alpha$, but this is out of the scope of this paper.

%%%%%%%%%%%%%%%%%%%%%%%%%%%%%%%%%%%
\smallskip\noindent\textbf{Step 2} 
%%%%%%%%%%%%%%%%%%%%%%%%%%%%%%%%%%%
In this step, we use the scaled estimators $\hat \alpha_t^{(\Delta)}$ from \eqref{dalph} to classify time points $t=1\dots,T$ into regimes $G_1,\dots,G_K$. Our classification algorithm aims to detect systematic differences in the empirical distances $\hat \Delta_{ts}:=||\hat \alpha_t^{(\Delta)}-\hat \alpha_s^{(\Delta)} ||_2^2$, where $||.||_2^2$ denotes the squared  $L^2$ norm defined as $||x||_2^2=\langle x,x\rangle$ for any $x\in L^2([0,1])$.

The algorithm detects regimes by iteratively searching for large differences $\hat \Delta_{ts}$. If $\hat \Delta_{ts}$ exceeds the value of a threshold parameter $\tau_{nT}>0$, it classifies time points $t$ and $s$ in different regimes. The procedure is initialized by setting $S^{(0)}:=\{1,\dots,T\}$ and iterates over $k = 0,1,2,\dots$ as follows:
%%%%%%%%%%%%%%%%%%%%% 
\begin{algorithmic}
\While {$\left|S^{(k)}\right|>0$}
\State select any $t\in S^{(k)}$, $\hat G_{k+1}\gets\emptyset$, $S^{(k+1)}\gets \emptyset$  
     \For {$s \in S^{(k)}$}
           \If {$\hat \Delta_{ts}\leq \tau_{nT}$}
            \State $\hat G_{k+1}\gets \hat G_{k+1} \cup \{s\}$  
            \Else \  $S^{(k+1)} \gets S^{(k+1)} \cup \{s\}$ 
      \EndIf
\EndFor
\EndWhile
\end{algorithmic}
%%%%%%%%%%%%%%%%%%%%%
where $\left|.\right|$ is used to denote the cardinality of a set. The algorithm stops as soon as all time points $t$ are classified into regimes and the total number $\hat K$ of estimated regimes $\hat G_1,\dots,\hat G_{\hat{K}}$ serves as a natural estimator for the true $K$. Our theoretical results show that this procedure consistently estimates the true regimes $G_k$ and the true number $K$. However, in order to improve the classification in finite samples, we suggest to set an upper bound $K_{\max}$ on $\hat K$, such that $\hat K \leq K_{\max}$. The practical choice of $K_{\max}$ is described in Section \ref{sec:THR}. In the case where $K_{\max}$ is binding, the algorithm is stopped after $K_{\max}-1$ iterations and all remaining time points $t$ are assigned to a final regime $\hat G_{K_{\max}}$. The structure of our unsupervised classification procedure allows to derive rigorous theoretical results. In applications, however, one may use any other well-performing unsupervised classification procedure too.\\

%%%%%%%%%%%%%%%%%%%%%%%%%%%%%%%%%%%%%%
\smallskip\noindent\textbf{Step 3.} 
%%%%%%%%%%%%%%%%%%%%%%%%%%%%%%%%%%%%%%
In this step, we build upon the regime structure determined in Step 2 in order to estimate $A_k$, $k =1,\dots,\hat K$. For a regime $k$ and any $t \in \hat G_k$, let $X_{it}^{cc}$ denote the regime specific centered functional regressor defined as $X_{it}^{cc}:=X_{it}-|\hat G_k|^{-1}\sum_{s\in \hat G_k}\bar{X}_{s}$. Further we define the corresponding $k$-specific empirical covariance operator $\tilde \Gamma_k$ by
\begin{align*}
(\tilde \Gamma_kx)(u)&:=\int_0^1\tilde{K}_{X,k}(u,v)x(v)\mathrm{d}v\quad\text{for all}\quad x\in L^2([0,1]),\\ 
\text{where}\quad\tilde{K}_{X,k}(u,v)&:=\frac{1}{n|\hat G_k|}\sum_{i=1}^n \sum_{t\in \hat G_k} X_{it}^{cc}(u)X_{it}^{cc}(v).
\end{align*}
We obtain our final estimator $\tilde A_k$ for $A_k$, in analogy to the pre-estimator $\hat{\alpha}_t$, as  
%%%%%%%%%%%%%%%
\begin{align*}
\tilde A_k&=\sum_{j=1}^{\tilde m_k} \tilde a_{j,k} \tilde \phi_{j,k}, \quad \text{with} \quad \tilde a_{j,k}= {\tilde \lambda_{j,k}}^{-1} \frac{1}{n|\hat G_k|}\sum_{i=1}^n \sum_{t\in \hat G_k}  \langle \tilde\phi_{j,k},X_{it}^{cc}\rangle (y_{it}^c-\hat \beta_t^\top z_{it}^c).
\end{align*}
%%%%%%%%%%%%%% 
Here $(\tilde{\lambda}_{j,k},\tilde{\phi}_{j,k})_{1 \leq j\leq n |\hat G_k|}$ denote the eigenvalue-eigenfunction pairs of the empirical covariance operator $\tilde \Gamma_k$, where $\tilde{\lambda}_{j,k}$  is the $j$-th largest eigenvalue. Again, for our theoretical analysis, we let $\tilde m_k= m_{k,nT}\to\infty$ as $n,T\to\infty$. In practice the cut-off parameter $\tilde m_k$ can be chosen, for instance, by CV or by a suitable information criterion as introduced in Section \ref{sec:THR}. Note that we do not re-estimate $\hat\beta_t$ in Step 3, since this can lead to biased estimates as the parameter $\beta_t$ is assumed $t$-specific. The assumption of a $t$-specific parameter $\beta_t$ is motivated from our real data application, where $z_{it}$ contains control variables for which a regime structure $G_k$ does not necessarily apply.

%%%%%%%%%%%%%%%%%%%%%%%%%%%%%%%%%%%%%%%%%%%%
\section{Asymptotic Theory}\label{sec:ASY}
%%%%%%%%%%%%%%%%%%%%%%%%%%%%%%%%%%%%%%%%%%%%
In the asymptotic analysis of our estimators we need to address two problems: first, there is a classification error contaminating the estimation of $A_k$. Second, the estimation of the $t$-specific parameters $\beta_t$ cannot be separated from the estimation of the regime specific parameter $A_k$. In the following we list our theoretical assumptions:

\begin{description}
\item[A1]
\begin{enumerate}
%%%%%%%%%%%%%%%%%
\item For every $1 \leq k \leq K$, the random variables $\{(X_{it},z_{it},\epsilon_{it}): \ 1\leq i \leq n, \ t\in G_k \}$ are strictly stationary and further independent over the index $i$ for any $t \in G_k$. Beyond that, the errors $\epsilon_{it}$ are centered and also independent over the index $1\leq t \leq T$.
%%%%%%%%%%%%%%%%%
\item For every $1\leq k \leq K$ and $1 \leq i \leq n$, the random variables $\{X_{it}: \ t\in G_k \}$ are $L^4$-m-approximable in the sense of Definition 2.1 in \cite{HK10}, which implies that $||E\left[ X_{it}^4 \right]||_2<\infty$. Furthermore, it is assumed that $E\left[z_{it}^4 \right]<\infty$, $E\left[\epsilon_{it}^4 \right]<\infty$ for any $1\leq i \leq n$ and $1\leq t \leq T$.
%%%%%%%%%%%%%%%%%
\item For every $1\leq k \leq K$ and $1 \leq i \leq n$, the random variables $\{z_{it}: \ t\in G_k \}$ are m-dependent. 
%%%%%%%%%%%%%%%%%
\item The error $\epsilon_{it}$ is independent of the covariates $X_{js}$ and $z_{js}$ for any $1\leq i,j \leq n$ and $1\leq t,s \leq T$.
\end{enumerate}

\item[A2] Suppose there exist constants $0< C_\lambda,C_\lambda',C_\theta,C_a ,C_{zX},C_\beta<\infty$, such that the following holds for every $1\leq k \leq K$:
\begin{enumerate}
\item $C_\lambda^{-1} j^{-\mu} \leq \lambda_{j,k} \leq C_\lambda j^{-\mu}$ and $\lambda_{j,k}-\lambda_{j+1,k}\geq C_\lambda' j^{-(\mu+1)}$, $j \geq 1$ for the eigenvalues $\lambda_{1,k}>\lambda_{2,k}>\dots$ of the covariance operator $\Gamma_k$ of $X_{it}$, $t\in G_k$ and a $\mu>1$,
%%%%%%%%%
\item  $E\left[ \langle X_{it}-E[X_{it}],\phi_{j,k} \rangle^4 \right]\leq C_\theta \lambda_{j,k}^2$ for the eigenfunction $\phi_{j,k}$ of $\Gamma_k$ corresponding to the $j$-th largest eigenvalue $\lambda_{j,k}$, $j \geq 1$,
%%%%%%%%%%
\item $|\langle A_k, \phi_{j,k} \rangle|\leq C_a j^{-\nu}$, $j\geq 1$, where $\nu$ is a strictly positive constant (see A4).
%%%%%%%%%%%
\item $|\langle { K}_{z_pX,k},\phi_{j,k} \rangle|\leq C_{zX}j^{-(\mu+\nu)}$, $j \geq 1$ for any $1\leq p \leq P$, where $K_{z_pX,k}:=E[(X_{it}-E[X_{it}])(z_{p,it}-E(z_{p,it}))]$ and
%%%%%%%%%
\item $\sup_{1\leq t \leq T}\beta_{p,t} \leq C_\beta$, for any $1\leq p \leq P$, with $\beta_{p,t}$ being the $p$-th coordinate in $\beta_{t}$.
\end{enumerate}

\item[A3]
Let $n\to\infty$ and $T\to\infty$ jointly, such that $T\propto n^\delta$ for some $0<\delta<1$ and $|G_k|\propto T$.
\item[A4] Suppose that $\nu>3\max\{r_1,r_2\}$, where $r_1:=1+\frac{1}{2}\mu$ and $r_2:=\frac{1+\mu (1+\delta)/3}{2(1-\delta)}$.
\item[A5] Suppose that $m_t=m_{t,nT}$ and $\tilde m_k=\tilde m_{k,nT}$ with $m_t\propto n^{\frac{1}{\mu+2 \nu}}$ and $\tilde m_k  \propto (n|G_k|)^{\frac{1}{\mu+2\nu}}$ for any $1\leq t \leq T$ and $1 \leq k \leq K$.
\item[A6] Consider the random vector $\mathbf{s}_{it}:=[s_{1,it},\dots,s_{P,it}]^\top$, defined according to
\begin{align*}
s_{p,it}:=(z_{p,it}-E[z_{p,it}]) - \int_0^1 (X_{it}(u)-E[X_{it}](u))\left(\sum_{j= 1}^\infty \frac{\langle K_{z_pX,k},\phi_{j,k} \rangle}{\lambda_{j,k}} \phi_{j,k}(u)\right)\mathrm{d}u,
\end{align*}
for $1\leq p \leq P$. Suppose that for any $1\leq k \leq K$ the random variables $\{\mathbf{s}_{it}: \ 1\leq i \leq n, \ t \in G_k\}$ are strictly stationary and further independent over the index $i$ for any $t \in G_k$. Also, suppose they are strictly stationary, ergodic and m-dependent over the index $t$ for any $1\leq i \leq n$. In addition, assume that $E[\mathbf{s}_{it}|\textbf{X}_k]=\mathbf{0}$, where $\mathbf{X}_k:=\{X_{it}: \ 1\leq i \leq n, \ t \in G_k\}$ and that the matrix $E[\mathbf{s}_{it}\mathbf{s}_{it}^\top]$ is positive semi-definite. 
\item[A7]
\begin{enumerate}
\item There exists some $C_{\Delta} > 0$ such that for any $1\leq k\leq K$ and any $t \in G_k$
\begin{align*}
\left|\left|\alpha_t^{(\Delta)}-\alpha_s^{(\Delta)} \right|\right|_2^2=:\Delta_{ts}
\begin{cases}
\geq C_{\Delta} & \text{if }  s \not\in G_k\\
= 0             & \text{if }  s \in G_k,
\end{cases}
\end{align*}
where $\alpha_r^{(\Delta)}:=\sigma_{\epsilon,l}^{-1} \sum_{j=1}^\infty \lambda_{j,l}^{1/2} \langle \alpha_r,\phi_{j,l} \rangle \phi_{j,l}$ and $\sigma_{\epsilon,l}^2 := E[\epsilon_{ir}^2]$ for $r\in G_l$.

\item The threshold parameter $\tau_{nT}\to 0$ satisfies $\prob \left( \max_{t,s \in G_k} \hat \Delta_{ts} \leq \tau_{nT} \right)\to 1$ as $n,T\to\infty$ for all $1\leq k \leq K$.
\end{enumerate}
\end{description}

Beyond the above assumptions we also suppose that the sign of the estimated eigenfunctions from Step 1 and Step 3 of our estimation procedure coincide with their population counterparts in the sense that $\int_{0}^1 \hat \phi_{j,t}(u)\phi_{j,k}(u)\mathrm{d}u\geq 0$, $1 \leq j \leq m_t$, and $\int_{0}^1 \tilde \phi_{j,k}(u)\phi_{j,k}(u)\mathrm{d}u\geq 0$, $1 \leq j \leq \tilde m_k$.

Assumptions A1-A6 correspond to the standard assumptions in the literature (see \citealp{HH07} and \citealp{S09}), adapted to our panel data version of the partial functional linear regression model. Assumption A1 postulates standard moment and exogeneity conditions and describes the dependence structure of the regressors over time. Assumptions A2 and A4 together govern, first, the complexity of the functional component and, second, the degree to which the multivariate model component affects the estimation problem. The first type of assumptions are postulated as usually in terms of the covariance structure of the functional regressor and its interplay with the parameter function $A_k$. The second type is formulated in terms of the interplay between between second moments of multivariate and functional regressor as well as the magnitude of the $t$-specific multivariate parameters $\beta_t$. Assumption A5 formulates the asymptotic behavior of the truncation parameters used in the first and third estimation steps. Assumption A6 contains further regularity assumptions on the relation between the functional and multivariate model components. Assumption A7 is a slightly modified version of Assumption $C_\tau$ in \cite{VL17}.

Our theoretical results establish the consistency of our classification procedure and the convergence rates for the proposed regression slope estimators. We provide convergence rates of the $t$-wise estimators $\hat\beta_t$ and $\hat\alpha_t$ from Step 1 of our estimation procedure in Theorem \ref{prop:twise}. Lemma \ref{lem:UNI} establishes uniform consistency of these estimators as well as the adjusted slope function estimator $\hat \alpha_t^{(\Delta)}$  over $t=1,\dots,T$. This is an important prerequisite for the consistency of our classification procedure, which is established in Theorem \ref{th:CCO}. Finally, Theorem \ref{th:CPE} establishes the convergence rate of our estimator $\tilde{A}_k$ from Step 3 of the estimation procedure.

%%%%%%%%%%%%%%
\begin{theorem}\label{prop:twise} 
Given Assumptions A1--A6 hold, it follows for all $1\leq t \leq T$ that
\begin{align*}
&\left|\left|\hat \beta_t-\beta_t\right|\right|^2 =O_p \left( n^{-1} \right) \\
\text{and}\quad &\left|\left|\hat \alpha_t -\alpha_t\right|\right|_2^2 =O_p \left( n^{\frac{1-2\nu}{\mu+2\nu}} \right),
\end{align*}
where $||.||$ denotes the Euclidean norm and $||.||_2$ the $L^2$ norm. The proof can be found in Appendix \ref{sec:APAL1} of the online supplement supporting this article. 
\end{theorem}
%%%%%%%%%%%%%%

Theorem \ref{prop:twise} is related to Theorems 3.1 and 3.2 in \cite{S09}, though our proof deviates from that in \cite{S09} at important instances. The above rates for $\hat \alpha_t$ correspond to the rates in the cross section context of \cite{HH07}. These pointwise rates provide a benchmark for the asymptotic properties of $\tilde A_k$, however the theorem is not sufficient for the consistency of our classification algorithm. For this, we need the following uniform consistency results:
%%%%%%%%%%%%%%
\begin{lemma}\label{lem:UNI} 
Given Assumptions A1--A6 hold, it follows that
\begin{align*}
\begin{array}{ll}
&\max_{1\leq t\leq T}\left|\left|\hat \beta_t-\beta_t\right|\right|^2 =o_p(1),\\
&\max_{1\leq t\leq T}\left|\left| \hat \alpha_t -\alpha_t \right|\right|_2^2 =o_p(1)\\
\text{and}\quad &\max_{1\leq t\leq T}\left|\left| \hat \alpha_t^{(\Delta)} -\alpha_t^{(\Delta)}\right|\right|_2^2 =o_p(1).
\end{array}
\end{align*}
The proof can be found in Appendix \ref{sec:APAL2} of the online supplement supporting this article. 
\end{lemma}
%%%%%%%%%%%%%%
Note that Lemma \ref{lem:UNI} is not a trivial consequence of Theorem \ref{prop:twise}, since $T$ tends to infinity with $n$ (see A3). The following theorem establishes consistency of our classification procedure and is based on our results in Lemma \ref{lem:UNI}: 
%%%%%%%%%%%%%%
\begin{theorem}\label{th:CCO} 
Given Assumptions A1--A7 hold, it follows that
\begin{align*} 
\prob\left(\{ \hat G_1,\dots,\hat G_{\hat K} \}\neq \{G_1,\dots,G_K\}  \right) = o(1).
\end{align*}
The proof can be found in Appendix \ref{sec:APACCO} of the online supplement supporting this article. 
\end{theorem}
%%%%%%%%%%%%%%

The statement of Theorem \ref{th:CCO} is twofold. First, it says that the number of regimes $K$ is asymptotically correctly determined. Second, it says that the estimators $\hat G_k$, $1\leq k \leq \hat K$ consistently estimate their population counterparts. This notion of classification consistency is sufficient to obtain the following asymptotic result for the corresponding estimators $\tilde A_k$, $1\leq k \leq \hat K$ from Step 3 of the estimation procedure:
%%%%%%%%%%%%%%
\begin{theorem}\label{th:CPE} Given Assumptions A1--A7 hold, it follows for all $1\leq k \leq \hat K$ that
\begin{align*} 
\quad \left|\left| \tilde A_k -A_k\right|\right|_2^2 &=\left\{\begin{array}{ll} O_p\left(n^{-1} \right) & \text{if}\quad \delta\geq \frac{1+\mu}{2\nu-1}\\
O_p\left((nT)^{\frac{1-2\nu}{\mu+2\nu}}\right) & \text{if}\quad \delta\leq  \frac{1+\mu}{2\nu-1}.\\ \end{array}\right. 
\end{align*}
The proof can be found in Appendix \ref{sec:APACPE} of the online supplement supporting this article. 
\end{theorem}
%%%%%%%%%%%%%

Theorem \ref{th:CPE} quantifies the extent to which the estimation error $||\hat \beta_t-\beta_t ||$ contaminates the estimation of $A_k$. In the first case ($\delta\geq (1+\mu)/(2\nu-1)$), $n$ diverges relatively slowly in comparison to $T$ and, therefore, the contamination due to estimating $\beta_t$ is not negligible. This results in the relatively slow convergence rate of $n^{-1/2}$, where the attribute ``slow'' has to be seen in relation to our panel context with $n\to\infty$ and $T\to\infty$. In the second case ($\delta\leq (1+\mu)/(2\nu-1)$), $n$ diverges sufficiently fast such that the contamination due to estimating $\beta_t$ becomes asymptotically negligible, which results in the faster convergence rate of $(nT)^{(1-2\nu)/(\mu+2\nu)}$. The latter rate coincides with the minimax optimal convergence result in \cite{HH07}.

%%%%%%%%%%%%%%%%%%%%%%%%%%%%%%%%%%%%%%%%%%%%
\section[]{Practical Choice of Tuning Parameters}\label{sec:THR}
%%%%%%%%%%%%%%%%%%%%%%%%%%%%%%%%%%%%%%%%%%%%
Inspired by the thresholding procedure in \cite{VL17}, we suggest choosing the threshold parameter $\tau_{nT}$ based on an approximate law for $\hat \Delta_{ts}=||\hat \alpha_t^{(\Delta)}-\hat \alpha_s^{(\Delta)} ||_2^2$ under the hypothesis that $t$ and $s$ belong to the same regime $G_k$. As argued in Appendix \ref{ssec:TC} of the online supplement supporting this article, the scaling of the estimators $\hat \alpha_t$ and $\hat \alpha_s$ as suggested in \eqref{dalph} leads, for large $n$, to 
%%%%%%%%%%%%%%%%%%%
\begin{align*}
\frac{n}{2}\hat \Delta_{ts} =\frac{n}{2}||\hat \alpha_t^{(\Delta)}-\hat \alpha_s^{(\Delta)} ||_2^2 \sim \chi^2_{\underline{m}} \quad \text{approximately}.
\end{align*}
%%%%%%%%%%%%%%%%%%%
Hence we recommend setting the threshold $\tau_{nT}$ to be $2/n$ times the $p_\tau$-quantile of a $\chi^2_{\underline{m}}$ distribution, where $p_\tau$ is close to one, for instance, $p_\tau=0.99$ or $p_\tau=0.999$. By scaling this quantile with $2/n$, the threshold converges, as required, to zero as $n$ tends to infinity (see also Section \ref{ssec:TC} in the appendix for more details). This is a simple ad-hoc solution ignoring autocorrelations between the estimators $\hat \alpha_t^{(\Delta)}$ and $\hat \alpha_s^{(\Delta)}$. The threshold works well in practice, however, in case of strong autocorrelations one may  use one of the well-known cluster algorithms applied to the vectors of the unscaled estimates $(\hat a_{1,t},\dots,\hat a_{m_t,t})$ from Step 1 of our estimation procedure. In our simulation study, for instance, we additionally investigate the performance the Gaussian mixture cluster approach of \cite{FR2002} as implemented in the \textsf{R}-package \texttt{mclust} of \cite{SFMR2016}. This cluster algorithm is very practical as it allows for an automatic choice of the number of clusters $K$. Alternatively, as already mentioned above, one might also derive a threshold using the theoretical results in \cite{GP17} and \cite{CR18}; however, this demands for some additional theoretical work which is out of the scope of this paper.

For selecting the truncation parameters $m_t$ and $\tilde m_k$ the literature offers two main strategies. The first strategy is to choose the truncation parameters in order to find an optimal prediction. A cross validation procedure is shown, e.g., in \cite{S09}. The second one is to choose the cut-off levels according to the covariance structure of the functional regressor. The latter is convenient from a computational point of view---particularly, for the typically large sample sizes in panel data. We thus suggest choosing $m_t$ and $\tilde m_k$ according to the following eigenvalue ratio criterion suggested in \cite{AH13}:
%%%%%%%%%%%%%%%%%
\begin{align*}
m_t&=\text{arg}\max_{1\leq l < n} \hat \lambda_{l,t}/\hat \lambda_{l+1,t}, &&\hspace*{-1cm} 1\leq t \leq T,\\
\text{and} \quad \tilde m_k&=\text{arg}\max_{1\leq l < n|\hat G_k|} \tilde \lambda_{l,k}/\tilde \lambda_{l+1,k}, &&\hspace*{-1cm} 1\leq k \leq \hat K.
\end{align*}
%%%%%%%%%%%%%%%%%

For selecting $K_{\max}$ we employ a standard estimate for the number of clusters from classical multivariate cluster analysis as introduced by \cite{CH74}. This translates to our context as follows. On an equidistant grid $0=s_1<s_2<\dots<s_L=1$ in $[0,1]$ we calculate the L-vectors $v_t:=[\hat \alpha_t^{(\Delta)}(s_l)]_{l=1,\dots,L}$ for $1 \leq t \leq T$. Based on these quantities we employ the maximizer 
%%%%%%%%%%%%%%%
\begin{align*} 
K_{\max}:=\text{arg}\max_{1 \leq k \leq (T-1)}\frac{tr\left(\sum_{j=1}^k |C_j|(v_t-\bar{v})(v_t-\bar{v})^{\top}\right)/(k-1)}{tr \left(\sum_{j=1}^k \sum_{t\in C_j} (v_t-c_j)(v_t-c_j)^{\top}\right)/(T-k)}
\end{align*}
%%%%%%%%%%%%%%% 
as an upper bound for $\hat{K}$. Here $C_j\subset\{1,\dots,T\}$ is the $j$-th cluster formed from a k-means algorithm with $c_j$ being the corresponding centroid. We further denote $\bar{v}:=T^{-1} \sum_{t=1}^T v_t$ and use $tr(\cdot)$ for the trace operator. Choosing $K_{\max}$ equal to the optimal number of clusters in the multivariate analogue of the functional classification problem leads to a comparably conservative choice of $\hat K \leq K_{\max}$. This guarantees a parsimonious parameterization of our model in finite samples and improves the interpretability of the estimates. 

We assess how this configuration of our estimation procedure performs in different finite-sample environments by means of a simulation study, which is described in the next section.

%%%%%%%%%%%%%%%%%%%%%%%%%%%%%%%%%%%%%%%%%%
\section{Simulations}\label{sec:SIM}
%%%%%%%%%%%%%%%%%%%%%%%%%%%%%%%%%%%%%%%%%%
In the following simulation study we consider two different data generating processes (Scenarios 1 and 2). In both scenarios there are $K=3$ parameter regimes and we set $\alpha_t=A_1$ if $t\in G_1=\{1,\dots,\lfloor T/3 \rfloor\}$, $\alpha_t=A_2$ if $t\in G_2=\{\lfloor T/3 \rfloor+1,\dots,\lfloor 2T/3 \rfloor\}$ and $\alpha_t=A_3$ if $t\in G_3=\{\lfloor 2T/3 \rfloor+1,\dots,T\}$, where
%%%%%%%%%%%%%%%%%
\begin{align*} 
A_1(u)&=\begin{cases}\sqrt{2} \sin(\pi u/2)-u^3/2+\sqrt{18}\sin(3\pi u/2) & \text{in Scenario 1 } \\ 8u-4u^2-5u^3+2 \sin(8u) & \text{in Scenario 2},\end{cases}\\
A_2(u)&=-2u+8u^2+5u^3+2\sin( 8 u )\hspace{2.65cm}\text{in Scenarios 1 and 2} \\
\text{and}\quad A_3(u)&=-2u+\cos(6u)   \hspace{5.12cm}\text{in Scenarios 1 and 2.}
\end{align*}
%%%%%%%%%%%%%%%%%

\begin{figure}[!hb]
\begin{center}
\includegraphics[width=0.9 \textwidth]{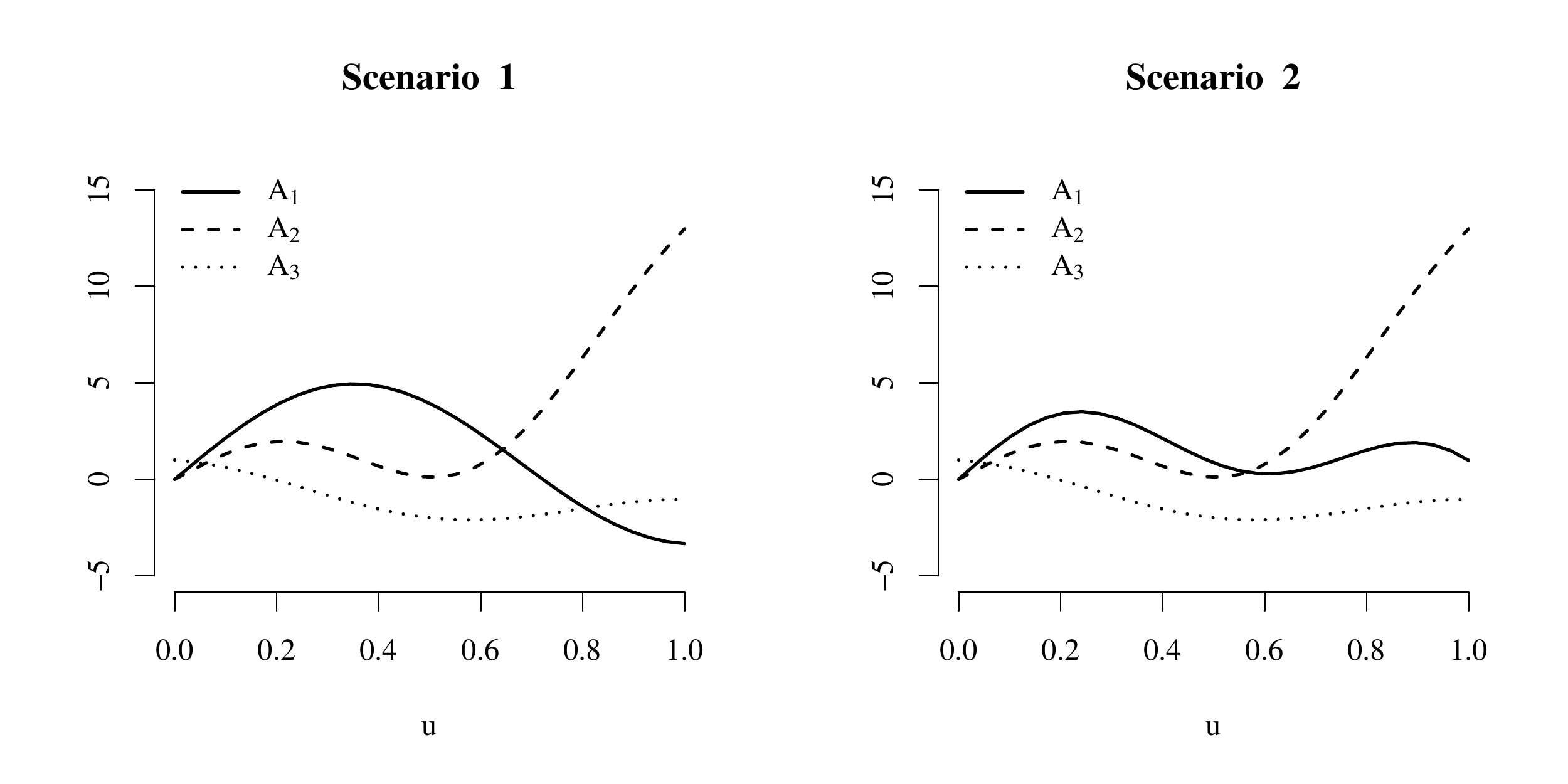}
\end{center}
\vspace*{-1cm}
\caption{Parameter functions $A_1$, $A_2$, and $A_3$ of Scenarios 1 and 2.}
\label{Fig_Scen}
\end{figure}

The graphs of the parameter functions $A_1$, $A_2$, and $A_3$ of Scenarios 1 and 2 are shown in Figure \ref{Fig_Scen}. Note that the distances between the regime specific slope functions are smaller in Scenario 2 than in Scenario 1, which makes Scenario 2 the more challenging one. For both scenarios we set $\beta_t=5\sin(t/\pi)$ and $\rho_t=5\cos(t/\pi)$. We simulate the regressor $z_{it}$ and the error term $\epsilon_{it}$ according to $z_{it}\sim\mathcal{N}(0,1)$ and $\epsilon_{it}\sim \mathcal{N}(0,1)$. The trajectories $X_{it}$ are obtained as $X_{it}(u)=\sum_{j=1}^{20} \theta_{it,j} \phi_j(u)$ with independent scores $\theta_{it,j}\sim\mathcal{N}\left(0,[(j-1/2)\pi]^{-2}\right)$ and eigenfunctions $\phi_j(u)=\sqrt{2}\sin((j-1/2)\pi u)$. Regarding the choice of the tuning parameters we proceed as described in Section \ref{sec:THR}. For selecting the threshold $\tau_{nT}$ we set $p_\tau=0.99$. As a practical alternative, we apply the Gaussian mixture cluster approach of \cite{FR2002} to the vectors of the unscaled estimates $(\hat a_{1,t},\dots,\hat a_{m_t,t})$, where we use the \texttt{mclust()} function of the \textsf{R}-package \texttt{mclust} \citep{SFMR2016}.
%%%%%%%%%%%%%%%%%%%%
%%%%%%%%%%%%%%%%%%%%

%%%%%%%%%%%%%%%%%%%%%%%%%%%%%%%%%%%%%%%%%%%%%%%%%%%%%    
\begin{table}[!ht]
\centering
\begin{tabular}{l c c c c c c c c c c c}
\toprule
$\mathbf{(n,T)=(50,50)}$ & \multicolumn{5}{c}{\textbf{Scenario 1}}  &       &  \multicolumn{5}{c}{\textbf{Scenario 2}}\\
& $q_{0.25}$ & $q_{0.5}$ & $\text{avg.}$  & $q_{0.75}$ & \text{sd.} & \quad & $q_{0.25}$ & $q_{0.5}$ & \text{avg.}  & $q_{0.75}$ & \text{sd.} \\ \cline{2-6} \cline{8-12}
%%%%%%%%%%%%  
$T^{-1}\sum_{t=1}^T(\hat\beta_t-\beta_t)^2\quad$& 0.02 & 0.02 & 0.02 & 0.03 & 0.00 & \quad & 0.02 & 0.02 & 0.02 & 0.03 & 0.00 \\ 
Class.~Error (THR)                             & 0.08 & 0.12 & 0.13 & 0.18 & 0.08 & \quad & 0.08 & 0.12 & 0.13 & 0.18 & 0.08 \\ 
Class.~Error (MCL)                             & 0.12 & 0.16 & 0.18 & 0.22 & 0.08 & \quad & 0.14 & 0.22 & 0.22 & 0.32 & 0.11 \\ 
$| | \tilde A_1-A_1 | |_2^2/||A_1||_2^2$        & 0.01 & 0.02 & 0.03 & 0.04 & 0.04& \quad & 0.06 & 0.08 & 0.11 & 0.12 & 0.11 \\ 
$| | \tilde A_2-A_2 | |_2^2/||A_2||_2^2$        & 0.02 & 0.03 & 0.04 & 0.04 & 0.05  & \quad & 0.02 & 0.03 & 0.04 & 0.05 & 0.05  \\ 
$| | \tilde A_3-A_3 | |_2^2/||A_3||_2^2$        & 0.05 & 0.09 & 0.11 & 0.16 & 0.08 & \quad & 0.05 & 0.09 & 0.11 & 0.16 & 0.08 \\ 
\midrule
\\[-2ex]
$\mathbf{(n,T)=(100,50)}$ & \multicolumn{5}{c}{\textbf{Scenario 1}}  &                &  \multicolumn{5}{c}{\textbf{Scenario 2}}\\
    & $q_{0.25}$ & $q_{0.5}$ & \text{avg.}  & $q_{0.75}$ & \text{sd.}  & \quad           & $q_{0.25}$ & $q_{0.5}$ & \text{avg.}  & $q_{0.75}$ & \text{sd.} \\
    \cline{2-6} \cline{8-12}
%%%%%%%%%%%%%%%%%%%%%%%%%%%%%%%    
$T^{-1}\sum_{t=1}^T(\hat\beta_t-\beta_t)^2\quad$&  0.01 & 0.01 & 0.01 & 0.01 & 0.00 & \quad & 0.01 & 0.01 & 0.01 & 0.01 & 0.00  \\ 
Class.~Error (THR)                             & 0.06 & 0.10 & 0.12 & 0.16 & 0.08 & \quad & 0.06 & 0.10 & 0.12 & 0.16 & 0.07 \\ 
Class.~Error (MCL)                             & 0.08 & 0.12 & 0.14 & 0.18 & 0.07 & \quad & 0.08 & 0.16 & 0.17 & 0.24 & 0.10 \\ 
$| | \tilde A_1-A_1 | |_2^2/||A_1||_2^2$        & 0.00 & 0.01 & 0.01 & 0.01 & 0.01 & \quad & 0.04 & 0.05 & 0.06 & 0.07 & 0.03 \\ 
$| | \tilde A_2-A_2 | |_2^2/||A_2||_2^2$        & 0.01 & 0.02 & 0.02 & 0.03 & 0.02 & \quad & 0.01 & 0.02 & 0.02 & 0.03 & 0.02 \\ 
$| | \tilde A_3-A_3 | |_2^2/||A_3||_2^2$        &  0.04 & 0.08 & 0.09 & 0.13 & 0.07 & \quad &  0.04 & 0.08 & 0.10 & 0.14 & 0.08 \\ 
\midrule
\\[-2ex]
$\mathbf{(n,T)=(150,80)}$ & \multicolumn{5}{c}{\textbf{Scenario 1}}  &                &  \multicolumn{5}{c}{\textbf{Scenario 2}}\\
    & $q_{0.25}$ & $q_{0.5}$ & \text{avg.}  & $q_{0.75}$ & \text{sd.}  &  \quad           & $q_{0.25}$ & $q_{0.5}$ & \text{avg.}  & $q_{0.75}$ & \text{sd.} \\
    \cline{2-6} \cline{8-12}    
$T^{-1}\sum_{t=1}^T(\hat\beta_t-\beta_t)^2\quad$& 0.01 & 0.01 & 0.01 & 0.01 & 0.00 & \quad & 0.01 & 0.01 & 0.01 & 0.01 & 0.00 \\ 
Class.~Error (THR)                             & 0.08 & 0.11 & 0.12 & 0.16 & 0.07 & \quad & 0.06 & 0.10 & 0.12 & 0.16 & 0.07\\ 
Class.~Error (MCL)                             &  0.05 & 0.10 & 0.11 & 0.14 & 0.06 & \quad & 0.08 & 0.14 & 0.15 & 0.19 & 0.10  \\ 
$| | \tilde A_1-A_1 | |_2^2/||A_1||_2^2$        & 0.00 & 0.00 & 0.01 & 0.01 & 0.01 & \quad & 0.04 & 0.04 & 0.05 & 0.05 & 0.01  \\ 
$| | \tilde A_2-A_2 | |_2^2/||A_2||_2^2$        & 0.01 & 0.01 & 0.02 & 0.02 & 0.01& \quad & 0.01 & 0.01 & 0.02 & 0.02 & 0.01 \\ 
$| | \tilde A_3-A_3 | |_2^2/||A_3||_2^2$        & 0.04 & 0.07 & 0.09 & 0.13 & 0.07  & \quad & 0.04 & 0.07 & 0.09 & 0.13 & 0.07  \\ 
\bottomrule
\end{tabular}
\caption{Simulation Results. The quantities $q_{0.25}$, $q_{0.5}$, $q_{0.75}$, avg., and sd. denote the 25\%, 50\% and 75\% quantiles, the arithmetic mean, and the standard deviation of the empirical distribution over Monte Carlo samples.}
\label{SimuTab}
\end{table}
%%%%%%%%%%%%%%%%%%%%%%%

In order to measure the precision of the classification procedure we calculate the classification error (Class.~Error) as the number of incorrectly classified time points $t$ divided by $T$. We consider the three different $(n,T)$-specifications, $(n,T)=(50,50)$, $(n,T)=(100,50)$ and $(n,T)=(150,80)$, and generate for each specification $1000$ Monte Carlo samples. The results are reported in Table \ref{SimuTab}. The classification errors of our threshold (THR) approach are at a low level in both scenarios and generally correspond to those of the Gaussian mixture cluster (MCL) approach, except for the small sample $(n,T)=(50,50)$, where our approach shows a better performance. Since, both cluster approaches show a very similar performance, we only report the estimation errors for our threshold approach in order to save space. The consistency of all parameter estimators as well as the accuracy of the classification procedure are well shown in our simulation results.

%%%%%%%%%%%%%%%%%%%%%%%%%%%%%%%%%%%%%%%%%%%%%%%%%%%%%%%%%%%%%%%%%%%%%%%%%
\section{Regime Dependent Pricing of Idiosyncratic Risk}\label{sec:APL}
%%%%%%%%%%%%%%%%%%%%%%%%%%%%%%%%%%%%%%%%%%%%%%%%%%%%%%%%%%%%%%%%%%%%%%%%%
Emerging from the influential work of \cite{AHXZ06} a considerable number of studies confirm that stock returns are negatively correlated with the idiosyncratic (i.e., non-systematic) volatility component of a stock (see, for instance, \citeauthor{FU09}, \citeyear{FU09}, and \citeauthor{HL16}, \citeyear{HL16}, and references therein). This finding is referred to as the ``idiosyncratic volatility puzzle'', since asset pricing theory suggests an opposite outcome. Investors can either hold well-diversified or underdiversified portfolios. In the case of well-diversified portfolios, the idiosyncratic volatility component is not a relevant pricing component and one expects no correlation between idiosyncratic volatility and returns. In the case of underdiversified portfolios, asset pricing theory predicts a positive correlation, since investors expect higher returns as a compensation for the additional risk. The observed negative correlations, however, cannot be explained by the asset pricing theory. As demonstrated in \cite{HL16} the idiosyncratic volatility puzzle has, to a substantial extent, remained unsolved.

In the literature, the idiosyncratic volatility puzzle is typically examined using aggregated monthly data. In contrast, we consider the relation between the returns $y_{it}\in\mathbb{R}$ and the disaggregated daily idiosyncratic volatility curve $X_{it}\in L^2([0,1])$ of asset $i=1,\dots,n$ at day $t=1,\dots,T$ using our functional linear panel regression model:
%%%%%%%%%%%%%
\begin{align} 
y_{it} =\rho_{t} + \int_{0}^{1} \alpha_t(s)X_{it}(s) \mathrm{d}s  + \beta_t^\top z_{it} + \epsilon_{it}.\label{emp_price}
\end{align}
%%%%%%%%%%%%%
Here $\rho_{t}\in\mathbb{R}$ is a daily fixed effect and $\alpha_t\in L^2([0,1])$ denotes the time-varying parameter function describing the effect of the idiosyncratic volatility curve $X_{it}\in L^2([0,1])$ at day $t$. The time-varying parameter vector $\beta_t\in\mathbb{R}^P$ describes the effect of additional control variables $z_{it}\in\mathbb{R}^P$. The term $\epsilon_{it}$ is a scalar error with zero mean and finite but potentially time heteroscedastic variances. We postulate that there are only $K<T$ different volatility pricing regimes $G_1,\dots,G_K$ collecting identical parameter functions $\alpha_t$. As above, the common slope function of regime $k$ is denoted by $A_k$. If a coefficient function $A_k$ is clearly negative over most of its domain, the corresponding regime appears to be non-conform with traditional asset pricing theory and thus constitutes a temporary idiosyncratic volatility puzzle. The advantage of our pricing Model \eqref{emp_price} is its capability to segment the set of trading days into puzzling and non-puzzling pricing regimes in a data-driven way.

Following \cite{FU09}, we define the dependent variable as the daytime log-return \[y_{it}:=\log(P_{it}(1)/P_{it}(0)),\] where $P_{it}(0)$ and $P_{it}(1)$ denote the opening price and closing price of asset $i$ at day $t$. As control variable $z_{it}\in\mathbb{R}$ we use the daily average bid-ask spread which serves as a proxy for liquidity risk---an important pricing-relevant factor as discussed for example in \cite{HL16}.

We consider intraday price data for $n=377$ stocks listed on the S\&P 500. Our sample consists of $T=136$ trading days between June $3$, $2016$, and December $15$, $2016$. Intraday stock prices are sampled every $\Delta=10$ minutes during the trading hours of the S\&P 500. More concretely, for asset $i$ at day $t$ we consider the last recorded transaction prices, $P_{it}(s_j)$, within 10-minute intervals, of which the $j$-th interval is denoted $[s_{j-1},s_{j}]$ with $s_0=0$, $s_J=1$, where $1\leq j \leq J=39$. For the construction of the idiosyncratic volatility curves $X_{it}(.)$, which is described below, we make use of three Fama-French factors. The Fama-French factors were downloaded from Kenneth French's homepage, while all other data were gathered from Bloomberg.

\begin{figure}[!ht]
\begin{center}
\includegraphics[width=1 \textwidth]{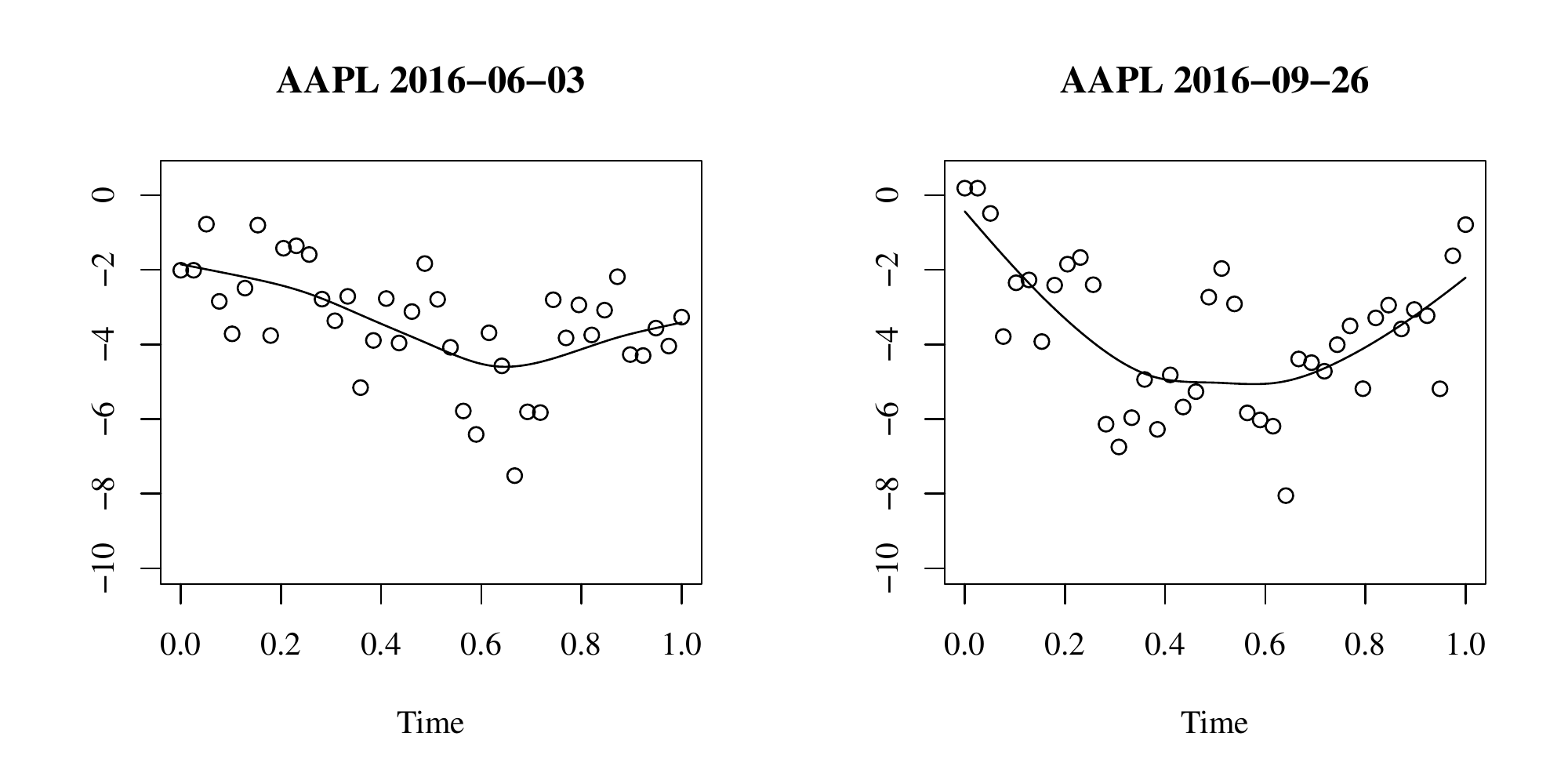}
\end{center}
\vspace*{-1cm}
\caption{Exemplary idiosyncratic volatility curves $X_{it}$ and raw scatter points for the Apple Inc.~stock (AAPL) at two randomly selected trading days.}
\label{Fig_aapl}
\end{figure}

For constructing the idiosyncratic volatility curves $X_{it}(.)$ we use the method proposed in \cite{MSS11} with a straightforward adaption to our context for estimating idiosyncratic volatility curves instead of total volatility curves. \cite{MSS11} propose an estimator of the total volatility curve of asset $i$ at day $t$, smoothing the points $(s_j,\tilde{Y}_{it,j})$, $j=1,\dots,J$ based on an algorithm, which allows to obtain functional principal components from discrete noisy data. Here $\tilde{Y}_{it,j}:=\log(\Delta^{-1}Y_{it}(s_j)^2)+q_0$ are scaled and logarithmized versions of the squared intraday returns $Y_{it}(s_j)^2$ with $Y_{it}(s_j):=\log(P_{it}(s_{j})/P_{it}(s_{j-1}))$. The points $\tilde{Y}_{it,j}$, $1\leq j \leq J$ can be interpreted as noisy evaluations of the random spot volatility curve of the underlying continuous time return process at the corresponding points $s_j$, $1\leq j \leq J$. The constant $q_0=1.27$ is necessary for re-centering the involved error term, whereas technical details can be found in \cite{MSS11}. We mainly proceed along the lines of their approach, however, instead of using the total intraday returns $Y_{it}(s_j)$, $j=1,\dots,J$, we employ the idiosyncratic intraday return components  $Y_{it}^*(s_j)$, $j=1,\dots,J$. This leads to a measure of the idiosyncratic volatility curve $X_{it}(.)$ rather than a measure of the total volatility curve. For computing the idiosyncratic intraday returns $Y_{it}^*(s_j)$, we follow the usual approach and correct the total intraday returns $Y_{it}(s_j)$ for their systematic market component by regressing them on three Fama-French factors (\citealp{FF95}). We do so by estimating the functional Fama-French regression model
%%%%%%%%%%%%%%%%%%%%%%%%%%%%%
\begin{align}\label{FFRegr}
Y_{it}(s_j)=b_{0,it}(s_j)+b_{1,it} \cdot M_t(s_j)+b_{2,it} \cdot S_t + b_{3,it} \cdot H_t + u_{it}(s_j),\quad j=1,\dots,J,
\end{align}
proposed by \cite{KMZ14}. The term $M_t(s_j)$ is the intraday S\&P 500 market return, $S_t$ denotes the ``small minus large'' factor and $H_t$ the ``high minus low'' factor. $S_t$ describes the difference in returns between portfolios of small and large stocks and $H_t$ describes the difference in returns between portfolios of high and low book-to-market value stocks. For estimating the model parameters in \eqref{FFRegr} we use the least-squares estimators proposed by \cite{KMZ14}. The idiosyncratic intraday returns are finally obtained as $Y_{it}^*(s_j)=\hat{b}_{0,it}(s_j)+\hat{u}_{it}(s_j)$, where $\hat{b}_{0,it}(s)$ denotes the fitted functional intercept parameter of the $(i,t)$-th regression and $\hat u_{it}(s)$ are the corresponding function-valued regression residuals. Table \ref{SummaryTab} provides summary statistics for our sample. Figure \ref{Fig_aapl} shows the idiosyncratic volatility curves $X_{it}$, along with their raw scatter points, for the Apple stock at two randomly selected trading days. 
%%%%%%%%%%%%%%%%%%%
%%%%%%%%%%%%

\bigskip
\noindent Remark. Applying the method of \cite{MSS11} in order to estimate the idiosyncratic volatility curves $X_{it}$ from the idiosyncratic intraday returns $Y_{it}^*(s_j)$, $j=1,\dots,J$, leads to ``volatility curves'', which are indeed logarithmic volatility curves (see Eq. (9) in \citealp{MSS11}). The log-transformation is monotonic and is thus sign-preserving what concerns the estimation of $\alpha_t$. This is of particular importance when assessing the idiosyncratic volatility puzzle. Working with log-transformed volatility objects is generally advisable, since the raw volatility measures are often heavily skewed (\citealp{HKLN16}).
\bigskip
%%%%%%%%%%%%%%%%%%%%%%%
\begin{table}[!ht]
\centering
\begin{tabular}{l  r@{.}l r@{.}l r@{.}l r@{.}l r@{.}l r@{.}l r@{.}l}
\toprule 
& \multicolumn{2}{c}{$q_{0.05}$} & \multicolumn{2}{c}{$q_{0.25}$} & \multicolumn{2}{c}{$q_{0.5}$} & \multicolumn{2}{c}{$q_{0.75}$} & \multicolumn{2}{c}{$q_{0.95}$} & \multicolumn{2}{c}{avg.} & \multicolumn{2}{c}{sd.}\\
\midrule
$y$ \ (in \%) &-1&74  &-0&56  &  0&01  &  0&60 &  1&79 &  0&02 & 1&16 \\ 
$\int_0^1 X(u) \mathrm{d}u$   &-4&42  &-3&72  & -3&20  & -2&63 & -1&64 & -3&14 & 0&85 \\ 
$||X||_2^2$                   & 3&46  & 7&67  & 11&02  & 14&67 & 20&50 & 11&39 & 5&19 \\ 
$z$  \ (in \%) & 0&02  & 0&02  &  0&03  &  0&05 &  0&09 &  0&04 & 0&03 \\ 
\bottomrule
\end{tabular}
\caption[Summary Statistics]{Quantiles, means, and standard deviations of the considered variables.}\label{SummaryTab}
\end{table}
%%%%%%%%%%%%%%%%%%%

\begin{figure}[!ht]
\begin{center}
\includegraphics[width=1.05 \textwidth]{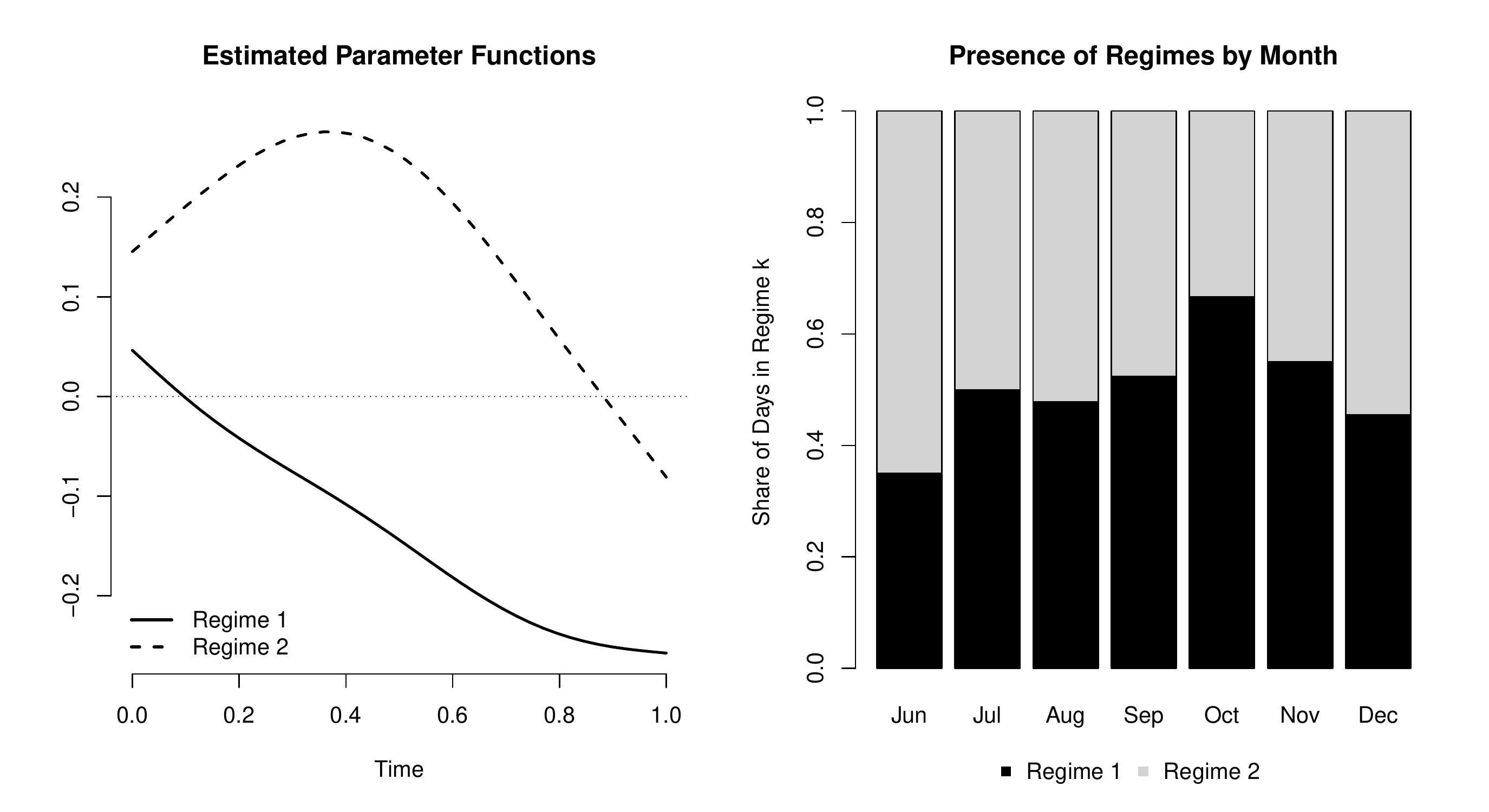}
\end{center}
\vspace*{-1cm}
\caption{{\sc Left Panel:} Estimated regime specific slope functions $\tilde A_1$ and $\tilde A_2$. {\sc Right Panel:} Monthly shares of days in Regimes 1 and 2.}
\label{AtildeEmp}
\end{figure}

The estimation of Model \eqref{emp_price} proceeds as described in Section \ref{sec:EST}. Using our classification algorithm, we find a number of $\hat K=2$ regimes, where each regime contains about the same number of days (see right panel in Figure \ref{AtildeEmp}). The left panel in Figure \ref{AtildeEmp} shows the estimated regime specific slope functions $\tilde A_1$ and $\tilde A_2$. In order to examine their daily effects on the returns, we consider the average effects $\int_0^1\tilde A_k(u)\mathrm{d}u$, $k=1,2$. For the first regime this average effect is clearly negative, for the second one clearly positive. Our classification thus separates trading days revealing an idiosyncratic volatility puzzle from days which are conform with asset pricing theory. Both parameter functions, however, indicate that the intensity of the pricing varies over the intaday trading time. 
%%%%%%%%%%%%%%%%%%%%%%%%
%%%%%%%%%%%%%%%%%%%%%%%%

In summary, our results indicate a rather complex pattern of puzzling and non-puzzling days. The high resolution of the time scale our novel model operates on, allows us to uncover a much more heterogeneous pricing of idiosyncratic volatility than one can infer from the usually considered monthly data (see, for example, \citealp{AHXZ06}). Hence, aggregating the data could thus misleadingly convolute the puzzling and non-puzzling pricing mechanisms. This in turn might contribute to the failure of current explanations of the idiosyncratic volatility puzzle (\citealp{HL16}).

%%%%%%%%%%%%%%%%%%%%%%%%%%%%
\section*{Acknowledgments}
%%%%%%%%%%%%%%%%%%%%%%%%%%%%
We want to thank Alois Kneip (University of Bonn) and Michael Vogt (University of Bonn) for fruitful discussions and valuable comments which helped to improve this research work. Furthermore, we are grateful to the referee and the editors for their constructive input.

%%%%%%%%%%%%%%%%%%%%%%%%%%%%%%%%%%%%%
\section{Conclusion}\label{sec:CON}
%%%%%%%%%%%%%%%%%%%%%%%%%%%%%%%%%%%%%
In this paper we present a novel regression framework, which allows us to examine regime specific effects of a random function on a scalar response in the presence of a multivariate regressor and time fixed effects. The suggested estimation procedure is designed for a panel data context. We prove consistency of the estimators including rates of convergence and address the practical choice of the tuning parameters involved. Our framework offers a flexible and data-driven way of assessing heterogeneity in large panels. The model can be extended in multiple directions for further research. For instance, establishing a connection to the work of \cite{SSP16} would allow us to identify group structures in the cross section in addition to identifying time-regime structures. 

The statistical model is motivated by our real data application, where we explore a phenomenon referred to as the idiosyncratic volatility puzzle. In an empirical study we search for the presence of such a puzzle in a large panel of US stock prices. Our method allows to separate puzzling days from non-puzzling days. The results suggest a much more heterogeneous pricing of idiosyncratic volatility than indicated by many existing analyses in the literature. 

%%%%%%%%%%%%%%%%%%%%%%%%%
\bibliographystyle{Chicago}
\bibliography{bibfile}
%%%%%%%%%%%%%%%%%%%%%%%%%%%%%%%%%%%%%%%%

%%%%%%%%%%%%%%%%%%%%%%%%%%%%%%%%%%%%%%%%
\newpage
\thispagestyle{empty}
\vspace*{.5cm}
%%%%%%%%%%%%%%%%%%%%%%%%%%%%%%%%%%%%%%%%%
\noindent{\large \bf Supplemental Paper for:}\\[2ex]

\if0\blind

\begin{center}
{\Large \bf Parameter Regimes in Partial Functional\\[1.5ex] Panel Regression}\\[2ex]

by Dominik Liebl and Fabian Walders
\end{center}

 \fi

\if1\blind
{
\begin{center}
    {\LARGE\bf Parameter Regimes in Partial Functional\\[1.5ex] Panel Regression}
\end{center}
  \medskip
} \fi

\bigskip

%%%%%%%%%%%%%%%%%%%%%%%%%%%%%
\appendix
\setcounter{page}{1}
\setcounter{section}{0}
\pagenumbering{Roman} 
%%%%%%%%%%%%%%%%%%%%%%%%%%%%%%

Throughout this appendix we use the symbols $C$ and $c$ to denote generic positive constants. 

\vspace*{4mm}
%%%%%%%%%%%%%%%%%%%%%%%%%%%%%%%%%%%%%
\section{Technical Appendix}\label{sec:APA}
%%%%%%%%%%%%%%%%%%%%%%%%%%%%%%%%%%%%%
In this part we use the following notation for norms in addition to the ones introduced in the main paper. Given a mapping $F_1:L^{2}([0,1])\to \mathbb{R}$, we use as norm of $F_1$ the operator norm $|| F_1 ||_{H'}:= \sup_{||f_1||_2=1}|F_1(f_1)|$. Further, for an integral operator $F_2:L^{2}([0,1])\to L^{2}([0,1])$ with kernel $f_2\in L^2([0,1]\times [0,1])$, denote its Hilbert-Schmidt norm as $||F_2||_\mathcal{S}:= ||f_2 ||_2$, where in this case $||\cdot ||_2$ is the $L^2$ norm in $L^2([0,1]\times [0,1])$.

For the following proofs we make use of the following closed form solutions of our least squares estimators $\hat a_{j,t}$ and $\hat\beta_t$: 
%%%%%%%%%%%%%%%%%%%%%%%%%%%%%%%
\begin{align*}
&\hat \alpha_t=\sum_{j=1}^{m_t} \hat a_{j,t} \hat \phi_{j,t} 
\quad\text{ with }\quad 
\hat a_{j,t}= \hat \lambda_{j,t}^{-1}\;\frac{1}{n}\sum_{i=1}^n \langle X_{it}^c,\hat \phi_{j,t} \rangle (y_{it}^c-\hat \beta_t^\top z_{it}^c)\quad\text{ and}\\%\quad\label{estim_alpha_t} 
&\hat \beta_t=\left[ \hat {\mathbf{K}}_{z,t}-\hat {\mathbf{\Phi}}_{t} (\hat  {\mathbf{K}}_{zX,t})  \right]^{-1} \left[ \hat  {\mathbf{K}}_{z y,t}-\hat \Phi_t \left(\hat{K}_{ y X,t}\right) \right],\notag 
\end{align*}
%%%%%%%%%%%%%%%%%%%%%%%%%%%%%%%%
where
%%%%%%%%%%%%%%%%%%%%%%%%%%%%%%%%
\begin{align*}
&\hat  {\mathbf{K}}_{z,t}:=\frac{1}{n} \sum_{i=1}^n z_{it}^c {z_{it}^c}^\top, 
 \;\;  
\hat  {\mathbf{K}}_{zX,t}(s):=[\hat  {{K}}_{z_1X,t}(s),\dots,\hat  {{K}}_{z_PX,t}(s)]^\top,  
 \;\;
 \hat  {\mathbf{K}}_{z y,t}:=[\hat  {{K}}_{z_1{ y},t},\dots,\hat  {{K}}_{z_P{ y},t}]^\top,\\
%%%%%%%%%%%%% 
&\hat K_{ y X,t}(s):=\frac{1}{n} \sum_{i=1}^n y_{it}^c X_{it}^c(s), 
 \quad
 \hat  {{K}}_{z_pX,t}(s):=\frac{1}{n} \sum_{i=1}^n z^c_{p,it}X_{it}^c(s), 
 \quad   
 \hat  {{K}}_{z_p y,t}:=\frac{1}{n} \sum_{i=1}^n z_{p,it}^c{y}^c_{it}, \\
%%%%%%%%%%%% 
&\hat \Phi_{t}(g):=  [\hat \Phi_{1,t}(g), \dots , \hat \Phi_{P,t}(g)]^\top, 
\quad 
\hat \Phi_{p,t}(g):= \sum_{j=1}^{m_t} \frac{\langle \hat K_{z_pX,t},\hat \phi_{j,t} \rangle \langle \hat \phi_{j,t},g \rangle}{\hat \lambda_{j,t} } \ \ \text{for any } g\in L^2([0,1]),\\
%%%%%%%%%%%%%
& \text{and} \quad 
\hat {\mathbf{\Phi}}_{t} (\hat  {\mathbf{K}}_{zX,t}) := [\hat \Phi_{p,t}(\hat  {{K}}_{z_qX,t})]_{1 \leq p \leq P, \ 1 \leq q \leq P};
\end{align*}
%%%%%%%%%%%%%%%%%%%%%%%%%%%%%%%%%%
see \cite{S09} for similar estimators in a cross section context.

For the sake of readability we will proof the lemma and theorems for $P=1$, while the generalization to $P>1$ is straightforward and does not add any additional insights. In this spirit we ease our notation by dropping boldface notation and the dependence on coordinate labels $p$. \\

Now, turning to a formal argumentation, we begin collecting a number of basic results readily available in the functional data literature. Provided Assumption 1 holds, the random variables $\{ (z_{it},X_{it},\epsilon_{it}): \ 1\leq i \leq n  \}$ are iid with finite fourth moments for every $1 \leq t \leq T$. Moment calculations as well as the results in \citeapp{HK10_app} imply for any $1\leq t \leq T$ as $n\to \infty$ that
%%%%%%%%%%%%%%%%%%%%
\begin{align}
&E\left[\left| \left| \hat K_{zX,t}-K_{zX,k} \right| \right|^2_2 \right] =O(n^{-1}) \label{kzxt} \\
& E\left[ \left| \hat K_{z,t}-K_{z,k}  \right|^2\right] =O(n^{-1}) \label{kzt} \\
&E\left[ \left| \left| \hat K_{X,t}-K_{X,k} \right| \right|_2^2 \right] =O(n^{-1}), \label{covX} 
\end{align}
where the index $k$ is such that $t\in G_k$, which we use in what follows without further reference. In Equation \eqref{covX} $K_{X,k}$ denotes the covariance function in the $k$-th regime, i.e. $K_{X,k}(u,v):=E[(X_{it}(u)-E[X_{it}](u))(X_{it}(v)-E[X_{it}](v))]$ and in analogy $K_{z,k}:=E[(z_{it}-E[z_{it}])^2]$. Further, it obviously holds that
%%%%%%%%%%%%%%%%%%
\begin{align*}
&E\left[|\bar z_t -E[z_{it}]|^2\right]=O(n^{-1})\\
& E\left[ \left| \left| \bar X_t-E[X_{it}] \right| \right|^2_2 \right]=O(n^{-1}),\\
&E\left[ ||\hat K_{X\epsilon,t}||_2^2\right]=O(n^{-1}) \\
&E\left[ |\hat K_{z\epsilon,t}|^2\right]=O(n^{-1})
 \end{align*}
where
\begin{align*}
&\hat K_{X\epsilon,t}:=n^{-1}\sum_{i=1}^n X_{it}^c \epsilon_{it}^c \\ 
\text{and }\quad &\hat K_{z\epsilon,t}:=n^{-1}\sum_{i=1}^n z_{it}^c \epsilon_{it}^c.
\end{align*}

%%%%%%%%%%%%%%%%%

Denote the Hilbert-Schmidt norm of the distance between t-wise empirical covariance operator and population covariance operator as $\mathcal{D}_t:=||\hat \Gamma_t-\Gamma_k ||_{\mathcal{S}}$. Note that for any $1 \leq j \leq n$, $|\hat \lambda_{j,t}-\lambda_{j,k}|\leq \mathcal{D}_t$ almost surely (see Theorem 1 in \citealpapp{HHN06_app} and references therein). Since $E[\mathcal{D}_t^q]=O(n^{-q/2})$ for $q=1,2,\dots$ (provided sufficiently high moments exist) it holds that 
%%%%%%%%%%%%%%%%%%%%%
\begin{align}
E\left[ \left| \hat \lambda_{j,t} - \lambda_{j,k}  \right|^q \right] = O(n^{-q/2}) \ \ q=1,2\dots  \label{lambdat}
\end{align}
%%%%%%%%%%%%%%%%%
for any $1 \leq j \leq m_t$ (cf.~Equation A.11 in \citealpapp{KPS16_app}). 

As a final observation, note that combining the results in \citeapp{S09_app} and \citeapp{HH07_app} allows to conclude that for any $1\leq t \leq T$
%%%%%%%%%%%%%%%%
\begin{align} 
||\hat \Phi_t-\Phi_k ||_{H'}^2&= \left|\left|\sum_{j=1}^m \frac{\langle \hat  K_{zX,t}, \hat \phi_{j,t} \rangle}{\hat \lambda_{j,t}}  \hat \phi_{j,t}  - \sum_{j=1}^\infty \frac{\langle K_{zX,k}, \phi_{j,k} \rangle}{\lambda_{j,k}}  \phi_{j,k} \right|\right|_2^2 \nonumber\\
 &=O_p\left(n^{\frac{1-2\nu}{\mu+2\nu}}\right), \label{phit} 
 \end{align} 
where we denote $m_t=m$ for simplicity, which we continue to do without further reference. The mapping $\Phi_k:L^2([0,1])\to \mathbb{R}$ is the population counterpart of $\hat \Phi_t$ and was implicitly used already in Assumption 6. It is formally defined according to 

\begin{align}
\Phi_k(g):=\sum_{j=1}^\infty \frac{\langle K_{zX,k}, \phi_{j,k} \rangle}{\lambda_{j,k}}  \langle \phi_{j,k},g \rangle \label{eq:popPhi}
\end{align}

for any $g\in L^2([0,1])$.

%%%%%%%%%%%%%%%%%%%%%%%%%%%%%%%%%%%%%%%%%%%%%%%%%%%%%%%%%%%%%%%%%%%%
\subsection{Proof of Theorem \ref{prop:twise}}\label{sec:APAL1}
%%%%%%%%%%%%%%%%%%%%%%%%%%%%%%%%%%%%%%%%%%%%%%%%%%%%%%%%%%%%%%%%%%%%

Consider any $1\leq t \leq T$, with $t$ in some regime $k$, i.e. $t\in G_k$. Note that the estimator $\hat \beta_t$ can be written as
\begin{align*}
\hat \beta_t &=\hat B_t^{-1} [\hat K_{z y,t}-\hat \Phi_t(\hat K_{ yX,t})] 
\end{align*}
with $\hat B_t :=[\hat K_{z,t}-\hat \Phi_t(\hat K_{zX,t})]$. Regarding the inverse in $\hat \beta_t$ note that it follows from \eqref{kzxt}, \eqref{kzt}, and \eqref{phit} in analogy to \citeapp{S09_app} that
\begin{align*}
\hat B_t&:=[\hat K_{z,t}-\hat \Phi_t(\hat K_{zX,t})] \overset{\prob}{\to} [ K_{z,k}-\Phi_k(K_{zX,k})]=:B_k>0
\end{align*}
as $n\to \infty$, which certainly implies $\hat B_t^{-1}=B_k^{-1}+o_p(1)$ by the continuous mapping theorem, whereas $B_k=E[\mathbf{s}_{it}^2]>0$ follows from Assumption 6. To see this also consider the decomposition shown in \eqref{eq:PHIDEC}. As in \citeapp{S09_app}, we assess the difference 
\begin{align*}
\hat \beta_t-\beta_t=\hat B_t^{-1} \left[n^{-1}\sum_{i=1}^n \left( z_{it}^c - \hat \Phi_t (X_{it}^c) \right) (\langle X_{it}^c , \alpha_t\rangle + \epsilon_{it}^c) \right]
\end{align*}
by splitting the term $n^{-1}\sum_{i=1}^n \left( z_{it}^c - \hat \Phi_t (X_{it}^c) \right) (\langle X_{it}^c , \alpha_t\rangle + \epsilon_{it}^c) $ according to

\begin{align*}
\left|n^{-1}\sum_{i=1}^n \left( z_{it}^c - \hat \Phi_t (X_{it}^c) \right) (\langle X_{it}^c , \alpha_t\rangle + \epsilon_{it}^c) \right| \leq | R_{0,1,t}|+|R_{0,2,t}|+|R_{0,3,t}|,
\end{align*}
where, in analogy to her work,
\begin{align*}
R_{0,1,t}&:=n^{-1}\sum_{i=1}^n (z_{it}^c-\Phi_k(X_{it}^c))\epsilon_{it}^c =O_p(n^{-1/2}) \\
R_{0,2,t}&:=n^{-1}\sum_{i=1}^n (\Phi_k(X_{it}^c)-\hat \Phi_t(X_{it}^c))\epsilon_{it}^c =O_p(n^{-1/2}).
\end{align*}
due to the exogeneity of the covariates and the assumed iid nature of the error term (cf. Assumption 1). However, the remaining term we approach in a different manner: 
\begin{align*}
|R_{0,3,t}|&:=\left|n^{-1}\sum_{i=1}^n (z_{it}^c-\hat \Phi_t(X_{it}^c))\langle X_{it}^c,\alpha_t \rangle \right| \\
&\leq \left|\langle \hat K_{zX,t}-K_{zX,k} , \alpha_t \rangle \right| + \left| \langle K_{zX,k} , \alpha_t \rangle  -  n^{-1}\sum_{i=1}^n \hat \Phi_t(X_{it}^c)\langle X_{it}^c,\alpha_t \rangle  \right| \\
&\leq R_{1,1,t}+R_{1,2,t}
\end{align*}
where for $R_{1,1,t}$
%%%%%%%%%%%%%%%%%%%
\begin{align*}
R_{1,1,t}&:=\left|\langle \hat K_{zX,t}-K_{zX,k} , \alpha_t \rangle \right| \\
& \leq ||\alpha_t ||_2 \cdot || \hat K_{zX,t}-K_{zX,k} ||_2 \\
&= O_p(n^{-1/2})
\end{align*}
%%%%%%%%%%%%%%%%%%%%
as a consequence of \eqref{kzxt}. The second term, $R_{1,2,t}$, in $R_{0,3,t}$ is defined as 
\begin{align*}
R_{1,2,t}&:= \left| \langle K_{zX,k} , \alpha_t \rangle  -  n^{-1}\sum_{i=1}^n \hat \Phi_t(X_{it}^c)\langle X_{it}^c,\alpha_t \rangle  \right| \\
&\leq R_{2,1}+R_{2,2,t},
\end{align*}
with
\begin{align*}
R_{2,1}&:=\left|\sum_{j=m+1}^\infty \langle K_{zX,k} , \phi_{j,k}  \rangle a_{j,t}^* \right| \\
R_{2,2,t}&:= \left| \sum_{j=1}^m \langle K_{zX,k} , \phi_{j,k}  \rangle a_{j,t}^* -\sum_{j=1}^m \langle \hat K_{zX,t} , \hat \phi_{j,t}  \rangle \langle \hat \phi_{j,t} , \alpha_t \rangle \right|,
\end{align*}
where we used $a_{j,t}^*:=\langle \alpha_t,\phi_{j,k} \rangle$ due to Assumptions 2, 4 and 5. For the first term observe $R_{2,1}=O\left(n^{\frac{1-\mu-2\nu}{\mu+2\nu}}\right)=O(n^{-1/2})$. The second one can be split in three parts
\begin{align*}
R_{2,2,t} \leq R_{3,1,t}+ R_{3,2,t}+R_{3,3,t}
\end{align*}
with
\begin{align*}
R_{3,1,t}&:= || \hat K_{zX,t}-K_{zX,k}||_2 \sum_{j=1}^m  \left(|| \hat \phi_{j,t}-\phi_{j,k} ||_2 \cdot ||  \alpha_t ||_2 +  |a_{j,t}^*|\right), \\
R_{3,2,t}& := ||\alpha_t ||_2\sum_{j=1}^m |\langle K_{zX,k},\phi_{j,k} \rangle| \cdot  ||\hat \phi_{j,t}-\phi_{j,k} ||_2 \\
&\text{and} \\
R_{3,3,t}&:= || K_{zX,k}||_2 \cdot || \alpha_t||_2 \sum_{j=1}^m || \hat \phi_{j,t}-\phi_{j,k} ||_2^2 + || K_{zX,k}||_2  \sum_{j=1}^m || \hat \phi_{j,t}-\phi_{j,k} ||_2 \cdot |a_{j,t}^*|.
 \end{align*}
% \cdot

An assessment of the asymptotic properties of $R_{3,1,t},R_{3,2,t}$ and $R_{3,3,t}$ requires to examine the asymptotic properties of $|| \hat \phi_{j,t}-\phi_{j,k} ||_2^2$ explicitly. Bounds can, for example, be obtained from Theorem 1 in \citeapp{HHN06_app} as
\begin{align}
|| \hat \phi_{t,j}-\phi_{j,k}||_2^q \leq \left[\frac{ 8^{1/2} \mathcal{D}_t }{\min_{1 \leq l \leq j}\{\lambda_{j,k}-\lambda_{j+1,k}\})}\right]^q \quad \text{almost surely} \label{PhiBoundAS}
\end{align}
which holds for $1 \leq j \leq m$, $q=1,2,\dots$ and any size $n$ of the cross section (see also Equation (5.2) in \citealpapp{HH07_app}). In the context of theory for functional linear regression, \citeapp{HH07_app} develop \textit{asymptotic} bounds on $|| \hat \phi_{j,t}-\phi_{j,k} ||_2^2$, $1 \leq j \leq m$, which are valid on events which occur with probability tending to one as $n\to \infty$. These bounds are particularly helpful, when addressing (weighted) sums over estimation errors as they appear e.g. in $R_{3,1,t}$--$R_{3,3,t}$. We will make use of these bounds, slightly adapting the arguments in \citeapp{HH07_app}, in order to formulate the result more explicitly. For this purpose we consider the three events 
\begin{enumerate}

\item $\mathcal{F}_{1,n,t}:= \left\{ C n^{\frac{2(1+\mu)}{\mu+2\nu}} \mathcal{D}_t^2 \leq 1/8 \right\}$

\item $\mathcal{F}_{2,n,t}:=\left\{ |\hat \lambda_{j,t}-\lambda_{l,k}|^{-2}\leq 2 | \lambda_{j,k}-\lambda_{l,k}|^{-2}\leq C n^{\frac{2(1+\mu)}{\mu+2\nu}}, \ 1 \leq j \leq m, j\neq l\in \mathbb{N} \right\}$.

\item $\mathcal{F}_{3,n,t}:=\mathcal{F}_{1,n,t} \cap \mathcal{F}_{2,n,t}$

\end{enumerate}
of which the second coincides with their work and the first one is a straightforward derivative of their arguments. Denoting the complement of a set $A$ as $A^c$, note that $\prob(\mathcal{F}_{1,n,t}^c)=o(1)$ as well as $\prob(\mathcal{F}_{2,n,t}^c)=o(1)$ due Assumptions 4--5 and root-$n$ consistency of the empirical covariance operator and its corresponding eigenvalues as well as assuming the constants in $\mathcal{F}_{1,n,t}$ and $\mathcal{F}_{2,n,t}$ to be appropriate. Since $\prob(\mathcal{F}_{3,n,t}^c)\leq \prob(\mathcal{F}_{1,n,t}^c)+\prob(\mathcal{F}_{2,n,t}^c)$, we conclude $\prob(\mathcal{F}_{3,n,t}^c)=o(1)$. We also show that this property holds uniformly over $1\leq t \leq T$ as $(n,T)\to \infty$ in the proof of Lemma \ref{lem:UNI} below. 
Equation (5.21) in \citeapp{HH07_app}, reads in our notation as 
\begin{align}
& || \hat \phi_{j,t}-\phi_{j,k} ||_2^2 \leq 8 \left( 1-4 C n^{\frac{2(1+\mu)}{\mu+2\nu}} \mathcal{D}_t^2 \right)^{-1}  R_{j,t}^{(\phi)}, \label{phi:1} \\
\text{where } & R_{j,t}^{(\phi)}:= \sum_{l:l\neq j} (\lambda_{j,k}-\lambda_{l,k})^{-2} \left[ \int_{0}^1\int_{0}^1 (\hat K_{X,t}(u,v)-K_{X,k}(u,v))\phi_{j,k}(u)\phi_{l,k}(v)\mathrm{d}u\mathrm{d}v \right]^2. \nonumber
\end{align}
The inequality in \eqref{phi:1} is valid on $\mathcal{F}_{2,n,t}$, whereas the constant $C$ on the right hand side is the constant in $\mathcal{F}_{1,n,t}$. On this event $\mathcal{F}_{1,n,t}$ it further holds that 
\begin{align*}
\left( 1-4 C n^{\frac{2(1+\mu)}{\mu+2\nu}} \mathcal{D}_t^2 \right)^{-1} \leq 2
\end{align*}
which implies, that on $\mathcal{F}_{3,n,t}$, it holds that 
\begin{align}
 || \hat \phi_{j,t}-\phi_{j,k} ||_2^2 \leq 16   R_{j,t}^{(\phi)}. \label{phi:2}
\end{align}
Note that Equation (5.22) in \citeapp{HH07_app} states that 
\begin{align}
E\left[ R_{j,t}^{(\phi)} \right] =O\left(j^2n^{-1}\right) \label{RPhi}
\end{align}
uniformly in $1\leq j \leq m$ (see also the corresponding proof of Equation (5.22) in Section 5.3 in \citealpapp{HH07_app}). Note that \eqref{phi:2} obviously implies that on $\mathcal{F}_{3,n,t}$,
\begin{align}
 || \hat \phi_{j,t}-\phi_{j,k} ||_2 \leq 4 {\left(R_{j,t}^{(\phi)}\right)}^{\frac{1}{2}} \label{phi:2.1}
\end{align}
of which the right hand side has the property $E\left[ \left( R_{j,t}^{(\phi)}\right)^{1/2} \right] \leq E\left[ R_{j,t}^{(\phi)} \right]^{1/2}=O\left( j n^{-1/2} \right)$ uniformly over $1 \leq j \leq m$, what follows from Jensen's inequality and \eqref{RPhi}. 

These observations imply that
\begin{align}
\prob \left(n m^{-3} \sum_{j=1}^m || \hat \phi_{j,t}-\phi_{j,k} ||_2^2>c \right) &\leq \prob \left( 16 n m^{-3} \sum_{j=1}^m R_{j,t}^{(\phi)} >c \right)+ \prob \left( \mathcal{F}_{3,n,t}^c \right) \nonumber \\
& \leq \frac{n m^{-3} \sum_{j=1}^m E\left[R_{j,t}^{(\phi)} \right]}{c/16} + o(1) \label{phi:3}
\end{align}
by the Markov inequality. The numerator on the right hand side of \eqref{phi:3} is bounded above as a consequence of \eqref{RPhi} and Assumptions 4 \& 5, and thus $ \sum_{j=1}^m|| \phi_{j,t}-\phi_{j,k}||_2^2=O_p\left( n^{-1}m^3 \right)$. From this and Assumptions 4 \& 5, of which the former is slightly stronger than in \citeapp{HH07_app} and \citeapp{S09_app}, we conclude for the first summand in $R_{3,3,t}$, 
\begin{align*}
|| K_{zX,k}||_2 \cdot || \alpha_t||_2 \sum_{j=1}^m || \hat \phi_{j,t}-\phi_{j,k} ||_2^2&=O_p(n^{-1}m^3) \\
&=O_p\left(n^{\frac{3-\mu-2\nu}{\mu+2\nu}}\right) \\
&=O_p(n^{-1/2})
\end{align*}
because $\nu>3-\mu/2$. Note that from our observations for \eqref{phi:2.1}, we can further conclude 
\begin{align*}
\sum_{j=1}^m || \hat \phi_{j,t}-\phi_{j,k} ||_2 = O_p(n^{-1/2}m^{2})
\end{align*}
using similar arguments as before. We have $n^{-1/2}m^{2}=n^{\frac{2-\mu/2-\nu}{\mu+2 \nu}}=o(1)$ by Assumption 4, which allows to conclude in combination with \eqref{kzxt} and Assumption 2, that $R_{3,1,t}=O_p(n^{-1/2})$. 

Using similar arguments as for \eqref{phi:3}, allows us to conclude for the second term in $R_{3,3,t}$:
\begin{align*}
\prob\left( n^{1/2}\sum_{j=1}^m || \hat \phi_{j,t}-\phi_{j,k} ||_2 \cdot |a_{j,t}^*|>c\right) &\leq \prob \left( 4 n^{1/2} \sum_{j=1}^m \left(R_{j,t}^{(\phi)}\right)^{1/2} C_a j^{-\nu} >c \right)+ \prob \left( \mathcal{F}_{3,n,t}^c \right) \nonumber \\
& \leq \frac{n^{1/2} \sum_{j=1}^m E\left[R_{j,t}^{(\phi)} \right]^{1/2} j^{-\nu}}{c/(4C_a)} + o(1),
\end{align*}
where the numerator on the right hand side of the last inequality is bounded above thanks to Assumptions 4--5 as well as our observation in \eqref{phi:2.1}. An analogue argument shows $R_{3,2,t}=O_p(n^{-1/2})$ (see also points 3 and 4 in Assumption 2 to see this).

Combining arguments implies $\hat \beta_t-\beta_t=O_p(n^{-1/2})$ for every $1\leq t \leq T$, which concludes the proof of the first result in Theorem \ref{prop:twise}. 
Turning to $\hat \alpha_t$ note that
\begin{align*}
||\hat \alpha_t-\alpha_t||_2^2\leq 3\sum_{j=1}^m (\hat a_{j,t}-a_{j,t}^*)^2 + 3 m\sum_{j=1}^m (a_{j,t}^*)^2|| \hat \phi_{j,t}-\phi_{j,k} ||_2^2+3 \sum_{j=m+1}^\infty (a_{j,t}^*)^2.
\end{align*}
The results in \citeapp{HH07_app} and \citeapp{S09_app} immediately translate to\\ $m\sum_{j=1}^m (a_{j,t}^*)^2|| \hat \phi_{j,t}-\phi_{j,k} ||_2^2$ and $\sum_{j=m+1}^\infty a_{j,t}^*$ which are both $O_p\left(n^{\frac{1-2\nu}{\mu+2\nu}}\right)$. The remaining term can be split according to 
\begin{align}
\sum_{j=1}^m (\hat a_{j,t}-a_{j,t}^*)^2 \leq & 2\sum_{j=1}^m(\hat \lambda_{j,t}^{-1} \langle \hat K_{yX,t}^{\#}-\hat \beta_t \hat K_{zX,t}^{\#},\hat \phi_{j,t} \rangle-a_{j,t}^*)^2+ 2\sum_{j=1}^m(\hat \lambda_{j,t}^{-1} \langle r_{y,t}r_{x,t}-\hat \beta_t  r_{z,t}r_{x,t},\hat \phi_{j,t} \rangle)^2 \label{eq:spl1}
\end{align}
with $\hat K_{yX,t}^{\#}:=n^{-1}\sum_{i=1}^n (y_{it}-E[y_{it}])(X_{it}-E[X_{it}])$, $\hat K_{zX,t}^{\#}:=n^{-1}\sum_{i=1}^n (z_{it}-E[z_{it}])(X_{it}-E[X_{it}])$, $r_{x,t}:=E[X_{it}]-\bar X_t$, $r_{y,t}:=E[y_{it}]-\bar y_t$ and $r_{z,t}:=E[z_{it}]-\bar z_t$. Note that $||r_{x,t}||_2,|r_{y,t}|$ and $|r_{z,t}|$ all correspond to errors from parametric estimation problems and are thus of order $n^{-1/2}$. Bounds on $\hat \lambda_{j,t}-\lambda_{j,k}$ as well as $||\hat \phi_{j,t}-\phi_{j,k}||_2$ are asymptotically equivalent for data centered around their arithmetic mean and data centered around their population expectation. Together with the above arguments it follows that the first term in \eqref{eq:spl1} is asymptotically equivalent to the corresponding term in \citeapp{S09_app}, implying $\sum_{j=1}^m(\hat \lambda_{j,t}^{-1} \langle \hat K_{yX,t}^{\#}-\hat \beta_t \hat K_{zX,t}^{\#},\hat \phi_{j,t} \rangle-a_{j,t}^*)^2=O_p\left(n^{\frac{1-2\nu}{\mu+2\nu}}\right)$. 
Now, define the event 
\begin{align*}
\mathcal{F}_{4,n,t}:=\{|\hat \lambda_{j,t}-\lambda_{j,k}|<\lambda_{j,k} / 2: \ 1\leq j \leq m   \}
\end{align*} 
for which we conclude $\prob( \mathcal{F}_{4,n,t}^c)=o(1)$ for any $1 \leq t \leq T$ as $n\to \infty$ as a consequence of \eqref{lambdat}. On this event the second term in \eqref{eq:spl1} can be bounded according to
\begin{align*}
\sum_{j=1}^m(\hat \lambda_{j,t}^{-1} \langle r_{y,t}r_{X,t}-\hat \beta_t  r_{z,t}r_{X,t},\hat \phi_{j,t}  \rangle)^2&\leq 8\sum_{j=1}^m\lambda_{j,k}^{-2} r_{y,t}^2||r_{X,t}||^2_2 + 8\sum_{j=1}^m\lambda_{j,k}^{-2} \hat \beta_t^2 r_{z,t}^2 ||r_{X,t}||^2_2 \\
&=O_p\left(n^{\frac{1+2\mu-2\mu-4\nu}{\mu+2\nu}}\right)=o_p\left(n^{\frac{1-2\nu}{\mu+2\nu}}\right).
\end{align*}
Finally combining arguments yields $||\hat \alpha_t-\alpha_t||_2^2=O_p(n^{\frac{1-2\nu}{\mu+2\nu}})$ for any $1\leq t \leq T$ as $n\to \infty$, which concludes the proof of the second part of Theorem \ref{prop:twise}. $\blacksquare$ \\

\subsection{Proof of Lemma \ref{lem:UNI}}\label{sec:APAL2}

In what follows we show that the quantities $\hat \alpha_t^{(\Delta)}$ are consistent for $\alpha_t^{(\Delta)}$ in the $L^2$ norm, uniformly over $1\leq t \leq T$. The remaining claims in the Lemma are required for this result to hold and are validated en route.

We begin introducing additional notation and listing a number of basic observations, which are a consequence of the iid sampling scheme in the cross section as well as stationarity of the regressors and the error over time within regimes. Note that since the random variables $\{(X_{it},z_{it},\epsilon_{it}) \ t\in G_k, \ 1\leq i \leq n\}$ are stationary, expectations of the below statistics calculated from these random variables do not vary over index $t$ for a given regime $k$. In order to reduce the complexity of our notation, however, we do not make this invariance explicit in every step. For the following properties we also use the results in \citeapp{HH07_app} and \citeapp{HK10_app}.

\begin{itemize}
\item Based on the above convention for our notation, we conclude, using the results in \citeapp{HK10_app}, our first observation:
\begin{align*}  
& \prob\left(\max_{1\leq t \leq T } \mathcal{D}_t^2>c\right)\leq \sum_{k=1}^K \sum_{t\in G_k} \prob\left(\mathcal{D}_t^2 >c \right)\leq K \max_{1 \leq k \leq K} |G_k|\frac{E \left[\mathcal{D}_t^2 \right]}{c}=O(n^{\delta-1}) =o(1), 
\end{align*}
since $|G_k|\propto T \propto n^\delta$ according to Assumption 3, which we will use in what follows without reference.

\item Further, empirical variances of $z_{it}$ and $\epsilon_{it}$ behave according to 
\begin{align}
&\prob\left(\max_{1\leq t \leq T}|\hat K_{z,t}-K_{z,k} |^2 >c \right) \nonumber \\
&\leq K \max_{1 \leq k \leq K} |G_k| \frac{n^{-1} E \left[ | (z_{it}-E[z_{it}])^2-K_{z,k}|^2 \right]}{c} + K \max_{1 \leq k \leq K} |G_k| \frac{E \left[  (\bar z_{t}-E[z_{it}])^4\right]}{c}  \nonumber \\ 
&=O(n^{\delta-1})+O(n^{\delta-2})  =o(1)\label{Uz} 
\end{align}
and similarly 
\begin{align*}
\prob\left(\max_{1\leq t \leq T}\left|n^{-1}\sum_{i=1}^n(\epsilon_{it}-\bar \epsilon_t)^2-\sigma_{\epsilon}^2 \right| >c \right)  &\leq K \max_{1 \leq k \leq K} |G_k| \frac{n^{-1} E \left[ (\epsilon_{it}^2-\sigma_{\epsilon}^2)^2 \right]}{c} + K \max_{1 \leq k \leq K} |G_k| \frac{E \left[ \bar \epsilon_t^4 \right]}{c} \\ 
&=O(n^{\delta-1})+O(n^{\delta-2})  =o(1).
\end{align*}

\item In analogy to before introduce $\hat K_{z \epsilon,t}^\#:=n^{-1}\sum_{i=1}^n (z_{it}-E[z_{it}])(\epsilon_{it}-E[\epsilon_{it}])$ and $\hat K_{X \epsilon ,t}^\#(u):=n^{-1}\sum_{i=1}^n (X_{it}(u)-E[X_{it}](u))\epsilon_{it}$ as well as $r_{\epsilon,t}:=\bar \epsilon_t $. It follows from simple moment calculations for the cross sectional empirical covariances between regressors and error that
\begin{align}
& \prob \left(\max_{1\leq t \leq T } ||\hat K_{X\epsilon,t}||_2^2>c \right)\leq \sum_{t=1}^T \prob\left(||\hat K_{X\epsilon,t}^\#||_2^2 >c/4 \right) + \sum_{t=1}^T \prob\left(|| r_{x,t}||_2^2r_{\epsilon,t}^2 >c/4 \right) \nonumber \\
&\leq K \max_{1 \leq k \leq K} |G_k|\frac{n^{-1}  \sigma_{\epsilon,k}^2 E \left[ ||X_{it}-E[X_{it}]||_2^2 \right]}{c} + K \max_{1 \leq k \leq K} |G_k|\frac{E[(\bar \epsilon_t)^2] E \left[ ||\bar X_{t}-E[X_{it}]||_2^2 \right]}{c}  \nonumber \\
&=O(n^{\delta-1})+O(n^{\delta-2})  =o(1)\label{Uxeps}
\end{align}
Similar arguments can be used to show 
\begin{align}
& \prob\left(\max_{1\leq t \leq T } |\hat K_{z\epsilon,t}|^2>c\right)=O(n^{\delta-1})+O(n^{\delta-2})  =o(1).\label{Uzeps}
\end{align}

\item Uniform consistency of the empirical covariance $\hat K_{zX,t}(u)$ can be shown with similar arguments according to 

\begin{align}
& \prob \left(\max_{1\leq t \leq T } ||\hat K_{z X,t}-K_{zX,k}||_2^2>c \right) \nonumber\\ 
\leq & \sum_{t=1}^T \prob\left(||\hat K_{z X,t}^\#-K_{zX,k}||_2^2 >c \right) + \sum_{t=1}^T \prob\left(|| r_{x,t}||_2^2r_{z,t}^2 >c \right)\nonumber\\
\leq &K \max_{1 \leq k \leq K} |G_k| \frac{n^{-1} E \left[ || (z_{it}-E[z_{it}])(X_{it}-E[X_{it}])-K_{zX,k}||_2^2 \right]}{c}\nonumber \\
&+ K \max_{1 \leq k \leq K} |G_k|\frac{E \left[ r_{z,t}^2 \right] E \left[ ||r_{x,t}||_2^2 \right]}{c} \nonumber  \\
&=O(n^{\delta-1})+O(n^{\delta-2})  =o(1).\label{Uzx}
\end{align}

\item Beyond the above observations, the following part of the proof requires the term $\sum_{j=1}^m \hat \lambda_{j,t}^{-2}||\phi_{j,k} - \hat \phi_{j,t}||_2^2$ to vanish in probability, uniformly over $1\leq t \leq T$. 

To see this, note that $E[R_{j,t}^{(\phi)}]$ as in \eqref{RPhi} does not vary over the index $t\in G_k$ within a regime $k$, but potentially across regimes $k=1,\dots,K$. This due to the stationarity of the functional regressor within regimes as postulated in Assumption 1. We thus conclude:
\begin{align}
&\prob\left( \max_{1\leq t \leq T } \sum_{j=1}^m \hat \lambda_{j,t}^{-2}||\phi_{j,k} - \hat \phi_{j,t}||_2^2 >c\right) \nonumber\\
\leq & \sum_{t=1}^T \prob\left( \sum_{j=1}^m \hat \lambda_{j,t}^{-2}||\phi_{j,k} - \hat \phi_{j,t}||_2^2 >c\right)\nonumber \\
\leq & \sum_{t=1}^T \left( \prob\left( 16\cdot 4C_\lambda^{-2} \sum_{j=1}^m j^{2\mu} R_{j,t}^{(\phi)} >c\right)+\prob \left( \mathcal{F}_{1,n,t}^c \right)+\prob \left( \mathcal{F}_{2,n,t}^c \right)+\prob \left( \mathcal{F}_{4,n,t}^c \right) \right) \nonumber\\
\leq & K \max_{1 \leq k \leq K} |G_k| \frac{ \sum_{j=1}^m E\left[R_{j,t}^{(\phi)} \right]j^{2\mu}}{c\cdot C_\lambda^2/(16\cdot 4)}+\sum_{t=1}^T \left(\prob \left( \mathcal{F}_{1,n,t}^c \right)+\prob \left( \mathcal{F}_{2,n,t}^c \right)+\prob \left( \mathcal{F}_{4,n,t}^c \right) \right) \nonumber \\
=& O \left( T n^{-1} m^{3+2\mu} \right)+\sum_{t=1}^T \left(\prob \left( \mathcal{F}_{1,n,t}^c \right)+\prob \left( \mathcal{F}_{2,n,t}^c \right)+\prob \left( \mathcal{F}_{4,n,t}^c \right) \right), \label{EndGegn}
\end{align}
where $C_{\lambda}$ is the constant from point 1 in Assumption 2. 
To obtain the second inequality, we used once more that $\hat \lambda_{j,t}\geq \lambda_{j,k}/2$ for $1\leq j \leq m$ on $\mathcal{F}_{4,n,t}$. The sequence in \eqref{EndGegn} is a null sequence because on the one hand
\begin{align*}
T n^{-1} m^{3+2\mu}=O\left(n^{\frac{3+(1+\delta)\mu-2(1-\delta)\nu}{\mu+2\nu}}\right)= o(1)
\end{align*}
thanks to Assumption 4 and on the other hand since 
\begin{align*}
\sum_{t=1}^T \prob \left( \mathcal{F}_{l,n,t}^c \right)\leq K \max_{1 \leq k \leq K} |G_k| \prob \left( \mathcal{F}_{l,n,t}^c \right)=o\left(1\right), \ l=1,2,4
\end{align*} 
as we argue next. First we observe
\begin{align*}
K \max_{1 \leq k \leq K} |G_k|\prob(\mathcal{F}_{1,n,t}^c)=&K \max_{1 \leq k \leq K} |G_k|\prob \left(C n^{\frac{2(1+\mu)}{\mu+2\nu}} \mathcal{D}_t^2 > 1/8 \right) \\
&\leq K \max_{1 \leq k \leq K} |G_k|8C n^{\frac{2(1+\mu)}{\mu+2\nu}} E[\mathcal{D}_t^2] \\
&=O\left(n^{\frac{2+(1+\delta )\mu-2(1-\delta)\nu)}{\mu+2\nu}}\right)=o(1)
\end{align*}
for any $C>0$ (cf. Assumption 4). Second, we argue that
\begin{align*}
K \max_{1 \leq k \leq K} |G_k|\prob(\mathcal{F}_{2,n,t}^c)&=K \max_{1 \leq k \leq K} |G_k|\prob(\exists 1 \leq j \leq m, \ j\neq l: \ |\hat \lambda_{j,t}-\lambda_{l,k}|^{-2}> 4 | \lambda_{j,k}-\lambda_{l,k}|^{-2}) \\
&= K \max_{1 \leq k \leq K} |G_k|\prob(\exists 1 \leq j \leq m, \ j\neq l: \ |\hat \lambda_{j,t}-\lambda_{l,k}|< \frac{1}{2} | \lambda_{j,k}-\lambda_{l,k}|) \\
&\leq K \max_{1 \leq k \leq K} |G_k|\prob(\exists 1 \leq j \leq m, \ j\neq l: \  |\hat \lambda_{j,t}-\lambda_{j,k}|> \frac{1}{2} | \lambda_{j,k}-\lambda_{l,k}|) \\
&\leq K \max_{1 \leq k \leq K} |G_k|\prob(\mathcal{D}_t> \frac{1}{2} \min\{\lambda_{j,k}-\lambda_{j+1,k}  ,\lambda_{j-1,k}-\lambda_{j,k} \}) \\
&\leq  K \max_{1 \leq k \leq K} |G_k| \frac{ E[\mathcal{D}_t^2]}{\min\{\lambda_{j,k}-\lambda_{j+1,k}  ,\lambda_{j-1,k}-\lambda_{j,k} \}^2/4}\\
&=O\left(T n^{-1}m^{2(1+\mu)}\right)\\
&=o(1)
\end{align*}
by the fact that $\mathcal{D}_t  \geq \sup_j|\hat \lambda_{j,t}-\lambda_{j,k}|$ almost surely as well as Assumptions 2--5. In lines of our arguments from the proof of Theorem \ref{prop:twise}, we conclude $K \max_{1\leq k \leq K} |G_k| \prob(\mathcal{F}_{3,n,t}^c)=o(1)$. Beyond that it holds
\begin{align*}
K\max_{1 \leq k \leq K}|G_k|\prob(\mathcal{F}_{4,n,t}^c)&\leq  K\max_{1\leq k \leq K} |G_k|\prob \left( \sup_{1 \leq j \leq m} |\hat \lambda_{j,t} - \lambda_{j,k} | >\frac{1}{2}\lambda_{m,k}\right) \\
&\leq K\max_{1\leq k \leq K} |G_k| \prob \left(\mathcal{D}_t>\frac{1}{2}\lambda_{m,k} \right) \\
&\leq K\max_{1\leq k \leq K} |G_k|  \frac{4 E \left[ \mathcal{D}_t^2 \right]}{\lambda_{m,k}^2} \\
&= O\left(n^{\delta} n^\frac{\mu-2\nu}{\mu+2\nu}\right) = o(1)
\end{align*}
again thanks to Assumption 4. 
%%%
Note that our result in \eqref{EndGegn} implies in particular that 
%%%
\begin{align*}
\prob\left(\max_{1\leq t\leq T} \sum_{j=1}^m ||\phi_{j,k} - \hat \phi_{j,t}||_2^2>c\right) =o(1),
\end{align*}
%%%
which will be used without further reference in what follows. \\

\item As a last observation, we note that $\max_{1 \leq t\leq T} ||\alpha_t ||_2 = \max_{1\leq k \leq K} || A_k ||_2$ is a constant and does not vary in $t$ and neither in $k$.

\end{itemize}

Now, turning to our concrete arguments for $\hat \alpha_{t}^{(\Delta)}$, we note that the scaling which distinguishes $\hat\alpha_{t}^{(\Delta)}$ from $\hat \alpha_t$ is composed by $\hat \sigma_{\epsilon,t}$ and the empirical eigenvalues $\hat \lambda_{j,t}$ ,$1 \leq j \leq m$. While the latter can be treated in a comparably simple way, $\hat \sigma_{\epsilon,t}$ requires closer attention. We thus begin focusing on this object and its constituents. For this purpose, define the event $\mathcal{S}_{n,t} $ for later use according to
%%%
\begin{align*}
\mathcal{S}_{n,t} &:= \left\{\left| \hat \sigma_{\epsilon,t}^2 - \sigma_{\epsilon,k}^2 \right|\leq \frac{1}{2}\sigma_{\epsilon,k}^2 \right\} .
\end{align*}
We show in a moment that $\sum_{t=1}^T \prob (\mathcal{S}_{n,t}^c )=o(1)$. However this requires some preparation since $\hat \sigma_{\epsilon,t}^2$ includes estimation errors from $\hat\beta_t$ and $\hat \alpha_t$. We thus start arguing that (i) $\prob (\max_{1\leq t \leq T} |\hat \beta_t-\beta_t|>c )=o(1)$ and (ii) $\prob (\max_{1\leq t \leq T} ||\hat \alpha_t-\alpha_t||_2>c )=o(1)$, as claimed in the lemma. Turning to the first point, note that the  estimator $\hat \beta_t$ makes multiple use of the operator $\hat \Phi_t$, which can, starting from the Riesz-Frechet representation Theorem (cf.~\citealpapp{S09_app}), be handled according to
%%%
\begin{align*}
\left| \left| \hat \Phi_t-\Phi_k \right| \right|^2_{H'} =3R_{4,1,t}+3R_{4,2,t}+3R_{4,3}.
\end{align*}
%%%
The last summand is defined as $R_{4,3}:=\left| \left| \sum_{j=m+1}^\infty \frac{\langle K_{zX,k}, \phi_{j,k} \rangle}{ \lambda_{j,k}} \phi_{j,k} \right| \right|^2_2$, which is independent of $t$ and $o(1)$ because the truncation parameter diverges at infinity and hence $R_{4,3}$ is arbitrarily small for $n$ large enough. The remaining summands are defined and handled as follows. For the first one we observe that
%%%
\begin{align*}
R_{4,1,t} := & \left| \left| \sum_{j=1}^m \left(\frac{\langle \hat K_{zX,t},\hat \phi_{j,t} \rangle}{\hat \lambda_{j,t}} - \frac{\langle  K_{zX,k}, \phi_{j,k} \rangle}{ \lambda_{j,k}} \right) \hat \phi_{j,t}  \right| \right|_2^2  \\
\leq&  2  \sum_{j=1}^m (\hat \lambda_{j,t} \lambda_{j,k})^{-2} \left[ \langle \lambda_{j,k} \hat K_{zX,t} - \hat \lambda_{j,t} K_{zX,k} , \phi_{j,k} \rangle +  \langle  \lambda_{j,k} \hat K_{zX,t} , (\hat \phi_{j,t} -  \phi_{j,k} )\rangle \right]^2 \\
\leq& 4 \sum_{j=1}^m (\hat \lambda_{j,t} \lambda_{j,k})^{-2} \left[ \langle K_{zX,k},\phi_{j,k}  \rangle^2 (\lambda_{j,k} - \hat \lambda_{j,t})^2 +  \langle \hat K_{zX,t} - K_{zX,k},\phi_{j,k}  \rangle^2 \lambda_{j,k}^2  \right] \\
& + 2 \sum_{j=1}^m (\hat \lambda_{j,t})^{-2} \langle \hat K_{zX,t} , (\hat \phi_{j,t} -  \phi_{j,k} )\rangle^2 \\
\leq & 4 \underbrace{\sum_{j=1}^m (\hat \lambda_{j,t} \lambda_{j,k})^{-2}   \langle K_{zX,k},\phi_{j,k}  \rangle^2 (\lambda_{j,k} - \hat \lambda_{j,t})^2}_{=:R_{5,1,t}} + 4 \underbrace{\sum_{j=1}^m || \hat K_{zX,t} - K_{zX,k} ||_2^2 {\hat \lambda_{j,t}}^{-2}}_{=:R_{5,2,t}}  \\
&+ 2\underbrace{\sum_{j=1}^m \hat \lambda_{j,t}^{-2} ||\hat K_{zX,t}||_2^2 || \hat\phi_{j,t}-\phi_{j,k}||^2_2}_{=:R_{5,3,t}}.
\end{align*}
For the three summands $R_{5,1,t},R_{5,2,t},R_{5,3,t}$ we use our above observation as well Assumptions 1--5 to conclude the following: \\

Ad $R_{5,1,t}$:
\begin{align*}
 &\sum_{t=1}^T \prob \left( \sum_{j=1}^m (\hat \lambda_{j,t} \lambda_{j,k})^{-2}  \langle K_{zX,k},\phi_{j,k}  \rangle^2 (\lambda_{j,k} - \hat \lambda_{j,t})^2  >c\right) \\
\leq & K\max_{1\leq k \leq K} |G_k| \prob \left( 4 \sum_{j=1}^m  \lambda_{j,k}^{-4}   \langle K_{zX,k},\phi_{j,k}  \rangle^2 (\lambda_{j,k} - \hat \lambda_{j,t})^2  >c\right) +K\max_{1\leq k \leq K} |G_k|\prob \left(\mathcal{F}_{4,n,t}^c \right)  \\
\leq & K \max_{1 \leq k \leq K} |G_k| \frac{E[\mathcal{D}_t^2] \sum_{j=1}^m \lambda_{j,k}^{-4} \langle K_{zX,k},\phi_{j,k}  \rangle^2   }{c/4} +o(1) \\
= & O(n^{\delta-1}) +o(1)=o(1).
\end{align*}
%%%
Ad $R_{5,2,t}$:
%%%
\begin{align*}
&\sum_{t=1}^T \prob \left( \sum_{j=1}^m || \hat K_{zX,t} - K_{zX,k} ||_2^2 {\hat \lambda_{j,t}}^{-2} >c \right) \\
&\leq K\max_{1\leq k \leq K} |G_k| \prob \left(4 \sum_{j=1}^m || \hat K_{zX,t} - K_{zX,k}  ||_2^2  \lambda_{j,k}^{-2}>c \right) +K\max_{1\leq k \leq K} |G_k| \prob (\mathcal{F}_{4,n,t}^c)\\
&\leq K\max_{1\leq k \leq K} |G_k|\frac{\sum_{j=1}^m \lambda_{j,k}^{-2} E\left[|| \hat K_{zX,t} - K_{zX,k}  ||_2^2 \right] }{c/4} +o(1)\\
&=O\left(n^\frac{1+(1+\delta)\mu-2(1-\delta)\nu}{\mu+2\nu}\right)+o(1) \\
&=o(1).
\end{align*}
%%%
Ad $R_{5,3,t}$:
\begin{align*}
&\sum_{t=1}^T  \prob \left(\sum_{j=1}^m \hat \lambda_{j,t}^{-2} ||\hat K_{zX,t}||_2^2 || \hat\phi_{j,t}-\phi_{j,k}||^2_2 >c\right)\\
\leq & K\max_{1\leq k \leq K} |G_k|  \prob \left( 2 || K_{zX,k}||_2^2 \sum_{j=1}^m \hat \lambda_{j,t}^{-2} || \hat\phi_{j,t}-\phi_{j,k}||^2_2 >c^{1/2}\right)\\
&+K\max_{1\leq k \leq K} |G_k| \prob \left( 2 ||\hat K_{zX,t} - K_{zX,k}||_2^2 >c^{1/4}\right)\\ 
&+K\max_{1\leq k \leq K} |G_k|  \prob \left( 2 \sum_{j=1}^m \hat \lambda_{j,t}^{-2} || \hat\phi_{j,t}-\phi_{j,k}||^2_2 >c^{1/4}\right) \\
=&o(1).
\end{align*}
Since $\prob \left(\max_{1 \leq t \leq T} R_{4,1,t} >c\right)\leq \sum_{l=1}^3\sum_{t=1}^T \prob \left(R_{5,l,t} >c/3\right)$, it follows that $\prob \left(\max_{1 \leq t \leq T} R_{4,1,t} >c\right)=o(1)$.
For $R_{4,2,t}$, defined as 
\begin{align*}
R_{4,2,t}:=\left| \left|  \sum_{j=1}^m \frac{\langle K_{zX,k}, \phi_{j,k} \rangle}{ \lambda_{j,k}} (\phi_{j,k}-\hat \phi_{j,t} ) \right| \right|_2^2,
\end{align*}
we note this expression can be most easily handled using the almost sure bound in \eqref{PhiBoundAS} according to
\begin{align*}
&K\max_{1\leq k \leq K} |G_k| \prob \left(\left| \left|  \sum_{j=1}^m \frac{\langle K_{zX,k}, \phi_{j,k} \rangle}{ \lambda_{j,k}} (\phi_{j,k}-\hat \phi_{j,t} ) \right| \right|_2^2  >c\right) \\
\leq & K\max_{1\leq k \leq K} |G_k| \prob \left( m \sum_{j=1}^m \frac{\langle K_{zX,k}, \phi_{j,k} \rangle^2}{ \lambda_{j,k}^2} \left| \left| \phi_{j,k}-\hat \phi_{j,t} \right| \right|_2^2  >c\right)\\
&K\max_{1\leq k \leq K} |G_k| \prob \left( m \mathcal{D}_t^2 \sum_{j=1}^m \frac{\langle K_{zX,k}, \phi_{j,k} \rangle^2}{ \lambda_{j,k}^2} \frac{j^{2(1+\mu)}}{C_\lambda^\prime}  >c\right)\\
\leq& K\max_{1\leq k \leq K} |G_k| \frac{m E[\mathcal{D}_t^2] \sum_{j=1}^m \langle K_{zX,k}, \phi_{j,k} \rangle^2 j^{2(1+\mu)}/ \lambda_{j,k}^2} {c \cdot  C_\lambda^\prime} \\
=& O\left(m n^{\delta-1} \right)=o(1).
\end{align*}
thanks to Assumption 2--5. In particular, these results imply
\begin{align}
\prob \left(\max_{1\leq t \leq T}  ||\hat \Phi_t -\Phi_k||_{H'}^2 >c \right)=o(1). \label{UPhi}
\end{align}

%%%%%%%%%%%%%%%%%%%%%%%

To proceed we work again on the differences $(\hat \beta_t - \beta_t)= \hat B_t^{-1} \left( R_{0,1,t}+R_{0,2,t}+R_{0,3,t} \right)$ with $R_{0,1,t}$, $R_{0,2,t}$ and $R_{0,3,t}$ as in the proof of Theorem \ref{prop:twise}. Addressing the inverse in these differences, define the $t$-wise event $Q_{n,t}:=\{|\hat B_t- B_k|\leq\frac{1}{2}B_k\}$. For this event, note that $\sum_{t=1}^T \prob \left( Q_{n,t}^c  \right) \leq R_{6,1}+R_{6,2}$, where $R_{6,1}:=\sum_{t=1}^T \prob \left( |\hat K_{z,t}-K_{z,k}|^2 >c \right)=o(1)$ as shown in \eqref{Uz}. For $R_{6,2}$ we use the arguments in \citeapp{S09_app} to obtain

%%%%%%%%%%%%%%%%%%%%%
\begin{align}
R_{6,2} := & \sum_{t=1}^T \prob \left( ||\hat \Phi_t-\Phi_k ||_{H'}^2 ||K_{zX,k} ||_2^2 + (||\hat\Phi_t-\Phi_k ||_{H'}+||\Phi_k ||_{H'})^2  ||\hat K_{zX,t}-K_{zX,k} ||_2^2 >c \right) \label{eq:PHIDEC}\\
\leq & 
\underbrace{K\max_{1\leq k \leq K} |G_k| \prob \left( ||\hat \Phi_t-\Phi_k ||_{H'}^2 ||K_{zX,k} ||_2^2 >c \right)}_{=:R_{7,1}}\nonumber\\
&+\underbrace{K\max_{1\leq k \leq K} |G_k|  \prob \left( 2 ||\hat\Phi_t-\Phi_k ||_{H'}^2 ||\hat K_{zX,t}-K_{zX,k} ||_2^2  >c^{1/2} \right)}_{=:R_{7,2}} \nonumber\\
&+\underbrace{K\max_{1\leq k \leq K} |G_k|  \prob \left(2 ||\Phi_k ||_{H'}^2 ||\hat K_{zX,t}-K_{zX,k} ||_2^2  >c^{1/2}\right) }_{=:R_{7,3}}.\nonumber
\end{align}
As shown before $R_{7,1},R_{7,3}=o(1)$. Further
\begin{align*}
R_{7,2} &\leq K\max_{1\leq k \leq K} |G_k|  \prob \left( ||\hat\Phi_t-\Phi_k||_{H'}^2 ||\hat K_{zX,t}-K_{zX,k} ||_2^2  >c^{1/2} \right) \\
& \leq K\max_{1\leq k \leq K} |G_k|  \prob \left(||\hat\Phi_t-\Phi_k||_{H'}^2 > c^{1/4}\right) + K\max_{1\leq k \leq K} |G_k|  \prob \left(||\hat K_{zX,t}-K_{zX,k} ||_2^2 > c^{1/4} \right) \\
&=o(1).
\end{align*}
For uniform consistency of $\hat \beta_t$ it remains to show that
\begin{itemize}
\item $\prob(\max_{1  \leq t \leq T} |R_{0,1,t}|>c )=o(1)$,
\item $\prob(\max_{1  \leq t \leq T} |R_{0,2,t}|>c )=o(1)$ and
\item $\prob(\max_{1  \leq t \leq T} |R_{0,3,t}|>c )=o(1)$.
\end{itemize}
For this we argue
\begin{align*}
\prob \left(\max_{1  \leq t \leq T} \left|R_{0,1,t} \right|>c \right) &=\prob\left(\max_{1  \leq t \leq T} \left|n^{-1}\sum_{i=1}^n (z_{it}^c-\Phi_k(X_{it}^c))\epsilon_{it}^c \right|>c \right) \\
 &\leq K \max_{1 \leq k \leq K}|G_k| \prob \left( \left|\hat K_{z\epsilon,t}\right|^2>c^2/4 \right)+  K \max_{1 \leq k \leq K}|G_k| \prob\left( || \Phi_k||_{H'}^2||\hat K_{\epsilon X,t}||_2^2>c^2/4 \right) \\
&=o(1)
\end{align*}
due to \eqref{Uxeps} and \eqref{Uzeps}. Further note for $R_{0,2,t}$
\begin{align*}
\prob \left(\max_{1  \leq t \leq T} \left|R_{0,2,t} \right|>c \right) &=\prob\left(\max_{1  \leq t \leq T} \left|\hat \Phi_t(\hat K_{\epsilon X,t})- \Phi_k(\hat K_{\epsilon X,t})\right|>c \right)  \\
&\leq K \max_{1 \leq k \leq K}  |G_k| \prob \left( ||\hat \Phi_t - \Phi_k||_{H'}||\hat K_{\epsilon X,t} ||_2>c \right) \\
\leq & K \max_{1 \leq k \leq K} |G_k| \prob \left(|| \hat K_{\epsilon X,t}||_2^2>c \right) \\
& + K \max_{1 \leq k \leq K} |G_k| \prob \left(||\hat \Phi_t - \Phi_k||_{H'}^2>c \right)  \\
=&o(1)
\end{align*}
as a consequence of \eqref{Uxeps} and \eqref{UPhi}. For the remaining terms, we argue along the same lines as in the proof of Theorem \ref{prop:twise}:
\begin{align*}
\prob \left(\max_{1  \leq t \leq T} |R_{0,3,t}|>c\right)&\leq\prob\left(\max_{1  \leq t \leq T} |R_{1,1,t}|>c\right)+\prob\left(\max_{1  \leq t \leq T} |R_{1,2,t}|>c\right)
\end{align*}
\begin{align*}
\prob\left(\max_{1  \leq t \leq T} |R_{1,1,t}|>c\right)&\leq \prob\left(\max_{1  \leq t \leq T} ||\alpha_t ||_2 \cdot || \hat K_{zX,t}-K_{zX} ||_2>c\right) \\
&=o(1)
\end{align*}
because of \eqref{Uzx}. The remaining term was shown to be bounded according to $R_{1,2}\leq R_{2,1,t}+R_{2,2,t}$, where the two summands are defined above. While $R_{2,1}=O(n^{-1/2})$ deterministically and independently of $t$, note for the second summand $ R_{2,2,t} \leq R_{3,1,t}+ R_{3,2,t}+R_{3,3,t} $ as before and further: 
\begin{align*}
&\prob \left(\max_{1 \leq t \leq T}|R_{3,1,t}|>c \right) \\
 \leq & K\max_{1 \leq k \leq K} |G_k| \prob\left(  \sum_{j=1}^m || \hat K_{zX,t}-K_{zX,k}||_2 (|| \hat \phi_{j,t}-\phi_{j,k} ||_2 \cdot ||  A_k ||_2 + |a_{j,t}^*|)  >c\right) \\
\leq & K\max_{1 \leq k \leq K} |G_k|  \prob\left( 2||\hat K_{zX,t}-K_{zX,k}||_2 \cdot ||  A_k ||_2  \sum_{j=1}^m   || \hat  \phi_{j,t}-\phi_{j,k} ||_2   >c\right) \\
 &+ K\max_{1 \leq k \leq K} |G_k|  \prob\left(2 ||\hat K_{zX,t}-K_{zX,k}||_2  \sum_{j=1}^m | a_{j,t}^*|  >c\right) \\
\leq & K\max_{1 \leq k \leq K} |G_k|  \prob\left( 2||\hat K_{zX,t}-K_{zX,k}||_2 \cdot ||  A_k ||_2 4  \sum_{j=1}^m  \left(R_{j,t}^{(\phi)}\right)^{1/2}  >c\right) \\
 &+ K\max_{1 \leq k \leq K} |G_k|  \prob\left(\mathcal{F}_{3,n,t}^c\right) + o(1) \\
\leq & K\max_{1 \leq k \leq K} |G_k| \frac{||A_k ||_2^2\cdot E\left[|| \hat K_{zX,t}-K_{zX,k}||_2^2\right]^{\frac{1}{2}} m^\frac{1}{2} \left( \sum_{j=1}^m E \left[ R_{j,t}^{(\phi)}\right]\right)^\frac{1}{2} }{c}+ o(1)+o(1) \\
=&O(n^{\delta-1}m^{2})+o(1)=o(1),
\end{align*}
due to Assumptions 2--5 and our above observations. Further for $R_{3,2,t}$ similar arguments yield:
\begin{align*}
\prob \left(\max_{1 \leq t \leq T}|R_{3,2,t}|>c \right) \leq & K\max_{1 \leq k \leq K} |G_k| \prob\left( ||A_k ||_2\sum_{j=1}^m |\langle K_{zX,k},\phi_{j,k} \rangle | \cdot ||\hat \phi_{j,t}-\phi_{j,k} ||_2 >c\right) \\
\leq & K\max_{1 \leq k \leq K} |G_k|  \frac{||A_k||_2^2 \cdot 16 \cdot C_{zX}^2 m  \sum_{j=1}^m j^{-2(\mu+\nu)} E\left[ R^{(\phi)}_{j,t}\right] }{c} + K\max_{1 \leq k \leq K} |G_k|  \prob\left(\mathcal{F}_{3,n,t}^c\right) \\
=& O(m n^{\delta-1})+o(1)=o(1).
\end{align*}
Similarly, we argue for $R_{3,3,t}$,
\begin{align*}
\prob\left(\max_{1 \leq t \leq T}|R_{3,3,t}|>c\right) \leq & K\max_{1 \leq k \leq K} |G_k|  \prob\left(|| K_{zX,k}||_2 ||  A_k ||_2 \sum_{j=1}^m || \hat \phi_{j,t}-\phi_{j,k} ||_2^2>c\right)  \\
& +K\max_{1 \leq k \leq K} |G_k|  \prob \left(|| K_{zX,k}||_2  \sum_{j=1}^m || \hat \phi_{j,t}-\phi_{j,k} ||_2 |a_{j,t}^*|>c\right) \\
&=o(1)
\end{align*}
where the first term is a null sequence as implied by  \eqref{EndGegn}. The second term is of the order $O(n^{\delta-1}m)=o(1)$ which follows from analogous arguments as used for $R_{3, 2,t}$.

Combining our above arguments, we conclude $\prob(\max_{1 \leq t \leq T} (\hat \beta_t - \beta_t)^2>c)=o(1)$ as claimed in the lemma. \\

Now, turning to the estimation error in $\hat \alpha_t$ we employ upper bounds 
\begin{align*}
\prob\left(\max_{1 \leq t \leq T} ||\hat \alpha_t-\alpha_t||_2^2>c\right)  &\leq K\max_{1 \leq k \leq K} |G_k| \prob\left( ||\hat \alpha_t-\alpha_t||_2^2 >c\right) \\
&\leq R_{8,1}+R_{8,2}+R_{8,3}+R_{8,4}.
\end{align*}
While the four summands on the right and side are defined below, the term $\sum_{j=m+1}^\infty {a_{j,t}^*}^2$ does not appear in the upper bound, as it is a null sequence and hence arbitrarily small for sufficiently large $n$ (cf. Assumptions 2,4 and 5).
The terms $R_{8,1}-R_{8,4}$ are as follows: \\
 
Ad $R_{8,1}$:
\begin{align*}
R_{8,1}&:=K\max_{1 \leq k \leq K} |G_k|  \prob\left( \sum_{j=1}^m {\hat \lambda_{j,t}}^{-2} \left(n^{-1}\sum_{i=1}^n  \langle X_{it}^c,\hat \phi_{j,t} \rangle \epsilon_{it}^c \right)^2  >c\right) \\
&\leq K\max_{1 \leq k \leq K} |G_k|  \frac{4 \sum_{j=1}^m \lambda_{j,k}^{-2} E\left[ ||\hat K_{X\epsilon,t} ||_2^2 \right]}{c}+ K\max_{1 \leq k \leq K} |G_k|  \prob\left( \mathcal{F}_{4,n,t}^c \right) \\
&=O \left( n^\frac{1+(1+\delta)\mu-2(1-\delta)\nu}{\mu+2\nu}\right)+o(1)=o(1)
\end{align*}
due to Assumptions 2--5.

Ad $R_{8,2}$:
\begin{align*}
R_{8,2} :=&K\max_{1 \leq k \leq K} |G_k|  \prob\left( \sum_{j=1}^m {\hat \lambda_{j,t}}^{-2} \left(n^{-1}\sum_{i=1}^n  \langle X_{it}^c,\hat \phi_{j,t} \rangle z_{it}^c \right)^2 (\hat \beta_t -\beta_t)^2  >c\right)\\
\leq & K\max_{1 \leq k \leq K} |G_k|  \prob\left( 4 \sum_{j=1}^m {\lambda^{-2}_{j,k}}  \langle \hat K_{zX,t},\hat \phi_{j,t} \rangle ^2 (\hat \beta_t -\beta_t)^2  >c\right)
 + K\max_{1 \leq k \leq K} |G_k|  \prob\left( \mathcal{F}_{4,n,t}^c \right) \\
 \leq & K\max_{1 \leq k \leq K} |G_k|  \prob\left( \sum_{j=1}^m {\lambda^{-2}_{j,k}}  \langle  K_{zX,k}, \phi_{j,k} \rangle ^2 (\hat \beta_t -\beta_t)^2  >c\right) +2K\max_{1 \leq k \leq K} |G_k|   \prob\left( (\hat \beta_t -\beta_t)^2>c\right) \\
& +K\max_{1 \leq k \leq K} |G_k|  \prob\left( ||K_{zX,k}||_2^2 \sum_{j=1}^m {\lambda^{-2}_{j,k}}  ||\phi_{j,k} - \hat \phi_{j,t}||_2^2>c\right) \\ 
&+ K\max_{1 \leq k \leq K} |G_k|  \prob\left( ||K_{zX,k}-\hat K_{zX,t}||_2^2>c\right) \\ &+K\max_{1 \leq k \leq K} |G_k| \prob\left( \mathcal{F}_{4,n,t}^c \right) \\
=&o(1)
\end{align*}
which follows from our above observations. \\

Ad $R_{8,3}$: \\

 With $a_{j,t}:=\langle \alpha_t,\hat \phi_{j,t} \rangle=\langle A_k,\hat \phi_{j,t} \rangle$, we obtain
\begin{align*}
R_{8,3}&:=K\max_{1 \leq k \leq K} |G_k| \prob\left( \left|\left|  \sum_{j=1}^m (a_{j,t}^*-a_{j,t}) \hat \phi_{j,t} \right|\right|_2^2 >c\right)  \\
&\leq K\max_{1 \leq k \leq K} |G_k| \prob\left(||A_k ||_2^2 \sum_{j=1}^m   \left|\left| \phi_{j,k} - \hat \phi_{j,t}\right|\right|_2^2 >c\right)\\
& =o(1)
\end{align*}
as a consequence of \eqref{EndGegn}. \\

Ad $R_{8,4}$:
\begin{align*}
R_{8,4}&:= K\max_{1 \leq k \leq K} |G_k| \prob\left( \left|\left|  \sum_{j=1}^m a_{j,t}^* (\hat \phi_{j,t}-\phi_{j,k}) \right|\right|_2^2>c\right) \\
&\leq K\max_{1 \leq k \leq K} |G_k| \prob\left( m  \sum_{j=1}^m {a_{j,t}^*}^2 \left|\left|    \hat \phi_{j,t}-\phi_{j,k} \right|\right|_2^2>c\right) \\
&= O \left( m n^{\delta-1} \right)=o(1),
\end{align*}
which follows from the arguments used for $R_{4,2,t}$, because $|\langle K_{zX,k}, \phi_{j,k} \rangle |/\lambda_{j,k}$ and $|a_{j,t}^*|$ are of the same order in $j$. Combining arguments yields $ \prob\left( \max_{1\leq t\leq T}||\hat \alpha_t-\alpha_t||_2^2 >c\right)=o(1)$ proving the second claim of the Lemma. \\

This would already justify classification on the distances $||\hat \alpha_t-\hat \alpha_s||_2^2$. However, as scaled versions of the estimators are employed the behavior of the scaling, which itself is random, needs to be explored. Contributing to this, now turn to the event $\mathcal{S}_{n,t}$, for which
\begin{align*}
K\max_{1 \leq k \leq K} |G_k| \prob \left( \mathcal{S}_{n,t}^c \right) \leq & K\max_{1 \leq k \leq K} |G_k|\prob \left( \left|  n^{-1}\sum_{i=1}^n \left( {\epsilon_{it}^c}^2- \sigma_{\epsilon,k}^2 + 2\epsilon_{it}^c\tilde r_{it} +  \tilde r_{it}^2 \right) \right| >\frac{1}{2}\sigma_{\epsilon,k}^2 \right) \\
\leq &K\max_{1 \leq k \leq K} |G_k|\prob \left( \left|  n^{-1}\sum_{i=1}^n \left( {\epsilon_{it}^c}^2- \sigma_{\epsilon,k}^2 + 2\epsilon_{it}^c\tilde r_{it} +  \tilde r_{it}^2 \right) \right| >\frac{1}{2} \min_{1 \leq k \leq K}\sigma_{\epsilon,k}^2 \right) \\
\leq & R_{9,1}+R_{9,2}+R_{9,3}
\end{align*}
where $\tilde r_{it} := z_{it}^c(\beta_t-\hat \beta_t) + \langle X_{it}^c ,\alpha_t-\hat \alpha_t \rangle$, $\min_{1 \leq k \leq K}\sigma_{\epsilon,k}^2$ a constant, and $R_{9,1}-R_{9,3}$ are as follows.\\

Ad $R_{9,1}$: 
\begin{align*}
R_{9,1}&:=K\max_{1 \leq k \leq K} |G_k| \prob\left( \left|  n^{-1}\sum_{i=1}^n \left( {\epsilon_{it}^c}^2- \sigma_{\epsilon,k}^2 \right) \right| >c \right) \\
& \leq K\max_{1 \leq k \leq K} |G_k| \prob\left( \left| n^{-1}\sum_{i=1}^n \left( {\epsilon_{it}}^2- \sigma_{\epsilon,k}^2 \right) \right| >c \right) + K\max_{1 \leq k \leq K} |G_k| \prob\left(   \left(n^{-1}\sum_{i=1}^n \epsilon_{it} \right)^2>c \right) \\ 
& \leq K\max_{1 \leq k \leq K} |G_k|  \frac{n^{-1}E\left[ \left( {\epsilon_{it}}^2- \sigma_{\epsilon,k}^2 \right)^2 \right]}{c} +K\max_{1 \leq k \leq K} |G_k| \frac{n^{-1}E\left[\epsilon_{it}^2 \right]}{c}  \\
&=o(1).
\end{align*}
Ad $R_{9,2}$:
\begin{align*}
R_{9,2}:=&K\max_{1 \leq k \leq K} |G_k| \prob\left(\left| n^{-1}\sum_{i=1}^n \left( {\epsilon}_{it}^c \tilde r_{it} \right) \right| >c \right) \\
&\leq R_{10,1}+R_{10,2}
\end{align*}
with $R_{10,1}-R_{10,2}$ as follows:
\begin{align*}
R_{10,1}&:=K\max_{1 \leq k \leq K} |G_k| \prob\left(\left| (\beta_t-\hat \beta_t) \hat K_{z\epsilon,t}  \right| >c \right) \\
&\leq  K\max_{1 \leq k \leq K} |G_k| \prob\left( \hat K_{z\epsilon,t}^2 >c \right) + K\max_{1 \leq k \leq K} |G_k| \prob\left( (\beta_t-\hat \beta_t)^2 >c \right) =o(1)
\end{align*}
by \eqref{Uzeps} and the above results. Further
\begin{align*}
R_{10,2}&:=K\max_{1 \leq k \leq K} |G_k| \prob\left(\left|  \langle \hat K_{X\epsilon,t},\alpha_t-\hat \alpha_t \rangle \right| >c \right) \\
&\leq  K\max_{1 \leq k \leq K} |G_k| \prob\left(||\hat K_{X\epsilon,t}||_2^2 >c \right)+K\max_{1 \leq k \leq K} |G_k| \prob\left(||\alpha_t-\hat \alpha_t||_2^2 >c \right) =o(1)
\end{align*}
by \eqref{Uxeps} and the above results on $\hat \alpha_t$.\\

Ad $R_{9,3}$:
\begin{align*}
R_{9,3}:= & K\max_{1 \leq k \leq K} |G_k| \prob\left( \left|  n^{-1}\sum_{i=1}^n  \tilde r_{it}^2 \right| >c \right) \\
\leq & \underbrace{K\max_{1 \leq k \leq K} |G_k| \prob\left( | \hat K_{z,t}| \cdot(\hat \beta_t -\beta_t)^2 > c \right)}_{=:R_{11,1}}\\
&+\underbrace{K\max_{1 \leq k \leq K} |G_k| \prob\left(\frac{1}{n}\sum_{i=1}^n|| X_{it}^c ||_2^2|| \hat \alpha_t -\alpha_t ||_2^2  >c\right)}_{=:R_{11,2}} 
\end{align*}
with $R_{11,1}$ and $R_{11,2}$ to be treated as follows.
\begin{align*}
R_{11,1}:= & K\max_{1 \leq k \leq K} |G_k| \prob\left( |K_{z,k}|\cdot(\hat \beta_t -\beta_t)^2   >c \right)\\
\leq & K\max_{1 \leq k \leq K} |G_k| \prob\left( |\hat K_{z,t}-K_{z,k}|(\hat \beta_t -\beta_t)^2   >c \right)+K\max_{1 \leq k \leq K} |G_k| \prob\left( |K_{z,k}|\cdot (\hat \beta_t -\beta_t)^2   >c \right) \\
\leq & K\max_{1 \leq k \leq K} |G_k|\prob\left( |\hat K_{z,t}-K_{z,k}|   >c \right)+K\max_{1 \leq k \leq K} |G_k| \prob\left( (\hat \beta_t -\beta_t)^2  >c\right) \\
&+K\max_{1 \leq k \leq K} |G_k| \prob\left( |K_{z,k}|(\hat \beta_t -\beta_t)^2   >c \right)=o(1)
\end{align*}
by \eqref{Uz} and the above results. Further it holds that
\begin{align*}
R_{11,2}:=& K\max_{1 \leq k \leq K} |G_k| \prob\left(\frac{1}{n}\sum_{i=1}^n|| X_{it}^c ||_2^2|| \hat \alpha_t -\alpha_t ||_2^2 >c  \right) \\
\leq & K\max_{1 \leq k \leq K} |G_k| \prob\left( \frac{1}{n}\sum_{i=1}^n \left| || X_{it}^c ||_2^2-E\left[ ||X_{it}^c||_2^2 \right] \right| \cdot  || \hat \alpha_t -\alpha_t ||_2^2 >c \right) \\
&+K\max_{1 \leq k \leq K} |G_k| \prob\left(E\left[ ||X_{it}^c||_2^2 \right] || \hat \alpha_t -\alpha_t ||_2^2 >c  \right)\\
\leq &  K \max_{1 \leq k \leq K} |G_k| \frac{n^{-1} E\left[\left( || X_{it}^c||_2^2-E\left[ ||X_{it}^c||_2^2 \right]  \right)^2 \right]}{c} +K\max_{1 \leq k \leq K} |G_k| \prob\left(  || \hat \alpha_t -\alpha_t ||_2^2 >c  \right) \\
&+K\max_{1 \leq k \leq K} |G_k|\prob\left(E\left[ ||X_{it}^c||_2^2 \right] || \hat \alpha_t -\alpha_t ||_2^2 >c \right)\\
=& O(n^{\delta-1})+o(1)+o(1)=o(1)
\end{align*}
in light of our above findings. Combining results yields $ K\max_{1 \leq k \leq K} |G_k|\prob \left(\mathcal{S}_{n,t}^c \right)=o(1)$. \\ 

Now, finally turning to $\hat \alpha_t^{(\Delta)}$, for sufficiently large $n$
\begin{align*}
\prob\left(\max_{1 \leq t \leq T} \left|\left| \hat \alpha_t^{(\Delta)} -\alpha_t^{(\Delta)} \right|\right|_2^2 > c \right) &\leq K\max_{1 \leq k \leq K} |G_k| \prob\left( \left|\left| \hat \alpha_t^{(\Delta)} -\alpha_t^{(\Delta)} \right|\right|_2^2 > c \right) \\
& \leq R_{12,1}+R_{12,2}
\end{align*}
with 
\begin{align*}
R_{12,1}&:=K\max_{1 \leq k \leq K} |G_k| \prob\left(  \sum_{j=1}^m (\hat a_{j,t}-a_{j,t})^2\frac{\hat \lambda_{j,t}}{\hat \sigma_{\epsilon,t}^2} > c  \right) \\
\text{and } \quad R_{12,2}&:=K\max_{1 \leq k \leq K} |G_k| \prob\left(  \left|\left|\sum_{j=1}^m \left(\frac{\hat \lambda_{j,t}^{1/2}}{\hat \sigma_{\epsilon,t}}\hat \phi_{j,t}a_{j,t}-\frac{\lambda_{j,k}^{1/2}}{\sigma_{\epsilon,k}} \phi_{j,k}a_{j,t}^*\right) \right|\right|_2^2 > c  \right).
\end{align*}
$R_{12,1}$ can be decomposed according to
\begin{align*}
R_{12,1}&\leq R_{13,1}+R_{13,2}+K\max_{1 \leq k \leq K} |G_k|\prob\left( \mathcal{S}_{n,t}^c\right)
\end{align*}
where
\begin{align*}
R_{13,1}&:= K\max_{1 \leq k \leq K} |G_k| \prob\left( \sigma_{\epsilon,k}^{-2}\sum_{j=1}^m\hat \lambda_{j,t}^{-1}\langle \hat K_{zX,t},\hat \phi_{j,t} \rangle^2 (\beta_{t}-\hat \beta_{t})^2> c  \right)\\
\text{and} \quad R_{13,2}&:=K\max_{1 \leq k \leq K} |G_k| \prob\left(\sigma_{\epsilon,k}^{-2}\sum_{j=1}^m \hat \lambda_{j,t}^{-1}\langle \hat K_{X\epsilon,t},\hat \phi_{j,t} \rangle^2 > c \right)
\end{align*}
because $\hat \sigma_{\epsilon,t}^{-2}\leq 2\sigma_{\epsilon,k}^{-2}$ on $\mathcal{S}_{n,t}$. Noting that $\sigma_{\epsilon,k}^{-2}$ is obviously bounded above by a constant, these terms in turn behave as follows:
\begin{align*}
R_{13,1}\leq & K\max_{1 \leq k \leq K} |G_k|\prob\left( \sum_{j=1}^m \lambda_{j,k}^{-1} \langle K_{zX,k},\phi_{j} \rangle^2 (\beta_{t}-\hat \beta_{t})^2  > c \right)+2 K\max_{1 \leq k \leq K} |G_k| \prob\left( (\beta_{t}-\hat \beta_{t})^2  > c \right) \\
&+ K\max_{1 \leq k \leq K} |G_k| \prob\left( \sum_{j=1}^m  \lambda_{j,k}^{-1} || \hat K_{zX,t}-K_{zX,k}||_2^2> c  \right)+ K\max_{1 \leq k \leq K} |G_k| \prob\left( \sum_{j=1}^m  \lambda_{j,k}^{-1} || \hat \phi_{j,t}-\phi_{j,k}||_2^2> c  \right)\\
&+ K\max_{1 \leq k \leq K} |G_k| \prob\left( \mathcal{F}_{4,n,t}^c  \right) \\
&=o(1)
\end{align*}
which follows from our above arguments. Further we conclude
\begin{align*}
R_{13,2}&\leq K\max_{1 \leq k \leq K} |G_k|\frac{\sum_{j=1}^m \lambda_{j,k}^{-1} E\left[ || \hat K_{X\epsilon,t} ||_2^2  \right]  }{c}+K\max_{1 \leq k \leq K} |G_k| \prob\left( \mathcal{F}_{4,n,t}^c  \right) \\
&=O(n^{\delta-1}m^{1+\mu})=o(1)
\end{align*}
as consequence of Assumptions 2--5. Now turning to $R_{12,2}$ note that
\begin{align*}
R_{12,2}\leq & R_{14,1}+ R_{14,2}
\end{align*}
where
\begin{align*}
R_{14,1}:=K\max_{1 \leq k \leq K} |G_k| \prob\left(  \left|\left|\sum_{j=1}^m  \left(\frac{\hat \lambda_{j,t}^{1/2}\sigma_{\epsilon,k}}{\sigma_{\epsilon,k}\hat \sigma_{\epsilon,t}}\hat \phi_{j,t}-\frac{\lambda_{j,k}^{1/2}\hat \sigma_{\epsilon,t}}{\sigma_{\epsilon,k}\hat \sigma_{\epsilon,t}} \phi_{j,k}\right)a_{j,t}^* \right|\right|_2^2 > c \right)
\end{align*}
and
\begin{align*}
R_{14,2}&=K\max_{1 \leq k \leq K} |G_k| \prob\left(  \left|\left|\sum_{j=1}^m  (a_{j,t}^*-a_{j,t})\frac{\hat \lambda_{j,t}^{1/2}}{\hat \sigma_{\epsilon,t}}\hat \phi_{j,k}\right|\right|_2^2 > c  \right).
\end{align*}
Note for $R_{14,1}$:
\begin{align*}
R_{14,1}&\leq R_{15,1}+R_{15,2}+R_{15,3}+K\max_{1 \leq k \leq K} |G_k|\prob(\mathcal{S}_{n,t}^c)
\end{align*}
with
\begin{align*}
R_{15,1}&:=K\max_{1 \leq k \leq K} |G_k|\prob\left( \sum_{j=1}^m (a_{j,t}^*)^2\hat \lambda_{j,t}(\sigma_{\epsilon,k}-\hat    \sigma_{\epsilon,t})^2 > c \right)\\
R_{15,2}&:=K\max_{1 \leq k \leq K} |G_k| \prob\left( m\sum_{j=1}^m (a_{j,t}^*)^2\hat \lambda_{j,t}\hat    \sigma_{\epsilon,t}^2 ||\phi_{j,k} - \hat \phi_{j,t} ||_2^2   > c\right)\\
R_{15,3}&:=K\max_{1 \leq k \leq K} |G_k| \prob\left( \sum_{j=1}^m (a_{j,t}^*)^2 \hat    \sigma_{\epsilon,t}^2 \left(\hat \lambda_{j,t}^{1/2} - \lambda_{j,k}^{1/2}\right)^2 > c \right).
\end{align*}
In order to assess the asymptotic behavior of these terms, we note that by the mean value theorem
\begin{itemize}
\item it holds on $\mathcal{S}_{n,t}$ that $|\hat \sigma_{\epsilon,t}-\sigma_{\epsilon,k}|\leq \frac{\sqrt{2}}{\sigma_{\epsilon,k}} |\hat \sigma_{\epsilon,t}^2-\sigma_{\epsilon,k}^2|$ and

\item it holds on $\mathcal{F}_{4,n,t}$ that $|\hat \lambda_{j,t}^{1/2}-\lambda_{j,k}^{1/2}|\leq \left(\frac{{2}}{\lambda_{j,k}}\right)^{\frac{1}{2}}|\hat \lambda_{j,t}-\lambda_{j,k}|$.

\end{itemize}
Adding these observations to the above allows us to conclude the following for $R_{15,1}-R_{15,3}$:\\

Ad $R_{15,1}$:
\begin{align*}
R_{15,1}=&K\max_{1 \leq k \leq K} |G_k| \prob\left( \sum_{j=1}^m (a_{j,t}^*)^2\hat \lambda_{j,t} |\sigma_{\epsilon,k}-\hat    \sigma_{\epsilon,t}|^2 > c  \right) \\
\leq & K\max_{1 \leq k \leq K} |G_k|\prob\left( \frac{2}{\sigma_{\epsilon,k}^2} \sum_{j=1}^m (a_{j,t}^*)^2 \lambda_{j,k} |\hat  \sigma_{\epsilon,t}^2  - \sigma_{\epsilon,k}^2|^2 > c \right)\\
&+ K\max_{1 \leq k \leq K} |G_k|\prob\left( \mathcal{F}_{4,n,t}^c\right)+K\max_{1 \leq k \leq K} |G_k| \prob\left( \mathcal{S}_{n,t}^c \right) \\
= & o(1).
\end{align*}
Ad $R_{15,2}$:
\begin{align*}
R_{15,2}\leq & K\max_{1 \leq k \leq K} |G_k| \prob\left( 2 m \sigma_{\epsilon,k}^2\sum_{j=1}^m (a_{j,t}^*)^2\hat \lambda_{j,t}  ||\phi_{j,k} - \hat \phi_{j,t} ||_2^2   > c \right)+K\max_{1 \leq k \leq K} |G_k| \prob\left( \mathcal{S}_{n,t}^c \right)\\
=& O(mn^{\delta-1})+o(1)=o(1)
\end{align*}
which follows from similar arguments as the ones used for $R_{4,2,t}$.  \\

Ad $R_{15,3}$:
\begin{align*}
R_{15,3}\leq & K\max_{1 \leq k \leq K} |G_k| \prob\left(  2 \sigma_{\epsilon,k}^2\sum_{j=1}^m (a_{j,t}^*)^2\lambda_{j,k}^{-1} |\hat \lambda_{j,t} - \lambda_{j,k}|^2  > c \right)+K\max_{1 \leq k \leq K} |G_k| \prob\left( \mathcal{S}_{n,t}^c \right)\\&+K\max_{1 \leq k \leq K} |G_k|\prob\left( \mathcal{F}_{4,n,t}^c\right) \\
\leq & K\max_{1 \leq k \leq K} |G_k|\frac{2 \sigma_{\epsilon,k}^2 \sum_{j=1}^m (a_{j,t}^*)^2 \lambda_{j,k}^{-1} E[\mathcal{D}_t^2]}{c}+K\max_{1 \leq k \leq K} |G_k| \prob\left( \mathcal{S}_{n,t}^c \right)+K\max_{1 \leq k \leq K} |G_k|\prob\left( \mathcal{F}_{4,n,t}^c\right)\\
=&o(1).
\end{align*}
It remains to show that $R_{14,2}=o(1)$. For this purpose note
\begin{align*}
R_{14,2}=& K\max_{1 \leq k \leq K} |G_k| \prob\left(  \sum_{j=1}^m  (a_{j,t}^*-a_{j,t})^2\frac{\hat \lambda_{j,t}}{\hat \sigma_{\epsilon,t}^2} > c  \right) \\
\leq & K\max_{1 \leq k \leq K} |G_k| \prob\left(  4 ||\alpha_t ||_2^2\sum_{j=1}^m   ||\hat \phi_{j,t} - \phi_{j,k} ||_2^2 \frac{\lambda_{j,t}}{\sigma_{\epsilon,k}^2} > c \right) + K\max_{1 \leq k \leq K} |G_k| \prob\left( \mathcal{F}_{4,n,t}^c\right) \\
& +K\max_{1 \leq k \leq K} |G_k| \prob\left( \mathcal{S}_{n,t}^c \right) \\
=& o(1)+o(1)= o(1),
\end{align*}
which follows from \eqref{EndGegn} and our above arguments. \\

Combining arguments implies the last statement in Lemma \ref{lem:UNI}. $\blacksquare$ \\

%%%%%%%%%%%%%%%%%%%%%%%%%%%%%%%%%%%%%%%%%%%%%%%%%%%%
\subsection{Proof of Theorem \ref{th:CCO}}\label{sec:APACCO}
%%%%%%%%%%%%%%%%%%%%%%%%%%%%%%%%%%%%%%%%%%%%%%%%%%%%
Using the results presented in the previous lemma it is possible to argue in analogy to the proof of Theorem 1 in \citeapp{VL17_app} to validate the classification consistency claimed in our Theorem \ref{th:CCO}. For this purpose consider the set $S^{(j)}=\{1,\dots,T\}\setminus \bigcup_{l<j}\hat G_l$ at an iteration step $1\leq j \leq \hat K-1$ of the algorithm described in Section \ref{sec:EST}. For a $t \in S^{(j)}$ denote the set of indexes corresponding to the ordered distances $\hat \Delta_{t (1)} \leq \dots \leq \hat \Delta_{t (|S^{(j)}|)}$ as $\{(1),\dots,(|S^{(j)}|)\}$. In analogy, the index set corresponding to the ordered population distances $ \Delta_{t [1]} \leq \dots \leq  \Delta_{t [|S^{(j)}|]}$ is denoted as $\{[1],\dots,[|S^{(j)}|]\}$, where $\Delta_{ts}$ is as in Assumption 7. Now, define the index $\hat \kappa$ according to $\hat \Delta_{t (\hat \kappa)}< \tau_{nT} < \hat \Delta_{t (\hat \kappa+1)}$. Its population counterpart, $\kappa$, obtains as $0= \Delta_{t [\kappa]}< \tau_{nT} < \Delta_{t [ \kappa+1]}$. It holds that
\begin{align} \prob\left( \left\{ (1),\dots,(\hat \kappa) \right\} \neq \left\{ ([1],\dots,[\kappa] \right\} \right) &\leq \prob\left( \left\{ (1),\dots,(\kappa) \right\} \neq \left\{ [1],\dots,[\kappa] \right\} \right)+\prob(\hat \kappa \neq \kappa) \label{Prob1} \\
&=o(1)+o(1). \nonumber
\end{align}
In order to prove that the first probability on the right hand side of \eqref{Prob1} is a null sequence, suppose that  $t\in G_k$, with $1 \leq k \leq K$. As indicated, there are $\kappa\geq 1$ indexes in $S^{(j)}$ being elements of $G_k$. For the corresponding distances  it holds that $\Delta_{t [1]} = \dots =  \Delta_{t [\kappa]}=0$ by definition. The remaining distances are bounded away from zero by $ 0<C_\Delta \leq \Delta_{t [\kappa+1]} \leq \dots \leq  \Delta_{t [|S^{(j)}|]}$ due to Assumption 7. \\
As stated in Lemma \ref{lem:UNI}, $\max_{1 \leq t\leq T}||\hat \alpha_t^{(\Delta)} -  \alpha_t^{(\Delta)} ||^2_2=o_p(1)$ implying that $\max_{1 \leq s\leq T}|\hat \Delta_{ts}- \Delta_{ts}|=o_p(1)$, which holds for any reference period $t$. Combining arguments allows to conclude $ \max_{1\leq s \leq \kappa } \hat \Delta_{t(s)} = o_p(1)$ and $\min_{\kappa <  s \leq |S^{(j)}|} \hat \Delta_{t(s)} \geq C_\Delta + o_p(1)$ as well as $\max_{1\leq s \leq \kappa } \hat \Delta_{t[s]} = o_p(1)$ and $\min_{\kappa <  s \leq |S^{(j)}|} \hat \Delta_{t[s]} \geq  C_\Delta + o_p(1)$.
This implies that the first probability on the right hand side of \eqref{Prob1} tends to zero. Further note that the specification of the threshold in Assumption 7 immediately implies $\prob\left(  \hat \Delta_{t[\kappa]} < \tau_{nT} \right) \to 1$ and $\prob\left( \hat \Delta_{t[\kappa+1]} > \tau_{nT} \right) \to 1$
as $n\to \infty$ in light of the preceding arguments. As a consequence of this $\prob\left( \hat \Delta_{t[\kappa]} < \tau_{nT} < \hat \Delta_{t[\kappa+1]} \right) \to 1$ as $n\to \infty$, implying that the second probability on the right hand side of \eqref{Prob1} is a null sequence. $\blacksquare$ \\

\noindent\textbf{Remark 1}\\
For the calculation of the convergence rate of our estimator $\tilde A_k$, the classification error is negligible as a consequence of Theorem \ref{th:CCO}. To see this note that an analogous argument as in \citeapp{VL17_app} holds in our context: let $s(n,T)$ be an arbitrary deterministic sequence such that $s(n,T)\to 0$ as $n,T\to \infty$. Now, note that for any constant $C>0$
\begin{align*}
&\prob\left((s(n,T))^{-1}||\tilde A_k - A_k ||_2^2>C \right) \\
&\leq \prob\left(\left\{(s(n,T))^{-1}||\tilde A_k - A_k ||_2^2>C \right\} \cap \left\{ \hat G_k = G_k \right\} \right) + \prob \left( \left\{ \hat G_k \neq G_k \right\}\right) \\
&= \prob\left( (s(n,T))^{-1}||\tilde A_k^* - A_k ||_2^2>C \right) +o(1),
\end{align*}
where the quantity $\tilde A_k^*$ denotes the estimator $\tilde A_k$ calculated from $\{(y_{it},X_{it},z_{it}): \ 1\leq i \leq n, \ t\in G_k\}$, i.e. from correctly classified periods. Note in particular that the time series dependence formulated in Assumption 1 does not affect this argument. \\

In light of this remark, the proof of Theorem \ref{th:CPE} starts from the ideal \textit{oracle} estimators $\tilde A_k^*$ rather than their contaminated counterparts. \\

%%%%%%%%%%%%%%%%%%%%%%%%%%%%%%%%%%%%%%%%
\noindent\textbf{Remark 2}\\
%%%%%%%%%%%%%%%%%%%%%%%%%%%%%%%%%%%%%%%%
For the proof of Theorem \ref{th:CPE}, we work with classification-error-free oracle variants of the estimators $\tilde \phi_{j,k},\tilde \lambda_{j,k},\tilde K_{X,k},\tilde K_{zX,k},\tilde K_{z,k}$ and $\tilde \Gamma_{k}$. Such estimators, calculated from $\{(z_{it},X_{it}): \ 1\leq i \leq n, \ t\in G_k \}$, are denoted $\tilde \phi_{j,k}^*,\tilde \lambda_{j,k}^*,\tilde K_{X,k}^*,\tilde K_{zX,k}^*,\tilde K_{z,k}^*$  and $\tilde \Gamma_{k}^*$. In analogy to before, we further denote the Hilbert Schmidt norm of the difference between $\tilde \Gamma_k^*$ and the population counterpart $\Gamma_k$ as $\widetilde {\mathcal{D}}_k^*:=||\tilde \Gamma_k^*- \Gamma_k ||_{\mathcal{S}}$. 
Beyond these quantities, the estimator $\tilde A_k^*$ makes implicitly use of the operator $\tilde \Phi_k^*$ which estimates, in analogy to $\hat \Phi_t$, the operator $\Phi_k$ as in \eqref{eq:popPhi}. $\tilde \Phi_k^*$ is defined according to 
\begin{align*}
\tilde \Phi_k^*(g) := \sum_{j=1}^{\tilde m} \frac{\langle \tilde K_{zX,k}^*, \tilde \phi_{j,k}^* \rangle}{\tilde \lambda_{j,k}^*}  \langle \tilde \phi_{j,k}^*,g \rangle
\end{align*}
for any $g\in L^2([0,1])$, where $\tilde m\equiv \tilde m_k$ for simplicity of notation.

Assessing the asymptotic properties of the classification-error-free estimators, note that due to Assumption 1 for every regime $G_k$, the random variables $\{X_{it}: \ 1 \leq i\leq n, \ t\in G_k\}$ are $L^4_m$-approximable. Thus, for suitably large constants, the following inequalities from \citeapp{HK10_app} hold:\footnote{Cf. Theorem 3.2 and the consequent discussion in \citeapp{HK10_app}.}
\begin{align}
&E\left[ \left(\widetilde {\mathcal{D}}_k^*\right)^2 \right] \leq C(n|G_k|)^{-1} \label{GamTilde} \\
&E\left[ \left| \left| \tilde K_{X,k}^*-K_{X,k} \right| \right|_2^2 \right] \leq C(n|G_k|)^{-1} \label{cov4} \\
&E\left[ \left| \tilde \lambda_{j,k}^* - \lambda_{j,k}  \right|^2 \right] \leq E\left[ \left(\widetilde {\mathcal{D}}_k^*\right)^2 \right] \leq  C(n|G_k|)^{-1} \label{cov5}
\end{align}
for $1\leq j \leq \tilde m$.
Further note that the dependence of the random variables $\{(z_{it},X_{it}): \ 1\leq i \leq n, \ t\in G_k \}$ is sufficiently weak, such that
\begin{align*}
&E\left[\left| \left| \tilde K_{zX,k}^*-K_{zX,k} \right| \right|^2_2 \right] =O((n|G_k|)^{-1}) 
\end{align*}
and further
\begin{align*}
& E\left[ \left| \tilde K_{z,k}^*-K_{z,k}  \right|^2\right] =O((n|G_k|)^{-1}), 
\end{align*}
which can be shown by straightforward moment calculations. In addition to that, bounds on $\left| \left| \tilde \phi_{j,k}^*-\phi_{j,k} \right| \right|_2^2$ can be obtained in analogy to the almost sure bound in \eqref{PhiBoundAS} and the asymptotic bound as in \eqref{phi:1}--\eqref{RPhi}. We make the latter precise defining the analogues to $\mathcal{F}_{1,n,t}$--$\mathcal{F}_{3,n,t}$ as
\begin{enumerate}
\item $\widetilde{\mathcal{F}}_{1,n,T,k}:= \left\{ C (n|G_k|)^{\frac{2(1+\mu)}{\mu+2\nu}} \left(\widetilde{\mathcal{D}}^*_k\right)^2 \leq 1/8 \right\}$
\item $\widetilde{\mathcal{F}}_{2,n,T,k}:=\left\{ |\tilde \lambda_{j,k}^*-\lambda_{l,k}|^{-2}\leq 2 | \lambda_{j,k}-\lambda_{l,k}|^{-2}\leq C (n|G_k|)^{\frac{2(1+\mu)}{\mu+2\nu}}, \ 1 \leq j \leq \tilde m, j\neq l\in \mathbb{N} \right\}$.
\item $\widetilde{\mathcal{F}}_{3,n,T,k}:=\widetilde{\mathcal{F}}_{1,n,T,k} \cap \widetilde{\mathcal{F}}_{2,n,T,k}$
\end{enumerate}
for which we note $\prob(\widetilde{\mathcal{F}}_{3,n,T,k}^c)\leq \prob(\widetilde{\mathcal{F}}_{1,n,T,k}^c)+\prob(\widetilde{\mathcal{F}}_{2,n,T,k}^c)=o(1)+o(1)$ as $(n,T)\to \infty$ from similar arguments as before. Also, as in our arguments for the $t$-wise estimators it holds on $\widetilde{\mathcal{F}}_{2,n,T,k} $ that 
\begin{align}
& || \tilde \phi_{j,k}^*-\phi_{j,k} ||_2^2 \leq 8 \left( 1-4 C  (n|G_k|)^{\frac{2(1+\mu)}{\mu+2\nu}} \left(\widetilde{\mathcal{D}}^*_k\right)^2 \right)^{-1}  \tilde R_{j,k}^{(\phi)}, \label{phitilde:1} \\
\text{where } & \tilde R_{j,k}^{(\phi)}:= \sum_{l:l\neq j} (\lambda_{j,k}-\lambda_{l,k})^{-2} \left[ \int_{0}^1\int_{0}^1 (\tilde K_{X,k}^*(u,v)-K_{X,k}(u,v))\phi_{j,k}(u)\phi_{l,k}(v)\mathrm{d}u\mathrm{d}v \right]^2, \nonumber
\end{align}
from which we conclude, that on $\widetilde {\mathcal{F}}_{3,n,T,k}$, it holds that 
\begin{align}
 || \tilde \phi_{j,k}^*-\phi_{j,k} ||_2^2 \leq 16   \tilde R_{j,k}^{(\phi)}. \label{phitilde:2}
\end{align}
The results in \citeapp{HH07_app} also allow to conclude $E\left[ \tilde R_{j,k}^{(\phi)} \right] =O\left(j^2(n | G_k|)^{-1}\right)$ uniformly in $1\leq j \leq \tilde m$ for weakly dependent random variables $\{X_{it}: \ 1\leq i \leq n, \ t\in G_k \}$. \\

As a further important observation we note that 
\begin{align*}
\left| \left| \tilde \Phi_{k}^* -\Phi_k\right| \right|^2_{H'} =O_p \left((n|G_k|)^\frac{1-2\nu}{\mu+2\nu} \right)
\end{align*}
given Assumptions 1-6 hold. This can be seen from a regression
\begin{align}
z_{it}-E[z_{it}] = \langle \zeta ,X_{it}-E[X_{it}] \rangle + s_{it} \label{eq:auxreg}
\end{align}
in the $k-th$ regime, where $1\leq t \leq  |G_k| $, $1 \leq i \leq n$ and $s_{it}$ as in Assumption 6.   Since the functional parameter $\zeta$ is formulated as being time invariant, it can be estimated as in \citeapp{HH07_app} from pooled data $(X_{j(i,t)},z_{j(i,t)})$, where $1\leq j(i,t):=(i-1)|G_k|+t \leq n|G_k|$. As noted by \citeapp{S09_app}, the resulting estimator, say $ \hat \zeta$, links to the operator $\tilde \Phi_k^*$ according to
\begin{align}
||\hat \zeta-\zeta ||_2^2=\left|\left|\tilde \Phi_{k}^* - \Phi_k\right|\right|_{H'}^2. \label{eq:zeta}
\end{align}
The argumentation in \citeapp{HH07_app} (cf. their Theorem 1 and corresponding proof) transfers mutatis mutandis to a setup with weakly dependent regressors ($L_m^4$ dependence) and weakly dependent errors (m-dependence) as is the case in our auxiliary regression \eqref{eq:auxreg}. This can be shown using the fundamental results formulated in \citeapp{HK10_app}. As $\hat \zeta$ is calculated from a sample of size $n |G_k|$, the results in \citeapp{HH07_app} together with \eqref{eq:zeta} thus imply
\begin{align*}
\left| \left| \tilde \Phi_{k}^* -\Phi_k\right| \right|^2_{H'} =O_p \left((n|G_k|)^\frac{1-2\nu}{\mu+2\nu} \right)
\end{align*}
as claimed before.

%%%%%%%%%%%%%%%%%%%%%%%%%%%%%%%%%%%%%%%%%%%%%%%%%%%%
\subsection{Proof of Theorem \ref{th:CPE}}\label{sec:APACPE}
%%%%%%%%%%%%%%%%%%%%%%%%%%%%%%%%%%%%%%%%%%%%%%%%%%%%

Note that on $\bigcap_{t\in G_k} Q_{n,t}$ it holds that ${\hat B_t}^{-1} \leq 2 {B_k}^{-1}$ for any $t \in G_k$, and so 
\begin{align*}
&\prob\left( n\left( | G_k| \right)^{-1}\left|\sum_{t\in  G_k} (\hat \beta_{t}-\beta_t)^2\right|>c\right)\\
 \leq & \prob\left(4 B_k^{-2} n\left( | G_k| \right)^{-1}\left| \sum_{t\in  G_k} \sum_{l=1}^3 R_{0,l,t}^2 \right|>c\right)+\prob\left( \bigcup_{t\in G_k} Q_{n,t} ^c \right) \\
  \leq & \prob\left(4 B_k^{-2} n\left( | G_k| \right)^{-1}  \sum_{t\in  G_k}\left( \sum_{l=1}^2 R_{0,l,t}^2+R_{1,1,t}^2+R_{2,1}^2+\sum_{j=1}^3 R_{3,j,t}^2 \right)>c\right)\\
& +  |G_k| \prob\left( Q_{n,t} ^c \right).
\end{align*}
In the proof of Lemma \ref{lem:UNI} it was shown that $\prob\left( Q_{n,t} ^c \right)=o(|G_k|^{-1})$. Regarding the remaining term, note that due to the exogeneity of the regressors and stationary distributions within the regimes the following holds: for any $c_j>0$, $j=1,2,3$, there exist constants $C_j=C_j(c_j)$, $j=1,2,3$ such that
\begin{align*}
\prob\left(n |G_{k}|^{-1}\sum_{t\in G_k}R_{0,1,t}^2>c_1 \right)&\leq \frac{n E[ R_{0,1,t}^2]}{c_1}\leq C_{1} \\
\prob\left(n |G_{k}|^{-1}\sum_{t\in G_k}R_{0,2,t}^2>c_2\right)&\leq \frac{n E[ R_{0,2,t}^2]}{c_2}\leq C_2 \\
\prob\left(n |G_{k}|^{-1}\sum_{t\in G_k}R_{1,1,t}^2>c_3 \right)&\leq \frac{n C E[ ||\hat K_{zX,t}-K_{zX} ||_2^2]}{c_3} \leq C_3.
\end{align*}
Further we observe that $R_{2,1}=o(n^{-1/2})$. Using that $\prob(\bigcup_{t \in G_k} \mathcal{F}_{3,n,t}^c)\leq \sum_{t=G_k} \prob(\mathcal{F}_{3,n,t}^c)=o(1)$ as shown in the proof of Lemma \ref{lem:UNI}, we argue that for any constant $c_4$, there exists a constant $C_4$, such that
\begin{align*}
&\prob\left(n |G_{k}|^{-1}\sum_{t\in G_k}R_{3,1,t}^2>c_4 \right) \\
\leq & \frac{ n ||A_k|| E\left[|| \hat K_{zX,t}-K_{zX,k}||_2^4\right]^{\frac{1}{2}} E\left[\mathcal{D}_t^4\right]^{\frac{1}{2}} {(C_\lambda^\prime)}^{-2} \left(\sum_{j=1}^m j^{1+\mu}\right)^2}{c\cdot c_4} \\
&+\frac{ nE\left[|| \hat K_{zX,t}-K_{zX,k}||_2^2\right]  \left(\sum_{j=1}^m |a_{j,t}^*|\right)^2}{c\cdot c_4} \\
\leq  & C_4,
\end{align*}
due to the stationarity of $(X_{it},z_{it})$, $t\in G_k$ and \eqref{PhiBoundAS}. Beyond that we argue for $R_{3,2,t}$ that for any $c_5>0$ it follows from similar arguments that there exists a $C_5>0$ such that 
\begin{align*}
&\prob\left(n |G_{k}|^{-1}\sum_{t\in G_k}R_{3,2,t}^2>c_5\right) \\
\leq & \frac{ n ||A_k|| \cdot E\left[\mathcal{D}_t^2\right] {(C_\lambda^\prime)}^{-2} C_{zX}^2 \left(\sum_{j=1}^m j^{1-\nu}\right)^2}{c\cdot c_5} \\
\leq & C_5.
\end{align*}
Finally, note that for any $c_6>0$, there exists a $C_6>0$ such that
\begin{align*}
&\prob \left(n |G_{k}|^{-1}\sum_{t\in G_k}R_{3,3,t}^2>c_6\right) \\
\leq &\frac{2n|| K_{zX,k}||_2^2 || A_k||_2^2 (C_\lambda^\prime)^{-2}
E\left[ \mathcal{D}_t^4  \right] \left(\sum_{j=1}^m j^{2(1+\mu)} \right)^2}{c_6}\\
&+\frac{2n|| K_{zX,k}||_2^2 (C_\lambda^\prime)^{-2}(C_a)^{2}
E\left[ \mathcal{D}_t^2  \right] \left(\sum_{j=1}^m j^{1+\mu-\nu} \right)^2}{c_6} \\
& \leq  C_6
 \end{align*}
thanks to Assumptions $2$--$5$, stationarity and once more the bound in \eqref{PhiBoundAS}. Combining arguments  allows us to conclude that $|G_k|^{-1}\sum_{t \in G_k} (\hat \beta_t-\beta_t)^2=O_p(n^{-1})$ as $n,T\to\infty$. We use this finding in a moment to obtain the convergence rate for $\tilde A_k$. To assess the underlying problem, we use the following notation:
\begin{itemize}

\item $X_{it}^{cc,*}:=X_{it}-\bar{\bar X}_k^*$ with $\bar{\bar X}_k^*:=\frac{1}{n|G_k|}\sum_{t\in G_k}\sum_{i=1}^n X_{it}$,

\item $\epsilon_{it}^{cc,*}:=\epsilon_{it}-\bar{\bar \epsilon}_k^*$ with $\bar{\bar \epsilon}_k^*:=\frac{1}{n|G_k|}\sum_{t\in G_k}\sum_{i=1}^n \epsilon_{it}$.

\end{itemize}
The classification-error-free oracle estimator for the regime specific parameter function reads as $\tilde A_k^*:=\sum_{j=1}^{\tilde m} \tilde a_{j,k}^*\tilde \phi_{j,k}^*$. The basis coefficients indexed $1 \leq j \leq \tilde m$ obtain as
\begin{align*}
\tilde a_{j,k}^{*}&:=(\tilde \lambda_{j,k}^*)^{-1}\frac{1}{n|G_k|}\sum_{t\in G_k}\sum_{i=1}^n \langle X_{it}^{cc,*},\tilde \phi_{j,k}^*\rangle (y_{it}^c-z_{it}^c \hat \beta_t) \\
&=\tilde a_{j,k}^{(1)}+\tilde a_{j,k}^{(2)},
\end{align*}
where 
\begin{align*}
\tilde a_{j,k}^{(1)}&:=(\tilde \lambda_{j,k}^*)^{-1}\frac{1}{n|G_k|}\sum_{t\in G_k}\sum_{i=1}^n \langle X_{it}^{cc,*},\tilde \phi_{j,k}^*\rangle \left( \langle  X_{it}^{cc,*},A_k \rangle +\epsilon_{it}^{cc,*} \right) 
\end{align*}
and
\begin{align*}
\tilde a_{j,k}^{(2)}&:= (\tilde \lambda_{j,k}^*)^{-1}\frac{1}{n|G_k|}\sum_{t\in G_k}\sum_{i=1}^n \langle X_{it}^{cc,*},\tilde \phi_{j,k}^*\rangle \left( z_{it}^c(\beta_t-\hat \beta_t)+ \langle \bar{\bar X}_k^*-{\bar X_t} ,A_k \rangle  + \bar{\bar \epsilon}_k^*-{\bar \epsilon_t}  \right). 
\end{align*}
The upper bound
\begin{align}
||\tilde A_k^*- A_k ||_2^2 &= \left| \left|\sum_{j=1}^{\tilde m} \left(\tilde a_{j,k}^{(1)}+\tilde a_{j,k}^{(2)} \right) \tilde  \phi_{j,k}^* -A_k\right|\right|^2_2 \\
&\leq  2\left| \left|\sum_{j=1}^{\tilde m} \tilde a_{j,k}^{(1)}\tilde  \phi_{j,k}^* -A_k\right|\right|^2_2+2\sum_{j=1}^{\tilde m}  \left(\tilde a_{j,k}^{(2)}\right)^2 \label{tildeAk}
\end{align}
can be obtained using the Cauchy Schwarz inequality. The first term is the estimator from \citeapp{HH07_app} in the case of $n|G_k|$ pooled observations and an $L^4_m$ approximable regressor function. Along the lines of our second remark and Assumptions 1-5, it holds that $\left| \left|\sum_{j=1}^{\tilde m} \tilde a_{j,k}^{(1)}\tilde  \phi_{j,k}^* -A_k \right|\right|^2_2=O_p\left( n^{\frac{(1+\delta)(1-2\nu)}{\mu+2\nu}} \right)$. The remaining term in \eqref{tildeAk} we split according to
\begin{align*}
\sum_{j=1}^{\tilde m}  \left(\tilde a_{j,k}^{(2)}\right)^2  \leq 3\cdot (R_{16,1}+R_{16,2}+R_{16,3}).
\end{align*}
where the terms $R_{16,1},R_{16,2}$ and $R_{16,3}$ are as follows: \\

Ad $R_{16,1}$: 
\begin{align*}
R_{16,1}:=& \sum_{j=1}^{\tilde m}(\tilde \lambda_{j,k}^*)^{-2} \left(  \frac{1}{|G_k|}\sum_{t\in G_k} \langle n^{-1}\sum_{i=1}^n X_{it}^{cc,*},\tilde \phi_{j,k}^*\rangle \left( \langle \bar{\bar X}_k^*-{\bar X_t} ,A_k \rangle   \right) \right)^2 \\ 
\leq & 4\sum_{j=1}^{\tilde m} ( \lambda_{j,k})^{-2} \left(\frac{1}{|G_k|}\sum_{t\in G_k} ||X_{t}-\bar {\bar X}_k^*  ||_2^2 ||A_k ||_2\right)^2 \\
=& O_p(\tilde m^{1+2\mu}n^{-2})
\end{align*}
on an event $\widetilde{\mathcal{F}}_{4,n,T,k}:=\{ |\tilde \lambda_{j,k}^*-\lambda_{j,k}|\leq \frac{1}{2}\lambda_{j,k} : \quad 1\leq j \leq \tilde m\}$. For this event in turn, note that $\prob(\widetilde{\mathcal{F}}_{4,n,T,k})\to 1$ which follows from analogous arguments, which lead to $\prob(\mathcal{F}_{{4,n,t}})\to 1$ above. \\ 

Ad $R_{16,2}$: 
\begin{align*}
R_{16,2}&:=\sum_{j=1}^{\tilde m}(\tilde \lambda_{j,k}^*)^{-2}\left( \frac{1}{n|G_k|}\sum_{t\in G_k}\sum_{i=1}^n \langle X_{it}^{cc,*},\tilde \phi_{j,k}^*\rangle \left( \bar{\bar \epsilon}_k^*-{\bar \epsilon_t}  \right) \right)^2 \\
&= \sum_{j=1}^{\tilde m}(\tilde \lambda_{j,k}^*)^{-2}\left( \frac{1}{|G_k|}\sum_{t\in G_k} \langle \bar X_t-\bar {\bar X}_k^*,\tilde \phi_{j,k}^*\rangle {\bar \epsilon_t}   \right)^2 \\
&\leq 4\sum_{j=1}^{\tilde m}(\lambda_{j,k})^{-2}\left( \frac{1}{|G_k|}\sum_{t\in G_k} || \bar X_t-\bar {\bar X}_k^*||_2 {\bar \epsilon_t}   \right)^2 \\
&=O_p(\tilde m^{1+2\mu}n^{-2})
\end{align*}
on $\widetilde{\mathcal{F}}_{4,n,T,k}$. \\

Ad $R_{16,3}$: 
\begin{align*}
R_{16,3}:=&\sum_{j=1}^{\tilde m}(\tilde \lambda_{j,k}^*)^{-2}\left( \frac{1}{n|G_k|}\sum_{t\in G_k}\sum_{i=1}^n \langle X_{it}^{cc,*},\tilde \phi_{j,k}^*\rangle  z_{it}^c\left(\beta_t-\hat \beta_t \right) \right)^2 \\
 =&\sum_{j=1}^{\tilde m}(\tilde \lambda_{j,k}^*)^{-2}\left( \frac{1}{|G_k|}\sum_{t\in G_k} \langle \hat K_{zX,t},\tilde \phi_{j,k}^*\rangle \left(\beta_t-\hat \beta_t \right) \right)^2 \\
\leq &
\sum_{j=1}^{\tilde m}(\tilde \lambda_{j,k}^*)^{-2}\left( \frac{1}{|G_k|}\sum_{t\in G_k} \langle \hat K_{zX,t},\tilde \phi_{j,k}^*\rangle^2\right) \left( \frac{1}{|G_k|}\sum_{t\in G_k}\left(\beta_t-\hat \beta_t \right)^2 \right),
\end{align*}
of which it is known from before that $\frac{1}{|G_k|}\sum_{t\in G_k}\left(\beta_t-\hat \beta_t \right)^2=O_p(n^{-1})$ and 
\begin{align*}
& \sum_{j=1}^{\tilde m}  (\tilde \lambda_{j,k}^*)^{-2} |G_k|^{-1}\sum_{t\in G_k} \langle\hat K_{zX,t},\tilde \phi_{j,k}^* \rangle ^2  \\&\leq \sum_{j=1}^{\tilde m}  (\tilde \lambda_{j,k}^*)^{-2}|G_k|^{-1}\sum_{t\in G_k}  3\left(  \langle K_{zX,k},\phi_{j,k} \rangle^2+\langle \hat K_{zX,t}-K_{zX,k},\tilde \phi_{j,k}^* \rangle^2 +\langle K_{zX,k},\phi_{j,k}-\tilde \phi_{j,k}^* \rangle^2   \right).
\end{align*}
We further conclude that on $\widetilde{\mathcal{F}}_{4,n,T,k}$
\begin{align*}
\sum_{j=1}^{\tilde m}  (\tilde \lambda_{j,k}^*)^{-2}  \langle K_{zX,k},\phi_{j,k} \rangle^2  \leq 4 \sum_{j=1}^{\tilde m}   \lambda_{j,k}^{-2}  \langle K_{zX,k},\phi_{j,k} \rangle^2 \propto \sum_{j=1}^{\tilde m} j^{2\mu-2(\mu+\nu)} =O(1)
\end{align*}
as well as
\begin{align*}
 |G_k|^{-1}\sum_{t\in G_k}\sum_{j=1}^{\tilde m}  (\tilde \lambda_{j,k}^*)^{-2}   \langle \hat K_{zX,t}-K_{zX,k},\tilde \phi_{j,k}^* \rangle^2  &\leq 2 |G_k|^{-1}\sum_{t\in G_k} ||\hat K_{zX,t}-K_{zX,k} ||^2_2 \sum_{j=1}^{\tilde m}   \lambda_{j,k}^{-2} \\
&=O_p\left(n^{-1}n^{\frac{(1+\delta)(1+2\mu)}{\mu+2\nu}}\right) \\
&=O_p \left(n^{\frac{(1+\delta)(1+2\mu)-\mu-2\nu}{\mu+2\nu}}\right) =o_p(1).
\end{align*}
Further, we use similar arguments as before (see the proof of Theorem \ref{prop:twise}) to obtain 
\begin{align*}
\sum_{j=1}^{\tilde m}  (\tilde \lambda_{j,k}^*)^{-2}  \langle K_{zX,k},\tilde \phi_{j,k}^*-\phi_{j,k} \rangle^2  & \leq 4||K_{zX,k} ||_2^2 \sum_{j=1}^{\tilde{m}}|| \tilde \phi_{j,k}^*-\phi_{j,k}||^2_2\lambda_{j,k}^{-2} =o_p(1)
\end{align*}
on $\widetilde{\mathcal{F}}_{3,n,T,k}\cap \widetilde{\mathcal{F}}_{4,n,T,k}$, which implies $\sum_{j=1}^{\tilde m}  (\tilde \lambda_{j,k}^*)^{-2}  \langle K_{zX,k},\tilde \phi_{j,k}^*-\phi_{j,k} \rangle^2=O_p(1)$.
Combining our above statements yields $\sum_{j=1}^{\tilde m}  \left(\tilde a_j^{(2)}\right)^2=O_p(n^{-1})$. Further, if $\nu>\frac{1+\mu+\delta}{2\delta}$, or equivalently $\delta> (1+\mu)/(2\nu-1)$, then $(nT)^{\frac{1-2\nu}{\mu+2\nu}}=o(n^{-1})$ and in case $\nu<\frac{1+\mu+\delta}{2\delta}$, $n^{-1}=o\left((nT)^{\frac{1-2\nu}{\mu+2\nu}}\right)$. Together with our Remark 1 on the classification error the result in the theorem follows. $\blacksquare$ \\

%%%%%%%%%%%%%%%%%%%%%%%%%%%%%%%%%%%%
\subsection{Threshold Choice}\label{ssec:TC}
%%%%%%%%%%%%%%%%%%%%%%%%%%%%%%%%%%%%

In order to illustrate the properties of the threshold $\tau_{nT}$ as suggested in Section \ref{sec:THR}, suppose for a moment that the truncation error in regime $k$ is negligible (i.e., $\lambda_{j,k}\approx 0,j\geq\underline{m}+1$) and that the eigenvalue-eigenfunction pairs $(\lambda_{j,k},\phi_{j,k})_{j\geq 1}$ as well as the error variance $\sigma_{\epsilon,k}^2$ of regime $k$ were known. In this case our estimation procedure yields variance adjusted estimators $\hat \alpha_t^{(\Delta^\ast)}:=\sum_{j=1}^{\underline{m}} \sigma_{\epsilon,k}^{-1}\lambda_{j,k}^{1/2}\hat a_{j,t}\phi_{j,k}$ and $\hat \alpha_s^{(\Delta^\ast)}:=\sum_{j=1}^{\underline{m}} \sigma_{\epsilon,k}^{-1}\lambda_{j,k}^{1/2}\hat a_{j,s}\phi_{j,k}$ where the appropriately scaled difference of their $j$-th components $(n/2)^{1/2} \sigma_{\epsilon,k}^{-1}\lambda_{j,k}^{1/2}(\hat a_{j,t}-\hat a_{j,s})$ is approximately standard normal (for large $n$ and small temporal correlations), such that for all $t,s\in G_k$
%%%%%%%%%%%%%%%
\begin{align*}
&\frac{n}{2}\Delta^\ast_{ts}
:=\frac{n}{2}||\hat \alpha_t^{(\Delta^\ast)}-\hat \alpha_s^{(\Delta^\ast)} ||_2^2
=\sum_{j=1}^{\underline{m}}\left(\left(\frac{n}{2}\right)^{1/2}\sigma_{\epsilon,k}^{-1}\lambda_{j,k}^{1/2}(\hat a_{j,t}-\hat a_{j,s})\right)^2=:Q_{ts}^{\underline{m}}\notag\\
\Rightarrow&\Delta^\ast_{ts}=\frac{2}{n}Q_{ts}^{\underline{m}},\quad\text{where (for large $n$)}\quad Q_{ts}^{\underline{m}}\sim \chi_{\underline{m}}^2\quad\text{if $t\neq s$\;and}\;\; Q_{ts}^{\underline{m}}\approx 0\;\text{if $t=s$}.
\end{align*} 
%%%%%%%%%%%%%%%
For accurate estimates and a small truncation error, we expect that  $||\hat\alpha_t^{(\Delta)} -\hat\alpha_s^{(\Delta)}||_2^2 \approx ||\hat\alpha_t^{(\Delta^\ast)}-\hat \alpha_s^{(\Delta^\ast)} ||_2^2$ and hence that $\hat \Delta_{ts}\approx \Delta^\ast_{ts}$. Note that neglecting the truncation error is often justified in practice, where a small number of eigencomponents is typically sufficient to explain virtually the total variance (see, for instance, \citealpapp{ANH15_app} who use an essentially equivalent practical approach and successfully approximate an infinite dimensional functional time-series using a finite dimensional VAR-model).

To achieve a consistent classification, it is necessary that the threshold parameter $\tau_{nT}\to 0$ as $n,T\to\infty$ since the distances $\Delta^\ast_{ts}$ are null sequences. However, $\tau_{nT}$ converges so fast that $\tau_{nT}$ remains slightly larger than the maximum within-regime distance $\max_{s\in G_k}\hat \Delta_{ts}$. That is, we need to require that $\prob\big(\max_{s\in G_k}\hat \Delta_{ts} \leq \tau_{nT} \big)\to 1$ or equivalently that $\prob\big(\max_{s\in G_k}\hat \Delta_{ts} \geq \tau_{nT} \big)\to 0$ for any $t\in G_k$. For finite samples this means requiring that $\prob\big(\max_{s\in G_k}\hat \Delta_{ts} \geq \tau_{nT} \big)\leq \varepsilon$ for some small $\varepsilon>0$. Next we use the approximation $\hat \Delta_{ts}\approx \Delta^\ast_{ts}$. Observe that for a given $t\in G_k$,
%%%%%%%%%%%%%%%%%
\begin{align*}
\prob\left(\max_{s\in G_k} \Delta_{ts}^*\geq \tau_{nT}  \right) =
\prob\left(\bigcup_{s\in G_k}\left\{ \Delta_{ts}^*\geq \tau_{nT}\right\}  \right)
\leq |G_k| \prob\left(Q_{ts}^{\underline{m}} \geq \frac{n}{2}\tau_{nT} \right),
\end{align*}
%%%%%%%%%%%%%%%%
where the latter inequality follows from Boole's inequality. From this upper bound we can learn about $\tau_{nT}$ according to
%%%%%%%%%%%%%%%%
\begin{align*}
|G_k| \prob\left(Q_{ts}^{\underline{m}} \geq \frac{n}{2}\tau_{nT} \right)=\varepsilon \quad
\Leftrightarrow\quad \tau_{nT} = \frac{2}{n}F^{-1}_{\underline{m}}\left(1-\frac{\varepsilon}{|G_k|}\right),
\end{align*}
%%%%%%%%%%%%%%%
where ${F}_{\underline{m}}^{-1}$ denotes the quantile function of the $\chi_{\underline{m}}^2$-distribution. As we consider a context where $|G_k|$ is large ($|G_k| \propto T$ in Assumption A3), we expect the value of $\varepsilon/|G_k|$ to be very close to zero. This motivates setting $\tau_{nT}=(2/n)F^{-1}_{\underline{m}} (p_\tau)$, for some $p_\tau$ very close to one as mentioned in Section \ref{sec:THR}. Note that according to Theorem A in \citeapp{I10_app} and our assumptions in Section \ref{sec:ASY}  
%%%%%%%%%%%%%%%
\begin{align*}
\tau_{nT} = \frac{2}{n}F_{\underline{m}}^{-1}\left(1-\frac{\varepsilon}{|G_k|}\right)&\leq \frac{2\underline{m}}{n}+\frac{4}{n}\left(\log \left(\frac{|G_k|}{\varepsilon}\right)+\sqrt{\underline{m}\log \left(\frac{|G_k|}{\varepsilon}\right)}\right) \to 0  
\end{align*} 
%%%%%%%%%%%%%%%%
as $n,T\to\infty$, which points at the large sample validity of the proposed threshold.

%%%%%%%%%%%%%%%%%%%%%%%%%%%%%%%%%%%%%%%%%%%%%%%%%%%
\bibliographystyleapp{Chicago}
\bibliographyapp{bibfile}
%%%%%%%%%%%%%%%%%%%%%%%%%%%%%%%%%%%%%%%%%%%%%%%%%%%

\end{document}